\PassOptionsToPackage{colorlinks,citecolor=blue,linkcolor=blue,urlcolor=blue, breaklinks=true}{hyperref}
\documentclass[pdflatex,sn-mathphys-num,footinbib,twocolumn]{sn-jnl}

\usepackage{graphicx}%
\usepackage{multirow}%
\usepackage{amsmath,amssymb,amsfonts}%
\usepackage{amsthm}%
\usepackage{mathrsfs}%
\usepackage[title]{appendix}%
\usepackage{xcolor}%
\usepackage{textcomp}%
\usepackage{manyfoot}%
\usepackage{booktabs}%
\usepackage{algorithm}%
\usepackage{algorithmicx}%
\usepackage{algpseudocode}%
\usepackage{listings}%
\usepackage{bm}
\usepackage{caption}
\usepackage{cuted}

\usepackage[offset=3ex]{simpler-wick}

\begin{document}


\title[Classical theories of gravity produce entanglement]{Classical theories of gravity produce entanglement}

\author[1]{\fnm{Joseph} \sur{Aziz}}

\author[1]{\fnm{Richard} \sur{Howl}}

\affil[1]{\orgdiv{Department of Physics}, \orgname{Royal Holloway, University of London}, \orgaddress{
\city{Egham}, \postcode{TW20 0EX}, \state{Surrey}, \country{United Kingdom}}}

\abstract{The unification of gravity and quantum mechanics remains one of the most profound open questions in science. With recent advances in quantum technology,  an experimental idea first proposed by Richard Feynman is now regarded as  a promising route to testing this unification for the first time. The experiment involves placing a massive object  in a quantum superposition of two locations and letting it gravitationally interact with another mass.  In modern versions of the experiment, if the two  objects  subsequently become entangled, this is considered unambiguous evidence that gravity obeys the laws of quantum mechanics.  This conclusion derives from theorems that treat a classical gravitational interaction as a local interaction capable of only transmitting classical, not quantum, information. Here, we argue that the classical gravitational interaction can transmit quantum information, and thus generate entanglement through physically  local processes. The effects are found to scale differently to the considered quantum gravity effect, providing information on the form of the experiment required to  evidence the  quantum nature of gravity.}

\maketitle

\section{Introduction}

While the other fundamental interactions - electromagnetism, and the strong and weak forces - have been successfully married to quantum theory, the standard methods of quantization  appear to fail for gravity \cite{rovelli2001notesbriefhistoryquantum}.  This has motivated alternative approaches to the unification of gravity with quantum theory, including string theory, loop quantum gravity, and proposals that gravity is not quantized at all but remains fundamentally classical \cite{Carlip_2008}. A decisive factor in determining which route is correct has so far been lacking: experimental evidence. At the 1957 Chapel Hill conference, Feynman proposed a thought experiment that could reveal the quantum nature of gravity \cite{FeynmanQG}, an idea now becoming feasible through rapid progress in quantum experiments \cite{bose2017spin,marletto2017gravitationallyinduced}. In Feynman's  proposal, an object of Planck mass ($0.02\,\mathrm{mg}$) is placed in a quantum superposition of two locations  before interacting gravitationally with another mass \cite{FeynmanQG}. While Feynman’s exact measurement prescription for then determining the  quantum nature of gravity is unclear from the original   conference transcript \cite{FeynmanQG}, today this is considered as the observation of entanglement between the massive objects, with several theorems and arguments for  how physically realistic (local) classical theories of gravity can never create entanglement between the massive objects  \cite{kafri2013noise,kafri2014classical,krisnanda2017revealing,bose2017spin,marletto2017gravitationallyinduced,marletto2020witnessing,Galley2022nogotheoremnatureof,ludescher2025gravity}. The experiment was also long regarded as essentially unfeasible until two proposals \cite{bose2017spin,marletto2017gravitationallyinduced}, developed independently of Feynman's original idea, considered both masses to start  in a quantum superposition of two locations, with  gravitationally-induced entanglement sought between them. The theorems for entanglement evidencing quantum gravity rest on the assumption that theories of classical gravity can only involve  local operations and  exchanges of classical information. This is because non-local, action-at-a-distance processes are  considered unphysical, and it seems natural  that a classical gravitational interaction cannot transmit quantum information. Under this assumption, the interaction falls into a class of processes that,  according to quantum information theory, or generalizations \cite{marletto2020witnessing,Galley2022nogotheoremnatureof,ludescher2025gravity}, cannot create entanglement,  formalized as  local operations and classical communication (LOCC) in quantum information theory \cite{kafri2014classical,bose2017spin}.   
 
A substantial number of experimental proposals have been developed for witnessing this gravitationally-induced entanglement \cite{bose2023massivequantumsystemsinterfaces,huggett2022quantum,Carney_2019} and initial work on such experiments is  underway \cite{bose2023massivequantumsystemsinterfaces,CoolingMicrosolid,westphal2021measurement,panda2024measuring,Folman}. Due to the fundamental significance of the experiments, there have been several works on whether  entanglement can really evidence quantum gravity, often inspired by  discussions at the   Chapel Hill conference \cite{FeynmanQG}. These works have focused on whether  classical gravity could  act through non-local operations  violating the LO part of LOCC and thus allowing for the generation of entanglement (for a review, see  \cite{christodoulou2022gravity,huggett2022quantum} and  Appendix \ref{sec:nonlocalCG}). However, this goes against our  understanding of interactions in nature acting locally, be it electromagnetism, the Standard Model, or general relativity, and is thus generally ruled out on physical grounds \cite{christodoulou2022gravity,kafri2013noise,kafri2014classical,bose2017spin,marletto2019answers,marletto2017gravitationallyinduced}.

Assuming the LO part of LOCC on  physical grounds,   leaves the CC part. The idea that a classical theory of gravity should only involve classical communication through the gravitational interaction seems  natural and has not been questioned. However, here we argue that a classical gravity interaction can  generate quantum communication, and thus entanglement. The arguments and theorems for classical gravity only operating as LOCC   treat matter in standard quantum mechanics or generalized probabilistic extensions \cite{kafri2013noise,kafri2014classical,krisnanda2017revealing,bose2017spin,marletto2017gravitationallyinduced,marletto2020witnessing,Galley2022nogotheoremnatureof,ludescher2025gravity}. However, to the best of our knowledge, matter  obeys quantum field theory (QFT), and when this is fully taken into account, we argue that there can be classical gravity interaction processes that can in principle give rise to quantum communication.

\begin{figure}
    \centering
\includegraphics[width=\columnwidth]{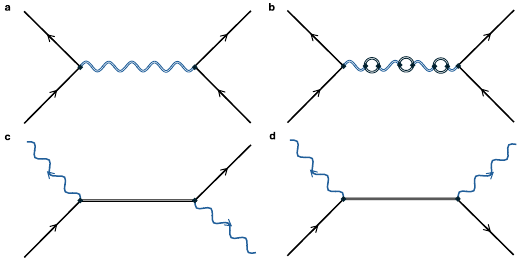}
\caption{Feynman diagrams for QED or linear quantum gravity. Wiggly blue lines  represent photons or gravitons; and black lines     represent electrons/positrons or general matter/antimatter particles. For ease of visualization, double lines  without arrows represent  virtual particles.}
\label{fig:Fig1}
\end{figure}

\section{Quantum electrodynamics}

To better understand how the classical gravity interaction can generate quantum communication, we first review quantum electrodynamics (QED). This is our best theory for how matter interacts through electromagnetism, and involves treating both matter and the electromagnetic field with QFT. The interaction Hamiltonian  is
\begin{align} \label{eq:HQED}
    \hat{H}^{QED}_{int} = \int d^3 \boldsymbol{x}\,  q \, \hat{A}_{\mu} (\boldsymbol{x}) \, \hat{\overline{\psi}}(\boldsymbol{x}) \,\gamma^{\mu}\, \hat{\psi} (\boldsymbol{x}),
\end{align}
where  $\hat{\psi} (\bm{x})$ is a charged fermionic field with $q$ its charge,   $\hat{\overline{\psi}} (\bm{x})$ is the Dirac adjoint field,  $\hat{A}_{\mu} (\boldsymbol{x})$ is the quantized 4-potential, and $\gamma^{\mu}$ are the gamma matrices. Calculations with QED are mostly performed perturbatively, where we can intuitively view interactions through Feynman diagrams. For example, Fig.\ \ref{fig:Fig1}\textcolor{blue}{a}, describes an interaction between two electrons at first-order in perturbation theory. In this diagram,  the electromagnetic interaction between the electrons is mediated by a virtual photon, which can be viewed as quantum communication \cite{kafri2013noise,Bose2022}.

However, a critical finding of QED compared to classical  electromagnetism is that the interaction processes need not be mediated by just the electromagnetic field, and thus virtual photons. For example, at higher order in the electron scattering process, there are diagrams such as Fig.\ \ref{fig:Fig1}\textcolor{blue}{b} where the virtual photon is accompanied by virtual electron particles. In fact, even at leading order there are processes that only involve virtual electron propagators, some of which are depicted in Figs \ref{fig:Fig1}\textcolor{blue}{c--d}. When viewed from a non-perturbative perspective, there is no way to separate virtual matter from virtual photons in QED, and all electromagnetic interactions should be viewed as a combination of matter and electromagnetic fields propagating the interaction \cite{Cao_1999}.

\begin{figure*}
    \centering
\includegraphics[width=1\textwidth]{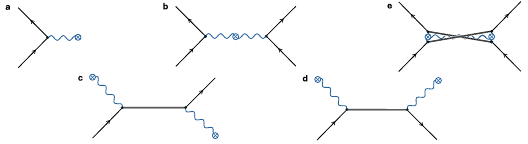}
 \caption{Feynman diagrams for QED with the approximation of classical electromagnetic fields, or linear classical gravity. The wiggly blue lines are classical electromagnetic or gravitational fields/potentials, with the crosses representing classical sources for the fields/potentials. As in Fig.\ \ref{fig:Fig1},  black lines represent electrons/positrons or general matter/antimatter particles; and for ease of visualization, double lines  without arrows represent  virtual particles.}
    \label{fig:Fig2}
\end{figure*}

\section{Perturbative quantum gravity}

While there is no consensus on a full theory of quantum gravity, at low energies it is widely accepted that such a theory should approximate perturbative quantum gravity \cite{Carney_2019,howl2021nongaussianity,LocallyHowl,wallace2022quantum,Donoghue:2017ovt}. This is an effective field theory that involves the quantization of linear general relativity with matter fields: the full spacetime metric $g_{\mu \nu}$ is broken up into $\eta_{\mu \nu} + h_{\mu \nu}$, with $\eta_{\mu \nu}$ the metric of a background classical spacetime  and $|h_{\mu \nu} | \ll 1$, which is quantized.   The interaction Hamiltonian for matter interacting with gravity  is
\begin{align} \label{eq:HQG}
    \hat{H}^{QG}_{int} = -\frac{1}{2}\int d^3 \boldsymbol{x}\,   \hat{h}^{\mu \nu}(\boldsymbol{x}) \, \hat{T}_{\mu \nu} (\boldsymbol{x}),
\end{align}
where $ \hat{T}_{\mu \nu} (\boldsymbol{x})$ is the quantized energy-momentum tensor for matter. For example, describing matter with a massive complex scalar field $\hat{\phi}$, $ \hat{T}_{\mu \nu} (\boldsymbol{x})$ reads
\begin{align}
    &\hat{T}_{\mu \nu} (\bm{x}) =  \hat{\mathcal{T}}_{\mu \nu} [\hat{\phi}^{\dagger} (\bm{x}) \hat{\phi} (\bm{x})]\\ \label{eq:Tmunu}
    &:=  \partial_{\{ \mu} \hat{\phi}^{\dagger}\partial_{\nu \} } \hat{\phi}  -   \eta_{\mu \nu}\partial_{\rho} \hat{\phi}^{\dagger} \partial^{\rho} \hat{\phi} -  \eta_{\mu \nu} \frac{m^2 c^2}{\hbar^2} \hat{\phi}^{\dagger}\hat{\phi},
\end{align}
where $m$ is the mass and $\partial_{\{ \mu} A \,\partial_{\nu \}} B := \partial_{\mu } A \,\partial_{\nu} B + \partial_{\nu } A\, \partial_{\mu} B$. The interaction Hamiltonian \eqref{eq:HQG} is of similar form to \eqref{eq:HQED}, and  perturbative quantum gravity parallels  QED closely \cite{christodoulou2022gravity}, with photons essentially  being replaced with gravitons. For example, for two matter particles interacting, at first order we  have a diagram such as Fig.\ \ref{fig:Fig1}\textcolor{blue}{a} but with a virtual graviton mediating the interaction rather than a virtual photon. Similarly, we also have all the other diagrams, Figs \ref{fig:Fig1}\textcolor{blue}{b--d}, where there are virtual matter and/or virtual graviton propagators, such as in Compton scattering \cite{Peskin:1995ev}.

\section{Perturbative classical gravity} \label{sec:PertCG}

In a  classical theory of gravity, the gravitational field is classical. Therefore, in the low-energy regime, most simply we would have an interaction Hamiltonian as \eqref{eq:HQG} but with  $h_{\mu \nu}$ not quantized:
\begin{align}\label{eq:HCQ}
    \hat{H}^{CG}_{int} = -\frac{1}{2}\int d^3 \boldsymbol{x}\,   h^{\mu \nu}(x) \, \hat{T}_{\mu \nu} (\boldsymbol{x}).
\end{align}
This is the interaction Hamiltonian of  
 QFT in linear curved spacetime, which, like any other QFT, is   a local theory \cite{Peskin:1995ev,dibiagio2023relativisticlocalityimplysubsystem,fewster2019algebraicquantumfieldtheory}.

The  interaction Hamiltonian \eqref{eq:HCQ} is analogous to that of an approximation sometimes performed in QED \cite{Peskin:1995ev,WeinbergVol1,mandl2013quantum,QFTClassSources}: for certain QED calculations, such as Rutherford scattering, a good approximation is to ignore the quantumness of the electromagnetic field such that we drop the hat from $A_{\mu}$ in \eqref{eq:HQED}. Feynman diagrams can  be drawn for this theory \cite{Peskin:1995ev}, where a cross is often used to denote  classical electromagnetic potentials and waves
or, equivalently, that there is a classical source for the electromagnetic field. For example, Fig.\ \ref{fig:Fig2}\textcolor{blue}{a} illustrates an electron interacting with a classical potential/wave, and Fig.\ \ref{fig:Fig2}\textcolor{blue}{b} depicts how two electrons would interact in this theory, where now the cross can be seen as breaking  quantum communication channels involving virtual photons in  QED.  However,  diagrams  are still present where virtual matter is propagating the interactions, such as Figs \ref{fig:Fig2}\textcolor{blue}{c--e}. 

Analogously, we can construct Feynman diagrams from the classical gravity interaction Hamiltonian \eqref{eq:HCQ},  where now wiggly lines with crosses in Fig.\ \ref{fig:Fig2} denote classical gravitational potentials and waves. However, just as in the QED approximation, although there are no gravitons, there are still virtual matter propagators (see e.g.\ Figs \ref{fig:Fig2}\textcolor{blue}{c--e}), and thus quantum communication. Then, since there is quantum communication, the classical gravity interaction can create  entanglement \cite{kafri2013noise}, getting round the theorems and arguments  discussed above \cite{bose2017spin,marletto2017gravitationallyinduced,marletto2020witnessing,Galley2022nogotheoremnatureof,krisnanda2017revealing,ludescher2025gravity}. This occurs because the theorems   take a more restrictive view  on what the  gravitational interaction consists of: they  consider that quantum  gravity  only involves virtual graviton propagators, but  at the field theory level, in general it can be argued that there will also  be  virtual matter propagators involved in this interaction. 

\begin{figure}
   \centering
\includegraphics[width=0.35\textwidth]{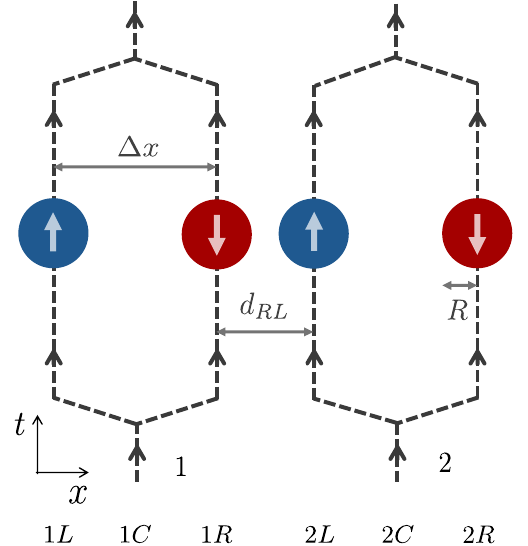}
\caption{Visualization of a version of Feynman's experiment. Two spherical mass distributions (1 and 2) of radius $R$ are placed in quantum superpositions of two locations as N00N states,  with blue and red denoting the   components separated by $\Delta x$. After gravitationally interacting for a short time, the paths are recombined  and entanglement is sought \cite{bose2017spin,marletto2017gravitationallyinduced}. While Stern-Gerlach interferometry with  internal  spins is illustrated \cite{bose2017spin}, alternative setups, such as parallel Mach-Zenhders are also possible \cite{marletto2017gravitationallyinduced}.  Here,  $\Delta x$ is depicted larger than the minimum separation $d_{RL}$, but a general configuration can be implemented, including    $\Delta x \ll d_{RL}$.}
   \label{fig:Fig3}
\end{figure}

\section{Experiment} \label{sec:exp}

To illustrate this further,  we now demonstrate how \eqref{eq:HCQ} can lead to entanglement in a simple version of the experiment ideology first introduced by Feynman. Two spherical mass distributions, each with total mass $M$ and radius $R$  are prepared in a quantum superposition of two  locations \cite{marletto2017gravitationallyinduced,bose2017spin}. This could be achieved by, for example, implementing matter-wave beam splitters \cite{marletto2017gravitationallyinduced}, manipulating  potentials \cite{Howl_2019} or exploiting internal  degrees of freedom, such as quantum spins  in Stern-Gerlach experiments \cite{bose2017spin} -- see  Fig.\ \ref{fig:Fig3}. Gravity is assumed the only interaction between  the particles,  and when quantized in the non-relativistic limit, and  describing matter  within first quantization, just acts a quantum phase $\varphi_{ij} := G M^2 t / (\hbar \, d_{ij})$ on each superposition branch \cite{bose2017spin,marletto2017gravitationallyinduced}, where $d_{ij}$ is the distance between the matter distributions in the branch labelled by $i,j \in \{L,R\}$, and $G M^2 / d_{ij}$ is  the Newtonian  potential energy. With the superposition size $\Delta x$ much greater than the smallest distance $d_{RL}$, only the quantum phase $\varphi := \varphi_{RL}$ is significant  such that the systems are clearly entangled, with  entanglement depending solely on $\varphi$ \cite{bose2017spin,marletto2017gravitationallyinduced}. In contrast, when $\Delta x \ll d_{RL}$, the relevant parameter for entanglement becomes essentially   $\overline{\varphi} := \varphi \, \Delta x^2 / d_{RL}^2$ \cite{Aspelmeyer2022}. To measure the entanglement, the superposed paths could be recombined and correlations  sought between the interferometer outputs \cite{marletto2017gravitationallyinduced} or  internal degrees of freedom \cite{bose2017spin}.

\subsection{Quantum gravity} \label{sec:QG}  

We now analyse this experiment using perturbative  quantum gravity with a QFT description of matter.  We treat perturbative quantum gravity as an effective quantum field theory, valid at low energies, and describe matter with a massive complex scalar field for simplicity.\footnote{For a study considering how   massless rather than massive scalar fields for matter (in this case describing light) interact in  linearized gravity, see e.g.\ \cite{lapponi2025gravitationalredshiftquantizedlinear}.} The full Hamiltonian of the system is then  written as
\begin{align}
    \hat{H} = \hat{H}_0 + \hat{H}^{QG}_{int},
\end{align}
where $\hat{H}_0 := \hat{H}_0^{M} +\hat{H}_0^{G} $, with  $\hat{H}_0^{M}$ and $\hat{H}_0^{G}$ representing, respectively, the free Hamiltonians of the matter  and gravitational fields - see e.g.\ \cite{Peskin:1995ev,QGLorentzGupta1,QGLorentzGupta2} - and $\hat{H}^{QG}_{int}$ describes the interaction between these two fields - see \eqref{eq:HQG}. As above,  gravity is assumed the dominating interaction between the particles. 
The evolution of the quantum state of the  system, in the Schr\"{o}dinger picture,  is then
\begin{align} \label{eq:Psit}
|\Psi(t) \rangle = \hat{U}_0 (t)\hat{U}_I (t)|\Psi\rangle,
\end{align}
where $|\Psi\rangle$ is the initial state of the system and
\begin{align} \label{eq:U0} 
    \hat{U}_0 (t)&:= e^{-i \hat{H}_0 t / \hbar},\\ \label{eq:Dyson} \hat{U}_I (t)&:= \hat{T} e^{- i \int^{t}_{0} d\tau \, \hat{H}_{I} (\tau) / \hbar },
\end{align}
with $\hat{T}$  the time-ordering operator and $\hat{H}_{I}$  the interaction Hamiltonian in the interaction picture: 
\begin{align} \label{eq:HIRel}
 \hat{H}_{I} := \hat{U}_0^{\dagger} \hat{H}^{QG}_{int} \hat{U}_0 =  
 -\frac{1}{2} \int d^3 \bm{x} \, \hat{h}^{\mu \nu} (x) \hat{T}_{\mu \nu} (x).
\end{align}
Here, $\hat{h}_{\mu \nu} (x)$ is the free gravitational  field operator in the Heisenberg picture, and  $\hat{T}^{\mu \nu}(x)$ is as \eqref{eq:Tmunu} but with  the free matter field $\hat{\phi}(x)$ in the Heisenberg picture:
\begin{align} \label{eq:phifree}
    \hat{\phi}(x) &:= c \sqrt{\hbar} \int \frac{d^3 \bm{k}}{(2 \pi)^3} \frac{1}{\sqrt{2 \omega_{\bm{k}}}} \left( \hat{a}_{\bm{k}} e^{i k.x} + \hat{b}^{\dagger}_{\bm{k}} e^{-i k.x}\right)\\
    &=: \hat{\phi}^{(+)}(x) + \hat{\phi}^{(-)}(x),
\end{align}
with $\hat{\phi}^{(+)}(x) := c \sqrt{\hbar} \int d^3 \bm{k}\,   \hat{a}_{\bm{k}} e^{i k.x}/ ((2 \pi)^3 \sqrt{2 \omega_{\bm{k}}})$ the positive frequency component of the field; $\hat{\phi}^{(-)}(x) := c \sqrt{\hbar} \int d^3 \bm{k} \,  \hat{b}^{\dagger}_{\bm{k}} e^{-i k.x}/ ((2 \pi)^3 \sqrt{2 \omega_{\bm{k}}})$ the negative frequency component;  $k.x := k^{\mu} x_{\mu}$;  $x^0 = c t$; and $\hat{a}_{\bm{k}}$ and $\hat{b}_{\bm{k}}$ are the annihilation operators for matter and antimatter particles respectively, such that $[\hat{a}_{\bm{k}}, \hat{a}^{\dagger}_{\bm{k}'}] = (2 \pi)^3 \delta^{(3)}(\bm{k} - \bm{k'})$, $[\hat{b}_{\bm{k}}, \hat{b}^{\dagger}_{\bm{k}'}] = (2 \pi)^3 \delta^{(3)}(\bm{k} - \bm{k'})$, and  $[\hat{a}_{\bm{k}}, \hat{b}^{\dagger}_{\bm{k}'}] = 0$. We assume the $(-,+,+,+)$ metric signature.

The unitary operation $\hat{U}_I$ can be expanded as the Dyson series
\begin{align}\nonumber
\hat{U}_I (t) &= 1 -  \frac{i}{\hbar} \int^{t}_{0} d\tau \, \hat{H}_{I} (\tau)\\ \nonumber  &- \frac{1}{2\hbar^2} \hat{T} \int^{t}_{0} d\tau  \, d\tau'\, \hat{H}_{I} (\tau)\, \hat{H}_{I} (\tau') + \cdots.
\end{align}

We take the initial state of the objects $|\Psi\rangle$ immediately after being placed in a quantum superposition as a product of N00N  states:
\begin{align}\nonumber
    |\Psi\rangle &= \frac{1}{2}  \left(|N\rangle_{1L} |0\rangle_{1R}| \uparrow\rangle_1 + |0\rangle_{1L} |N\rangle_{1R} | \downarrow\rangle_1\right) 
   \\ \label{eq:PsiInitial} &\otimes  \left( |N\rangle_{1L} |0\rangle_{2R}  |\uparrow\rangle_2 + |0\rangle_{1L} |N\rangle_{2R} |\downarrow\rangle_2\right),
\end{align}
where we have included spin degrees of freedom that could be used to generate the spatial superpositions as in \cite{bose2017spin} - in this case the above state would be that of the system just after Stern-Gerlach experiments - see Fig.\ \ref{fig:Fig3}. In writing $|\Psi\rangle$, we have ignored possible configuration states of the gravitational field and the anti-matter particles since these are not important to the following discussion. The state $|N\rangle_{\kappa i}$, with $\kappa \in \{1,2\}$ and $i \in \{L,R\}$, is an N-particle independent position state \cite{pavvsivc2018localized}, which is  defined as
\begin{align}\label{eq:Nstate}
    |N\rangle_{\kappa i} := \frac{1}{\sqrt{N!}} \int \prod^N_{j=1} d^3 \bm{x}_j \tilde{\phi}_{\kappa i}(\bm{x}_j) |\bm{x}_j \rangle, 
\end{align}
where $|\bm{x}_j\rangle$ is a single-particle position state defined below; $\tilde{\phi}_{\kappa i}(\bm{x}):= \theta_{\kappa i}(\bm{x})/\sqrt{V}$, with $\theta_{\kappa i}(\bm{x}) := \theta (R - |\bm{x} - \bm{X}_{\kappa i}|)$; $R$ is the radius of the matter spheres in the experiment; $V = 4 \pi R^3 / 3$; and $\bm{X}_{\kappa i}$ is the centre-of-mass coordinate for the sphere $\kappa$ in branch $i$.  This definition of the initial matter state follows from how wavepackets are used to model  particles in QFT:  in QFT single-particle wavepackets are defined, in general as \cite{Peskin:1995ev}:
\begin{align} \label{eq:wavepacket}
    |\psi\rangle =  \int  \frac{d^3 \bm{k}}{(2 \pi)^3} \frac{1}{\sqrt{2 \omega_{\bm{k}}}} \phi(\bm{k}) |\bm{k}\rangle,
\end{align}
where $|\bm{k}\rangle := \sqrt{2 \omega_{\bm{k}}} \hat{a}^{\dagger}_{\bm{k}} |0\rangle$ and $\phi(\bm{k})$  is the Fourier transform of the spatial wavefunction $
    \phi(\bm{k}) := \int d^3 \bm{x} \tilde{\phi}(\bm{x}) e^{-i\bm{k}.\bm{x}}$, with $\int d^3 \bm{x} |\tilde{\phi}(\bm{x})|^2 = 1$. For a localized particle, its wavepacket $|\bm{x}\rangle$ is then
\begin{align} \label{eq:x}
    |\bm{x} \rangle = \int \frac{d^3 \bm{k}}{(2 \pi)^3} \frac{1}{\sqrt{2 \omega_{\bm{k}}}} e^{-i\bm{k}.\bm{x}} |\bm{k} \rangle,
\end{align}
and so $\langle \bm{y} |\bm{x} \rangle = \delta^{(3)}(\bm{x}-\bm{y})$. Using this,  we  define an N-particle wavepacket  as \eqref{eq:Nstate} with $\tilde{\phi}_{\kappa i}(\bm{x})$ being the general spatial wavefunction.  Taking the atoms as part of a spherical object, then  $ \tilde{\phi}_{\kappa i}(\bm{x}) := \theta_{\kappa i} (\bm{x})/\sqrt{V}$ as described above, where $\int d^3 \bm{x} |\tilde{\phi}_{\kappa i}(\bm{x})|^2 = 1 $ and also $  \,_{\kappa i}{\langle}N| N \rangle_{\lambda j} = 0 $ when the different spheres are non-overlapping $d_{ij} := |\bm{X}_{\kappa i} - \bm{X}_{\lambda j}| > 2 R$, with $\kappa j \neq \lambda j$, which is satisfied in the experiment. We, therefore, have ``orthonormal'' initial states, and such states will remain orthonormal under unitary evolution.  

After the matter systems have interacted through gravity, for our considered experiment  we will be interested in   states of the form \cite{bose2017spin,fn6}:
\begin{align}\nonumber
    |\Psi(t)\rangle = \frac{1}{\mathcal{N}}  \Bigg(&\alpha_{LL} |N\rangle_{1L}  |N\rangle_{2L} |\uparrow\rangle_1 |\uparrow\rangle_2  \\ \nonumber &+  \alpha_{LR} |N\rangle_{1L} |N\rangle_{2R} |\uparrow\rangle_1 |\downarrow\rangle_2 \\  \nonumber 
    &+ \alpha_{RL} |N\rangle_{1R} |N\rangle_{2L} |\downarrow\rangle_1 |\uparrow\rangle_2 \\  \label{eq:stateAfterG} &+ \alpha_{RR} |N\rangle_{1R} |N\rangle_{2R} |\downarrow\rangle_1 |\downarrow\rangle_2\Bigg),
\end{align}
where $\mathcal{N}$ is the normalization constant and we have ignored vacuum states  for simplicity. This is just the second-quantized, N00N state version of the final state considered in  modern interpretations of Feynman's experiment, see e.g.\ \cite{bose2017spin,marletto2017gravitationallyinduced}.

\begin{figure*}
    \centering
\includegraphics[width=0.5\textwidth]{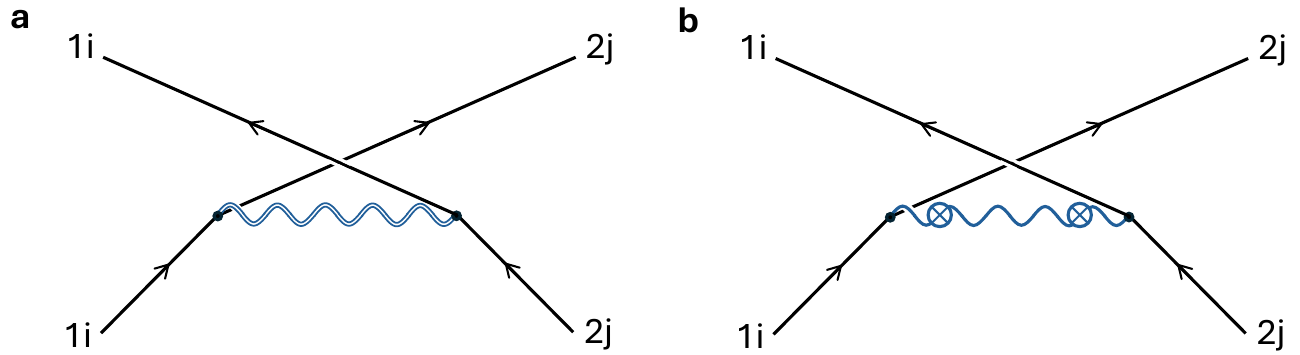}
    \caption{a) Feynman diagram corresponding to Wick contraction \eqref{eq:WickCrissCross}. The $1i$ and $2j$ label the first and second objects, with $i,j \in \{L,R\}$. b) The corresponding diagram when there is a classical gravity interaction (the two circles with crosses indicate the two classical sources of gravity, i.e.\ the two matter objects). The amplitudes of both diagrams are found to be vanishing. In contrast to standard perturbative QFT  diagrams, the external legs here represent position-like states rather than definite momentum states, as detailed in the main  text, with the arrows indicating time evolution.     }
    \label{fig:CrissCross}
\end{figure*}

After the interaction, the matter systems are brought back together and interfered \cite{bose2017spin,marletto2017gravitationallyinduced}. For example, in the case of embedded spins,  reverse Stern-Gerlach devices could in principle be used \cite{bose2017spin} such that the above state becomes 
\begin{align}\nonumber
    |\Psi(t)\rangle = \frac{1}{\mathcal{N}}   &\Big(\alpha_{LL}  |\uparrow\rangle_1 |\uparrow\rangle_2  +  \alpha_{LR}  |\uparrow\rangle_1 |\downarrow\rangle_2 \\\nonumber &+ \alpha_{RL}  |\downarrow\rangle_1 |\uparrow\rangle_2  + \alpha_{RR} |\downarrow\rangle_1 |\downarrow\rangle_2\Big)\\ \label{eq:psiAfterSG}&\hspace{0.5cm}\otimes |N\rangle_{1C} |N\rangle_{2C},
\end{align}
where $\kappa C$ is the position of the matter object $\kappa$ \cite{bose2017spin}.    As described in the  proposal \cite{bose2017spin}, the spins of the particles can be measured to determine whether the matter systems are entangled. This entanglement for the state of interest is   dependent on the values that the amplitudes  $\alpha_{LL}, \alpha_{LR}, \alpha_{RL}$ and $\alpha_{RR}$ take, and since the amplitudes in \eqref{eq:psiAfterSG} are the same as in \eqref{eq:stateAfterG}, we can just use \eqref{eq:stateAfterG} in calculating them. We do this through the standard perturbative QFT technique of acting the expected final states on the evolved initial states, such that
\begin{align} \nonumber
    \alpha_{i j}  &= \, _{1i}{\langle}N|\, _{2j}{\langle} N| \hat{U}_0 \hat{U}_I |\Psi\rangle \\  \label{eq:U0UI}  & = \, _{1i}{\langle}N|\, _{2j}{\langle} N| \hat{U}_0 \hat{U}_I |N\rangle_{1i} |N\rangle_{2j},
\end{align}
where we have used the orthonormality property of $|N\rangle_{\kappa i}$. As we show  below, see \eqref{eq:globalphase},  the unitary operator $\hat{U}_0$ just acts the same phase on each superposition branch so that it only provides a global phase on the full state. This is due to the fact that, within the approximations of the experiment, the  objects  stay fixed under free evolution. 
We can, therefore, ignore the action of $\hat{U}_0$ and just concentrate on $\hat{U}_I$, which matches previous works where  the free evolution is also ignored \cite{bose2017spin}. In Appendix \ref{app:Gaussian}, we go beyond these approximations and consider the gravitational interaction when free evolution of the wavepackets is also taken into account. Considering that $\hat{U}_I$ can be expanded in the Dyson series above, we write $\alpha_{ij}$ as $\alpha_{ij} = \alpha^{(0)}_{ij} + \alpha^{(1)}_{ij}+\alpha^{(2)}_{ij} + \cdots $, where $\alpha^{(n)}_{ij}$ (with $n \in \{0,1,2,\ldots\}$) corresponds to the particular order of the Dyson series. From \eqref{eq:Dyson}, at zeroth order there is just unity and so $\alpha^{(0)}_{ij} =1$ for all $i,j$, which is analogous to  the trivial part of the usual S-matrix \cite{Peskin:1995ev}. At first order in \eqref{eq:Dyson}, there are no corresponding Feynman diagrams that involve virtual gravitons such that we can ignore this order,  as detailed in Appendix \ref{app:MoreQG}. However, at second order, we have
\begin{strip}
\begin{align}
    \alpha^{(2)}_{i j}  &= -\frac{1}{2 \hbar^2}  \, _{1i}{\langle}N|\, _{2j}{\langle} N| \hat{T} \int^t_0 d \tau d \tau' \hat{H}_I(\tau) \hat{H}_I(\tau') |N\rangle_{1i} |N\rangle_{2j},\\ \nonumber
    &=-\frac{1}{8 \hbar^2 c^2}  \, _{1i}{\langle}N|\, _{2j}{\langle} N| \hat{T} \int_t d^4 x \, d^4 y \, \hat{h}^{\mu \nu} (x) \hat{T}_{\mu \nu} (x) \\ \label{eq:alphaQG} 
    &\times \hat{h}^{\rho \sigma} (y) \hat{T}_{\rho \sigma} (y)   |N\rangle_{1i} |N\rangle_{2j},
\end{align}
where $\int_t d^4 x := \int^{ct}_0 d x^0 \int d^3 \bm{x}$. The above can be computed using Wick contractions as per standard QFT \cite{Peskin:1995ev}. The relevant contractions are of the form: 
\begin{align} 
 \nonumber
    &\gamma_{ij}^{(2)} :=  -\frac{1}{4 \hbar^2 c^2}   \int_t d^4 x \int_t d^4  y \, \times   \\ \label{eq:alphaQGCont}  
    &\,_{1i}{\langle} 
    \wick{\c1 N| \, \,_{2j}{\langle} \c2 N| \hat{\mathcal{T}}_{\mu \nu}[\c1{\hat{\phi}}^{\dagger} (x) \c4{\hat{\phi}}(x)]  \hat{\mathcal{T}}_{\rho \sigma}[ \c2{\hat{\phi}}^{\dagger} (y) \c5{\hat{\phi}}(y) ] \c3{\hat{h}}^{\mu \nu}(x)  \c3{\hat{h}}^{\rho \sigma}(y)   |\c4 N \rangle_{1i} \,| \c5 N \rangle_{2j}},
\end{align}
\end{strip}
This contraction  corresponds to the Feynman diagram \ref{fig:Fig1}\textcolor{blue}{a}. All other Wick contractions either correspond to unconnected bubble diagrams, which we can ignore \cite{Peskin:1995ev}, or do not contribute to entanglement. Of note is the contraction corresponding to Feynman diagram \ref{fig:CrissCross}\textcolor{blue}{a}, which  gives a vanishing contribution in the approximation we are working - see Appendix \ref{app:MoreQG}. Note that, although we refer to `Feynman diagrams' here, they should be seen more as a visualization of the process in an analogous way to standard Feynman diagrams rather than strictly proper Feynman diagrams. This is because we are assuming position-like in and out states - Equation \eqref{eq:Nstate} - and so the momentum of the external legs can be zero. The arrow on the external legs should then be considered more as representing a flow in time  than a flow in space.

In \eqref{eq:alphaQGCont}, the contraction of the gravitational fields is the graviton Feynman propagator. In the Lorenz gauge, this is \cite{GWBook,PhysRev.162.1195}:
\begin{align} \nonumber
    \wick{ \c1{\hat{h}}^{\mu \nu}(x)  \c1{\hat{h}}^{\rho \sigma}(y)} &= \frac{16 \pi G \hbar}{ c^3} \left(\eta^{\mu \rho} \eta^{\nu \sigma} + \eta^{\mu \sigma} \eta^{\nu \rho} - \eta^{\mu \nu} \eta^{\rho \sigma}\right)\\ \label{eq:Wickh}
    &\times \int \frac{d^4 k}{(2\pi)^4} \frac{-i}{k^2 - i \epsilon} e^{i k. (x - y)},
\end{align}
which leaves the contraction of the matter field on our ``in'' and ``out'' states. Following the usual definition of contracting matter fields on in and out momentum states \cite{Peskin:1995ev},  the contraction on position states is:
\begin{align}
    &\wick{  \c{\hat{\phi}}(x)  |\c N \rangle_{\kappa i}} = \frac{1}{\sqrt{N!}} \int \prod^N_{j=1} d^3 \bm{x}_j \tilde{\phi}_{\kappa i}(\bm{x}_j) 
  \hat{\phi}^{(+)} (x) | \bm{x}_j \rangle\\
  &= c \sqrt{\hbar} \frac{1}{\sqrt{N!}} \int \prod^N_{j=1} d^3 \bm{x}_j \tilde{\phi}_{\kappa i}(\bm{x}_j)  \int \frac{d^3 \bm{k}}{(2 \pi)^3} \frac{1}{\sqrt{2 \omega_{\bm{k}}}} e^{i k.x}  
 \hat{a}_{\bm{k}} | \bm{x}_j \rangle\\ \label{eq:NwickRel}
 &= c \sqrt{\hbar} \sqrt{N} \int d^3 \bm{y}\, \tilde{\phi}_{\kappa i}(\bm{y})  \int \frac{d^3 \bm{k}}{(2 \pi)^3} \frac{1}{\sqrt{2 \omega_{\bm{k}}}} e^{i k_0 x^0} e^{i \bm{k}.(\bm{y} - \bm{x})}  
 |N-1\rangle_{\kappa i} \\ \label{eq:NonRelArgument}
  &= c \sqrt{\hbar} \sqrt{N}  \int \frac{d^3 \bm{k}}{(2 \pi)^3} \frac{1}{\sqrt{2 \omega_{\bm{k}}}} e^{i k. x} \tilde{\phi}_{\kappa i} (\bm{k}) |N-1 \rangle_{\kappa i},
\end{align}
where $\tilde{\phi}_{\kappa i}(\bm{k}) := \int d^3 \bm{x} e^{- \bm{k}.\bm{x}} \tilde{\phi}_{\kappa i}(\bm{x})$ is the Fourier transform of  $\tilde{\phi}_{\kappa i}(\bm{x})$. Given that $\tilde{\phi}_{\kappa i}(\bm{x}):= \theta_{\kappa i}(\bm{x})/\sqrt{V}$, its Fourier transform is $  \tilde{\phi}_{\kappa i}(\bm{k}) = 4 \pi \left(\sin(|\bm{k}| R) - |\bm{k}| R \cos(|\bm{k}| R) \right) \exp(-i\bm{k}.\bm{X}_{\kappa i}) / (\sqrt{V} |\bm{k}|^3)$. As long as $R \gg \hbar / (m c) $, which we would expect  in a realistic experiment, then $\tilde{\phi}_{\kappa i}(\bm{k})$ rapidly drops off as $|\bm{k}|$ increases and is approximately zero before $|\bm{k}|$ gets close to $m c / \hbar$. This all follows from the fact that the in and out states we have chosen, \eqref{eq:Nstate}, are non-relativistic as long as $R \gg \hbar / (m c) $. In this non-relativistic  approximation, since $\tilde{\phi}_{\kappa i}(\bm{k})$ is almost vanishing before  $|\bm{k}| \approx m c / \hbar$, we can approximate $\omega_{k}$ in \eqref{eq:NwickRel} with $m c^2 / \hbar$, which follows the usual non-relativistic definition of in and out momentum states used in standard perturbative  QFT calculations, such as the derivation of the Coulomb potential \cite{Peskin:1995ev}. This then results in
\begin{align} \label{eq:WickN}
    \wick{  \c{\hat{\phi}}(x)  |\c N \rangle_{\kappa i}} &\approx \frac{\hbar}{\sqrt{2m}} \sqrt{N}  e^{-i m c x^0 / \hbar} \tilde{\phi}_{\kappa i}(\bm{x})  |N-1 \rangle_{\kappa i},
 \end{align}
 \begin{strip}
 -----------------------------------------------------------------------------\\
 where the factor $e^{-i m c x^0 / \hbar} \tilde{\phi}_{\kappa i}(\bm{x})$ comes from the fact that we have  essentially assumed stationary, single-particle matter waves for the in and out matter states \cite{fn10}.   In the same approximation, the free unitary operator $\hat{U}_0(t) = \hat{U}_{ 0}^{ G }(t)\exp(-i \int d^3 \bm{k}  \, \omega_k \hat{a}^{\dagger}_{\bm{k}} \hat{a}_{\bm{k}} t / (2 \pi)^3 ) $ in \eqref{eq:U0UI} acts on the final state a global phase $2 M c^2 t / \hbar$, where $\hat{U}_{ 0}^{ G } (t)$ is  free evolution associated with the gravitational field \cite{QGLorentzGupta1,QGLorentzGupta2}:
{\small
\begin{align}\nonumber
    &\int \frac{d^3 \bm{p}}{(2 \pi)^3} \hbar \omega_{\bm{p}} \hat{a}^{\dagger}_{\bm{p}} \hat{a}_{\bm{p}} |N\rangle_{1i} |N\rangle_{2j} 
    \\\nonumber&= 
     \frac{1}{N!} \int \frac{d^3 \bm{p}}{(2 \pi)^3} \prod^N_{s} d^3 \bm{x}_s \, d^3 \bm{y}_s \tilde{\phi}_{1i} (\bm{x}_s)  \tilde{\phi}_{2j} (\bm{y}_s)  \hbar \omega_{\bm{p}} \hat{a}^{\dagger}_{\bm{p}} \hat{a}_{\bm{p}} |\bm{x}_s\rangle |\bm{y}_s\rangle
    \\\nonumber
    &=\frac{1}{N!} \int  \prod^N_{s} d^3 \bm{x}_s \, d^3 \bm{y}_s  \frac{d^3 \bm{k}_s \,d^3 \bm{q}_s \, d^3 \bm{p}}{(2 \pi)^{3(1+ 2s)}} \tilde{\phi}_{1i} (\bm{x}_s)  \tilde{\phi}_{2j} (\bm{y}_s) e^{-i \bm{k}_s.\bm{x}_s} e^{-i \bm{q}_s.\bm{y}_s} \hbar \omega_{\bm{p}} \hat{a}^{\dagger}_{\bm{p}} \hat{a}_{\bm{p}} \hat{a}_{\bm{k}_s}^{\dagger} \hat{a}_{\bm{q}_s}^{\dagger} |0\rangle 
    \\\nonumber&= \frac{1}{N!} \int \left(\prod^N_{s} d^3 \bm{x}_s \, d^3 \bm{y}_s \frac{d^3 \bm{k}_s \,d^3 \bm{q}_s }{(2 \pi)^{6s}} \tilde{\phi}_{1i} (\bm{x}_s)  \tilde{\phi}_{2j} (\bm{y}_s) e^{-i \bm{k}_s.\bm{x}_s} e^{-i \bm{q}_s.\bm{y}_s} \hat{a}_{\bm{k}_s}^{\dagger} \hat{a}_{\bm{q}_s}^{\dagger}\right) \sum_{t}^N \hbar (\omega_{\bm{k}_t} + \omega_{\bm{q}_t})    |0\rangle
    \\\nonumber&= \frac{1}{N!}    \int \frac{d^3 \bm{k}}{(2 \pi)^3} \hbar \omega_{\bm{k}} \hat{a}_{\bm{k}}^{\dagger} \sum^N_{n}   \left( \prod^{N-1}_{s\neq n} \prod^N_{t} \int d^3 \bm{x}_n \tilde{\phi}_{1i}(\bm{x}_n) e^{-i \bm{k}.\bm{x}_n} + \prod^{N}_{s} \prod^{N-1}_{t\neq n}\int d^3 \bm{y}_n \tilde{\phi}_{2j}(\bm{y}_n) e^{-i \bm{k}.\bm{y}_n}\right)\\\nonumber
    &\hspace{8cm}\times
    \int d^3 \bm{x}_s\, d^3 \bm{y}_t \,\tilde{\phi}_{1i} (\bm{x}_s)  \tilde{\phi}_{2j} (\bm{y}_t)
    |\bm{x}_s\rangle |\bm{y}_t\rangle 
    \\\nonumber&= \frac{N}{N!}    \int \frac{d^3 \bm{k}}{(2 \pi)^3} \hbar \omega_{\bm{k}} \hat{a}_{\bm{k}}^{\dagger}   \left( \prod^{N-1}_{s\neq n} \prod^N_{t}  \tilde{\phi}_{1i}(\bm{k})  + \prod^{N}_{s} \prod^{N-1}_{t\neq n} \tilde{\phi}_{2j}(\bm{k}) \right)
    \int d^3 \bm{x}_s\, d^3 \bm{y}_t \,\tilde{\phi}_{1i} (\bm{x}_s)  \tilde{\phi}_{2j} (\bm{y}_t)
    |\bm{x}_s\rangle |\bm{y}_t\rangle
\\\nonumber&\approx \frac{N m c^2}{N!}    \int \frac{d^3 \bm{k}}{(2 \pi)^3} \hat{a}_{\bm{k}}^{\dagger}   \left( \prod^{N-1}_{s\neq n} \prod^N_{t}  \tilde{\phi}_{1i}(\bm{k})  + \prod^{N}_{s} \prod^{N-1}_{t\neq n} \tilde{\phi}_{2j}(\bm{k}) \right)
    \int d^3 \bm{x}_s\, d^3 \bm{y}_t \,\tilde{\phi}_{1i} (\bm{x}_s)  \tilde{\phi}_{2j} (\bm{y}_t)
    |\bm{x}_s\rangle |\bm{y}_t\rangle
\\\nonumber&= \frac{N m c^2}{N!}    \int \frac{d^3 \bm{k}}{(2 \pi)^3} \hat{a}_{\bm{k}}^{\dagger} \sum^N_{n}   \left( \prod^{N-1}_{s\neq n} \prod^N_{t} \int d^3 \bm{x}_n \tilde{\phi}_{1i}(\bm{x}_n) e^{-i \bm{k}.\bm{x}_n} + \prod^{N}_{s} \prod^{N-1}_{t\neq n}\int d^3 \bm{y}_n \tilde{\phi}_{2j}(\bm{y}_n) e^{-i \bm{k}.\bm{y}_n}\right)\\\nonumber
    &\hspace{8cm}\times
    \int d^3 \bm{x}_s\, d^3 \bm{y}_t \,\tilde{\phi}_{1i} (\bm{x}_s)  \tilde{\phi}_{2j} (\bm{y}_t)
    |\bm{x}_s\rangle |\bm{y}_t\rangle
 \\\nonumber&= \frac{2 M c^2}{N!}   \prod^N_{s} \int d^3 \bm{x}_s  d^3 \bm{y}_s  \tilde{\phi}_{1i} (\bm{x}_s)  \tilde{\phi}_{2j} (\bm{y}_s)  |\bm{x}_s\rangle  |\bm{y}_s\rangle 
 \\\nonumber&= 2 M c^2  |N\rangle_{1i} |N\rangle_{2j}  \\ \label{eq:globalphase}
 &\implies \exp(-i \int d^3 \bm{k}  \, \omega_k \hat{a}^{\dagger}_{\bm{k}} \hat{a}_{\bm{k}} t / (2 \pi)^3 ) |N\rangle_{1i} |N\rangle_{2j} \approx e^{2 M i c^2 t / \hbar} |N\rangle_{1i} |N\rangle_{2j}.
\end{align} 
}
\end{strip}
 Since the free evolution by itself just contributes approximately a global phase, it can be ignored.    That is, within the approximations of the assumed experiment, the  objects  stay fixed under free evolution - the two matter objects $|N\rangle_{1i}$ and $|N\rangle_{2j}$ are prepared spacelike separated  and are held approximately fixed in position modes for the assumed duration of the experiment \cite{fn15}.

 In addition to the simple contraction  $\wick{  \c{\hat{\phi}}(x)  |\c N \rangle_{\kappa i}}$, we also have contractions of derivatives of the field coming from the energy-momentum tensor operator $\hat{\mathcal{T}}_{\mu \nu}$. For example,  $ \partial_{\mu}  \wick{  \c{\hat{\phi}}(x)  |\c N \rangle_{\kappa i}}$. However, given the above non-relativistic approximation, the only relevant terms in this case are those where the derivative is in the time coordinate:   $ \partial_{0}  \wick{  \c{\hat{\phi}}(x)  |\c N \rangle_{\kappa i}}$, which is just the time derivative of the right-hand side of \eqref{eq:WickN} in our non-relativistic approximation. 

 With the above contraction \eqref{eq:WickN}, we are operating in a low-energy regime suitable to adequately describe the experiment, which also  justifies the use of linearized quantum gravity as an effective field theory. Using this contraction, \eqref{eq:WickN},  with the gravitational contraction \eqref{eq:Wickh}, we can now  compute the amplitude $\gamma_{ij}^{(2)}$. This leaves us with  
\begin{align} \nonumber
\gamma^{(2)}_{ij} &= \frac{4 i \pi G M^2}{ \hbar c} \int_t d^4 x \int_t d^4 y  \\ \label{eq:alphaQG2} &\times  \int \frac{d^4 k}{(2 \pi)^4 } \frac{1}{k^2 - i \epsilon} e^{ik.(x-y)} \, |\tilde{\phi}_{1i} (\bm{x})|^2 |\tilde{\phi}_{2j} (\bm{y})|^2.
\end{align}
We first integrate over $\bm{k}$ using:
\begin{align} \label{eq:intkQG}
    \int \frac{d^3 \bm{k}}{(2 \pi)^3} \frac{1}{-(k^0)^2+\bm{k}^2} e^{i \bm{k}.(\bm{x}-\bm{y})} = \frac{1}{4 \pi} \frac{1}{|\bm{x} - \bm{y}|} e^{-i k^0 |\bm{x} - \bm{y}|}.
\end{align}
Next we  integrate over $k^0$:
\begin{align}
    \frac{1}{4 \pi |\bm{x} - \bm{y}|} \int \frac{d  k^0}{2 \pi} e^{-i k^0 [(|\bm{x} - \bm{y}| - (x^0 - y^0)] }\\ = \frac{1}{4 \pi |\bm{x} - \bm{y}|} \delta (|\bm{x} - \bm{y}| - (x^0 - y^0)),
\end{align}
and then we can integrate over $x^0$ and $y^0$ from $0$ to $ c t$, finding
\begin{align}\nonumber
    \frac{1}{4 \pi |\bm{x} - \bm{y}|} \int^{c t}_0 \int^{c t}_0 d x^0 d y^0 \, &\delta (|\bm{x} - \bm{y}| - (x^0 - y^0)) \\ &= \frac{1}{4 \pi} \left(\frac{c t}{|\bm{x} - \bm{y}|}  - 1\right) \theta(c t - |\bm{x} - \bm{y}|).
\end{align}
Plugging this back into \eqref{eq:alphaQG2}, we have
 \begin{align}\nonumber
    \gamma^{(2)}_{ij} &= \frac{ i G M^2}{  \hbar c} \int d^3 \bm{x} \int d^3 \bm{y}     |\tilde{\phi}_{1i} (\bm{x})|^2 |\tilde{\phi}_{2j} (\bm{y})|^2 \\&\left(\frac{c t}{|\bm{x} - \bm{y}|}  - 1\right) \theta(c t - |\bm{x} - \bm{y}|) \\ \nonumber 
    &= \frac{ i G M^2 }{  \hbar c V^2} \int d^3 \bm{x} \int d^3 \bm{y} \,    \theta_{1i}(\bm{x}) \, \theta_{2j}(\bm{y})  \\ \label{eq:gammarel} &\left(\frac{c t}{|\bm{x} - \bm{y}|}  - 1\right) \theta(c t - |\bm{x} - \bm{y}|).
    \end{align}
This is  the  relativistic expression (taking into account the finite speed of gravity) of the quantum phases for each superposition branch. It upgrades the relativistic expression derived in \cite{LocallyHowl} from non-relativistic point particles to spherical objects. We can re-derive the expression for point particles using the approximation $d_{ij} \gg R$ (for example, moving the $\bm{x}$ and $\bm{y}$ coordinates to the centres of the respective spheres and using $d_{ij} \gg R$, noting the integration bounds of the new coordinates when considering the integration of  $|\bm{x} - \bm{y}|$), which results in:
\begin{align} 
    \gamma^{(2)}_{ij} \approx 
    \frac{i G M^2}{ \hbar c} \left( \frac{c t}{d_{ij}} - 1\right) \theta(c t - d_{ij}).
\end{align}
As discussed in  \cite{LocallyHowl}, this demonstrates that entanglement between the initially spacelike separated matter objects, which we have assumed here, is only generated once the initial light cone of one object contains the other object.  However, note that if the graviton were slightly massive \cite{OGIEVETSKY1965167,PhysRevD.82.044020}, the equivalent amplitude to  \eqref{eq:gammarel} would be non-zero (but incredibly small) even for times $ c t < |\bm{x} - \bm{y}|$, despite being derived from a  relativistically local theory. 

Finally, when $c t \gg d_{ij}$, which is very much going to be the case for a realistic experiment, we arrive at the originally derived, fully non-relativistic version of the quantum phases \cite{bose2017spin,marletto2017gravitationallyinduced} $\gamma^{(2)}_{ij} = i \varphi_{ij} := i G M^2 t / (\hbar d_{ij})$, where $d_{ij} := |\bm{X}_{1i} - \bm{X}_{2j}|$. We could have also got here immediately after using \eqref{eq:intkQG} and then assuming  $ c t \gg 1$ (specifically, $ c t \gg d_{ij}$ and $ c t \gg \hbar / ( m c)$), since after \eqref{eq:intkQG}  we could have first performed the time integrals rather than the $k^0$ integral:
\begin{align}\nonumber
    \gamma^{(2)}_{ij} &= \frac{i G M^2}{ \hbar c} \int d^3 \bm{x} \int d^3 \bm{y} \frac{\tilde{\phi}^2_{1i}(\bm{x}) \tilde{\phi}^2_{2j}(\bm{y}) }{|\bm{x} - \bm{y}|} \, \times \\ \label{eq:QGdelta} & \int^{ct}_0 dx^0 e^{i x^0 k_0} \int^{ct}_0 dy^0 e^{-i y^0 k_0}\int \frac{d k^0}{2 \pi} e^{-i k^0 |\bm{x} - \bm{y}|} \\ \nonumber
    &= \frac{i G M^2}{ \hbar c} \int d^3 \bm{x} \int d^3 \bm{y} \frac{\tilde{\phi}^2_{1i}(\bm{x}) \tilde{\phi}^2_{2j}(\bm{y})}{|\bm{x} - \bm{y}|} \, \times \\ &
    \int d k^0  \int^{ct}_0 dy^0 e^{-i y^0 k_0} e^{-i k^0 |\bm{x} - \bm{y}|} e^{i c t k_0 / 2} \frac{\sin( c t k_0 / 2)}{\pi k_0}\\ \nonumber
    &\approx \frac{i G M^2}{ \hbar c } \int d^3 \bm{x} \int d^3 \bm{y} \frac{\tilde{\phi}^2_{1i}(\bm{x}) \tilde{\phi}^2_{2j}(\bm{y})}{|\bm{x} - \bm{y}|} \, \times \\ &\int d k^0\int^{ct}_0 dy^0 e^{-i y^0 k_0} e^{-i k^0 |\bm{x} - \bm{y}|}  e^{i c t k_0 / 2}  \delta(k_0)\\ \label{eq:alphaTheta}
    &= \frac{i G M^2}{ \hbar c V^2} \int d^3 \bm{x} \int d^3 \bm{y} \frac{\theta_{1i}(\bm{x}) \, \theta_{2j}(\bm{y})}{|\bm{x} - \bm{y}|}  \int^{ct}_0 dy^0 = \frac{i G M^2}{\hbar d_{ij}} \\&\equiv i\varphi_{ij},
\end{align}
where we have used $\lim_{\gamma \rightarrow \infty} \sin(\gamma x)/(\pi x) = \delta(x)$. Alternatively, we could have also used the Fourier transform of  $\mathrm{sinc}$:  $\int^{\infty}_{-\infty} dx\,  y \,\mathrm{sinc}( x y/2) e^{-i x s} = 2 \pi\,\theta(y/2 - |s|)$. 

Ignoring the other Wick contraction (Fig.\  \ref{fig:CrissCross}\textcolor{blue}{a}) and the first order contributions for now, as well as any second-order contributions that just result in a global phase \cite{fn1},  when considering the full amplitudes $\alpha_{ij} \approx \alpha_{ij}^{(0)}   + \alpha_{ij}^{(2)}$, we have $\alpha_{ij} \approx 1 + i \varphi_{ij}$, which is just the first order expansion of the quantum phase $e^{i \varphi_{ij}}$. Therefore, with $d_{RL} \gg \Delta x$, the final state in the Stern-Gerlach version of the experiment would be  of the form:
\begin{align}\nonumber
    |\Psi(t)\rangle &= \frac{1}{2}   \Big( |\uparrow\rangle_1 |\uparrow\rangle_2  \\ \label{eq:finalStateQG}&+    |\uparrow\rangle_1 |\downarrow\rangle_2 + (1 + i \varphi)|\downarrow\rangle_1 |\uparrow\rangle_2  +   |\downarrow\rangle_1 |\downarrow\rangle_2\Big),
\end{align}
to second order in the Dyson series, which is an entangled state. As this is a perturbative calculation, where we have ignored terms higher than second order in the Dyson series, the result \eqref{eq:finalStateQG} is only valid for $\varphi := \varphi_{RL} \ll 1$ and is just the first-order expansion of the quantum phase $e^{i \varphi}$ that was previously derived in first-quantization works \cite{bose2017spin,marletto2017gravitationallyinduced}. However, the full non-perturbative expression $e^{i \varphi}$ can straightforwardly be obtained by considering that at each even order in the Dyson series we have essentially the Feynman diagram \ref{fig:Fig1}\textcolor{blue}{a}  again but with an extra graviton propagator and additional in and out states. The amplitude $\gamma_{ij}^{(2)}$ is then just taken to an extra power together with the corresponding factorial from the Dyson series, providing the Taylor expansion of $e^{i \varphi}$. It is  common  in perturbative QFT that a low-order calculation can be extrapolated to a non-perturbative result \cite{Peskin:1995ev}.

With $ \Delta x \ll d_{RL}$, each phase $\varphi_{ij}$ is approximately the same and, rather than entanglement only depending on $\varphi$, the relevant parameter is now $\varphi\, \Delta x^2 / d_{RL}^2$ as discussed above. This can be derived by considering the overlap of the superposition states \cite{Aspelmeyer2022} or, for example, considering the expression for the negativity of the final state \cite{EntNegativity}. 

From \eqref{eq:alphaTheta}, we can see that the  quantum gravity effect depends strongly on the mass and time chosen for the experiment. A range of mass values and times  have been considered for different versions of Feynman's experiment \cite{FeynmanQG,bose2017spin,marletto2017gravitationallyinduced,bose2023massivequantumsystemsinterfaces,krisnanda2020observable,howl2023gravitationallyinducedentanglementcoldatoms}, with the Planck mass, $M_p \approx 10^{-8}\,\mathrm{kg}$, thought to play a significant role \cite{FeynmanQG,christodoulou2019possibility,Howl_2019}. For example, in  \cite{bose2017spin}, relatively small masses  are suggested, $M \approx 10^{-14}\,\mathrm{kg}$, at the expense of relatively long coherence times $t \approx 2\,\mathrm{s}$ and large superpositions $\Delta x \approx 0.1\mathrm{mm}$, and with $d_{RL} = 200\,\mathrm{\mu m}$ such that quantum gravity would be an order of magnitude greater than residual electromagnetic interactions (alternatively, a conducting screen can be placed between the objects to eliminate  electromagnetic interactions \cite{PhysRevA.102.062807}).  However,  long coherence times  present  a significant experimental challenge due to expected decoherence mechanisms \cite{Aspelmeyer2022}. For instance, the source of decoherence that is thought to be most dominant \cite{bose2017spin}, scattering of molecules from an imperfect vacuum, requires extremely low pressures of  $10^{-15}\,\mathrm{Pa}$ for $M = 10^{-14}\,\mathrm{kg}$ and $t = 2\,\mathrm{s}$, which presents a formidable  challenge \cite{bose2017spin,Howl_2019,Aspelmeyer2022}. The rate of decoherence from scattering scales linearly with pressure but as $M^{2/3}$ with mass, which is far weaker than the mass scaling $M^{2}$ of $\varphi$ \cite{Aspelmeyer2022}.  Therefore, since it allows for much lower pressures, larger masses and smaller times may be experimentally preferable \cite{Aspelmeyer2022}, which has  been considered in equivalent  tests \cite{marletto2017gravitationallyinduced,EquivPlenio,MariosEquiv}, with times such as $1\,\mathrm{\mu s}$ and masses ranging from $10^{-12}\,\mathrm{kg}$ to $1\,\mathrm{kg}$ \cite{marletto2017gravitationallyinduced,krisnanda2020observable,FeynmanQG,Aspelmeyer2022,bose2023massivequantumsystemsinterfaces,howl2023gravitationallyinducedentanglementcoldatoms}. Smaller superposition sizes have also been considered since creating large $\Delta x$ has proven experimentally challenging thus far \cite{Aspelmeyer2022}.

\subsection{Classical gravity} \label{sec:CG}

We now consider the above experiment within the context of classical gravity interactions. The calculation follows the previous section, but  with the interaction Hamiltonian \eqref{eq:HCQ} rather than \eqref{eq:HQG} for the matter objects. At second order in the Dyson series there are  no non-vanishing Wick contractions corresponding to Feynman diagrams that contain quantum communication between the matter objects, and the diagram responsible for entanglement in quantum gravity, Fig.\ \ref{fig:Fig1}\textcolor{blue}{a},  becomes Fig.\ \ref{fig:Fig2}\textcolor{blue}{b}. This diagram  represents the two matter objects sitting in their combined classical 
 gravitational field,  with the  amplitude just contributing a local relative quantum phase between the branches of  each matter object, which does not lead to entanglement \cite{Carney_2019}. The amplitude for the diagram is the following:
\begin{align} \nonumber
    &\beta^{(2)}_{ij} = -\frac{1}{8 \hbar^2 c^2}   \int_t d^4 x \int_t d^4 y  \, h^{\mu \nu}(x) \,  h^{\rho \sigma}(y) \, \times \\ \label{eq:beta2nd} &\,_{1i}{\langle} \wick{\c1 N| \, \,_{2j}{\langle} \c2 N| \hat{\mathcal{T}}_{\mu \nu}[\c1{\hat{\phi}}^{\dagger} (x) \c4{\hat{\phi}}(x) ]   \hat{\mathcal{T}}_{\rho \sigma}[ \c2{\hat{\phi}}^{\dagger} (y)  \c3{\hat{\phi}}(y)] |\c4 N \rangle_{1i} \,| \c3 N \rangle_{2j}}.
\end{align}
 Here, $h_{\mu \nu}(x)$ is  the classical gravitational field of the matter objects, satisfying $|h_{\mu \nu}| \ll 1$. Since it is classical, there is no Wick contraction for it (there is no associated non-commutativity). Crucially, since it is not associated with a quantum operator, it takes the \emph{same} value in each superposition branch. If this were not the case, then  the gravitational field would be in a quantum superposition, and thus not classical. This has caused confusion in the literature where gravity is considered classical but the field or the Newtonian force is still allowed to go into a superposition such that \eqref{eq:beta2nd} can result in the same amplitude and thus entanglement as \eqref{eq:alphaQG}  (see below for more detail) \cite{anastopoulos2018commentaspinentanglement,Anastopoulos_2021,Fragkos2022}. We do not assume such a scenario here, keeping to the notion that quantum superposition is a purely quantum-mechanical phenomena.

\begin{figure*}
   \centering
\includegraphics[width=0.8\textwidth]{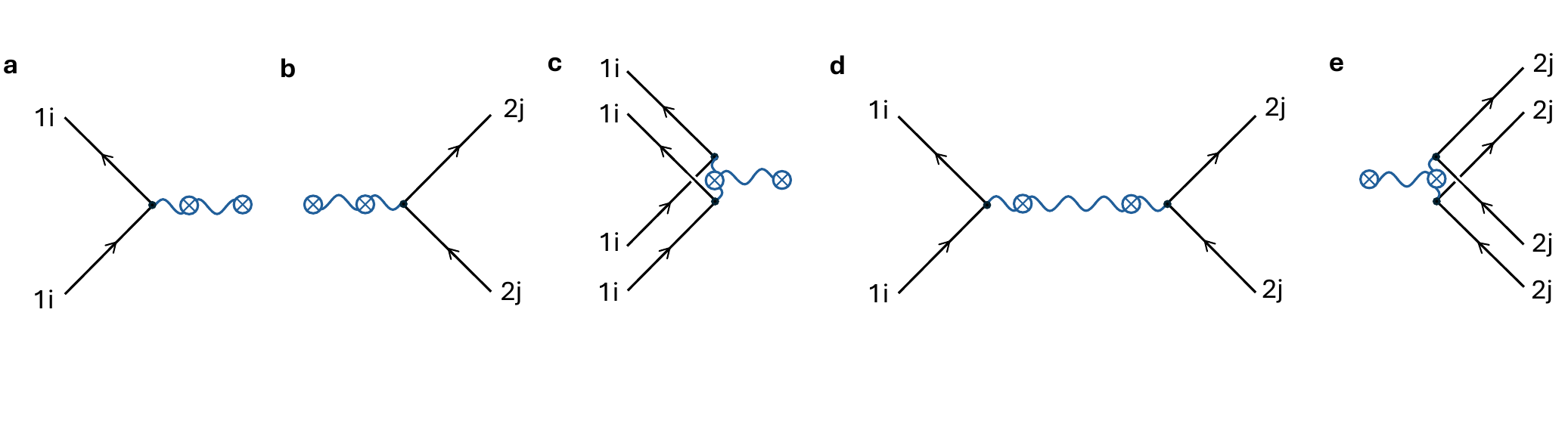}
    \caption{First and second order Feynman diagrams that contribute towards the relative quantum phases $\Delta \varphi_1$ and $\Delta \varphi_2$ in \eqref{eq:PsiRelativeClassicalPhase} but not entanglement. The $1i$ and $2j$ label the first and second matter distributions, with $i,j \in \{L,R\}$. The two circles with crosses indicate the two classical sources of gravity, i.e.\ the two matter distributions. As stated also in Fig.\ \ref{fig:CrissCross}, in contrast to standard perturbative QFT  diagrams, the external legs represent position-like states, with the arrows indicating time evolution.}
    \label{fig:ClassicalPhase}
\end{figure*}

As in the previous section, in computing \eqref{eq:beta2nd} we can consider a non-relativistic approximation for the in and out states (see \eqref{eq:WickN}), such that we only need  consider  contractions of the field and its time derivative on the in and out states, Eq.\ \eqref{eq:WickN}. As in the quantum gravity case, this means that the free evolution generated by $\hat{H}_0$ only contributes a global phase and thus no entanglement - see  \eqref{eq:globalphase}. That is, we can ignore the free evolution generated by $\hat{H}_0$ as above.  Furthermore, since  we know that non-relativistic gravity is a good approximation for the experiment, we can further  assume that $h_{\mu \nu} (\bm{x}) = - 2 \Phi(\bm{x}) \delta_{\mu \nu} / c^2 $, with $\Phi(\bm{x})$ the Newtonian potential of the matter objects, which  is  assumed spatially varying and time-independent for simplicity. In this  approximation, the interaction Hamiltonian \eqref{eq:HCQ} simplifies to
\begin{align} \label{eq:HintNonRel}
  \hat{H}^{CG}_{int}  &= \frac{4}{c^2} \int d^3 \bm{x}  \, \Phi(\bm{x})  \Big(   \hat{\pi}(\bm{x}) \hat{\pi}^{\dagger}(\bm{x})   - \frac{m^2 c^2}{2\hbar^2} \hat{\phi}^{\dagger}(\bm{x}) \hat{\phi}(\bm{x})\Big),
\end{align}
which in the interaction picture is simply
\begin{align} \label{eq:HIonRel}
  \hat{H}_{I}  &= \frac{4}{c^2} \int d^3 \bm{x}  \, \Phi(\bm{x})  \Big(   \hat{\pi}(x) \hat{\pi}^{\dagger}(x)   - \frac{m^2 c^2}{2\hbar^2} \hat{\phi}^{\dagger}(x) \hat{\phi}(x)\Big),
\end{align}
where $\hat{\pi} := \partial_0 \hat{\phi}^{\dagger}$. Now we can immediately  see that we only have to worry about the contraction of the field and its time-derivative on the in and out states, rather than any spatial derivatives, such as $\partial_x \wick{ \c{\hat{\phi}}(x) |\c N\rangle_{\kappa i}}$.  Using     $\wick{  \c{\hat{\pi}}^{\dagger}(x)  |\c N \rangle_{\kappa i}} \approx -i \sqrt{m N/2} \,c \,  e^{-i m c x^0 / \hbar} \tilde{\phi}_{\kappa i}(\bm{x})  |N-1 \rangle_{\kappa i}$, we find:
\begin{align}
    \beta^{(2)}_{ij} = - \varphi_{1i} \varphi_{2j}, 
\end{align}
where:
\begin{align}
    \varphi_{\kappa i} &= \frac{M t}{\hbar} \int d^3 \bm{x} \, |\tilde{\phi}_{\kappa i}(\bm{x})|^2 \Phi(\bm{x})\\ \label{eq:phiki}
    &=\frac{M t}{\hbar V} \int d^3 \bm{x} \, \theta_{\kappa i}(\bm{x}) \Phi(\bm{x}),
\end{align} 
with $i,j \in \{L,R\}$. Since the two matter objects are no longer quantum-mechanically connected, the amplitude \eqref{eq:beta2nd} does not contribute to entanglement. Instead, it combines with the first-order and second-order processes in Fig.\ \ref{fig:ClassicalPhase} to generate the amplitudes
\begin{align} \label{eq:alphaCG}
\alpha_{ij} = \alpha_{ij}^{(0)} + \alpha_{ij}^{(1)} + \alpha_{ij}^{(2)} = 1 - i \left(\varphi_{1i} + \varphi_{2j}\right) - \frac{1}{2} \left(\varphi_{1i} + \varphi_{2j}\right)^2, 
\end{align}
to second order, such that the final state  would be
\begin{align}
    |\Psi(t) \rangle &= \frac{1}{2} \Bigg( \Big[ |N\rangle_{1L} \left(1 + i \varphi_{1L} - \frac{1}{2} \varphi^2_{1L}\right) \\&+ |N\rangle_{1R} \left(1 + i \varphi_{1R} - \frac{1}{2} \varphi^2_{1R}\right) \Big]\\
    &\otimes \Big[ |N\rangle_{2L} \left(1 + i \varphi_{2L} - \frac{1}{2} \varphi^2_{2L}\right) \\&+ |N\rangle_{2R} \left(1 + i \varphi_{2R} - \frac{1}{2} \varphi^2_{2R}\right) \Big]\Bigg),
\end{align}
which is  the second-order approximation to
\begin{align} \nonumber
    |\Psi(t) \rangle &= \frac{1}{2} \Bigg( \left( e^{i \varphi_{1L}} |N\rangle_{1L}  + e^{i \varphi_{1R}} |N\rangle_{1R}  \right)
    \\&\otimes \left( e^{i \varphi_{2L}} |N\rangle_{2L}  + e^{i \varphi_{2R}} |N\rangle_{2R} \right)\Bigg)\\ \nonumber
    &\equiv\frac{1}{2} \Bigg( \left(  |N\rangle_{1L}  + e^{i \Delta \varphi_{1}} |N\rangle_{1R}  \right)
    \\ \label{eq:PsiRelativeClassicalPhase} &\otimes \left(  |N\rangle_{2L}  + e^{i\Delta \varphi_{2}} |N\rangle_{2R} \right)\Bigg),
\end{align}
with $\Delta \varphi_{1}:= \varphi_{1R} - \varphi_{1L}$ and $\Delta \varphi_{2}:= \varphi_{2R} - \varphi_{2L}$. This has the same form as the state of two matter spheres in a quantum superposition of two locations sitting in an external classical gravitational potential, and is a separable state.  

Therefore, at second order our final state \eqref{eq:stateAfterG} is a separable state. This is because there is no possibility to generate a quantum propagator between the two objects up to this order with the initial state \eqref{eq:PsiInitial} and final state  \eqref{eq:stateAfterG}. The same applies also at third order, as can be simply  deduced  by considering the possible connected Feynman diagrams at this order. Instead, we have to move to the fourth order  for there to be an allowed   quantum propagator between the objects, and the corresponding diagram for this process is Fig.\  \ref{fig:Fig2}\textcolor{blue}{e}. Before taking the non-relativistic gravity limit, the amplitude for this diagram takes the form:
\begin{strip}
    \begin{align}\nonumber
        &\beta^{(4)}_{ij}  = \frac{1}{16 \, \hbar^4 c^4} \int_t d^4 x \, \int_t d^4  y \,  \int_t d^4 z\, \int_t d^4 w\, h^{\mu \nu}(w)\, h^{\rho \sigma}(z)\, h^{\gamma \delta}(y)\, h^{\kappa \lambda}(x) \,\times \\ \label{eq:beta4Comp} &\,_{1i}\langle \wick{\c1 N| \, _{2j}{\langle} \c2 N|  \hat{\mathcal{T}}_{\mu \nu}[\c2{\hat{\phi}}^{\dagger} (w) \c3{\hat{\phi}} (w)]      \hat{\mathcal{T}}_{\rho \sigma}[\c1{\hat{\phi}}^{\dagger} (z) \c4{\hat{\phi}} (z)] \hat{\mathcal{T}}_{\gamma \delta}[\c3{\hat{\phi}}^{\dagger}(y) \c5{\hat{\phi}} (y)]
      \hat{\mathcal{T}}_{\kappa \lambda}[\c4{\hat{\phi}}^{\dagger} (x) \c6{\hat{\phi}} (x) ] |\c5 N \rangle_{1i} \,| \c6 N} \rangle_{2j}.
\end{align}
\end{strip}
With the  non-relativistic approximation \eqref{eq:HIonRel}, the amplitude involving just the field $\hat{\phi}$ and not the momentum conjugate $\hat{\pi}$ is
\begin{align}\nonumber
  &\frac{ 4 m^6 N^2}{  \hbar^6 c^2} \int_t d^4 x \int_t d^4  y \int_t d^4 w \int_t d^4  z \\ \nonumber &\times \int \frac{d^4 k}{(2 \pi)^4} \frac{1}{k^2 + m^2 c^2 / \hbar^2 + i \epsilon} e^{i k.(x-z)}  \,  \\ \nonumber &\times\int \frac{d^4 l}{(2 \pi)^4} \frac{1}{l^2 + m^2 c^2 / \hbar^2 + i \epsilon} e^{i l.(w-y)} \\ \nonumber &\times \tilde{\phi}_{{2j}} (\bm{x}) \tilde{\phi}_{{1i}} (\bm{z}) \tilde{\phi}_{{1i}} (\bm{y})  \tilde{\phi}_{{2j}} (\bm{w})  \,  \\ \label{eq:fullCG} &\times \Phi(\bm{x}) \Phi(\bm{y}) \Phi(\bm{z}) \Phi(\bm{w}) e^{i m c (z^0 - x^0)/ \hbar} e^{i m c (y^0 - w^0)/ \hbar},
    \end{align}
    where the integrals over $k$ and $l$ are coming from the virtual matter Feynman propagator for complex scalar fields \cite{Peskin:1995ev}.  
The contractions also involving the momentum conjugate $\hat{\pi}$ give the expression \eqref{eq:beta4Comp}  above but with time derivatives on the phase factors $\exp(i m c x^0 / \hbar)$ (due to contractions of $\hat{\pi}$ with the in or out states), and/or the Feynman propagators.  
    
    Following the previous section, we first integrate over $\bm{k}$ (and $\bm{l}$) using: 
\begin{align} \label{eq:Intk}
    \int \frac{d^3 \bm{k}}{(2 \pi)^3} \frac{1}{\bm{k}^2 - (k^0)^2 + \gamma^2 } e^{i \bm{k}.(\bm{x} - \bm{y})} = \frac{1}{4 \pi^2 i r} \int^{\infty}_{-\infty} dk\frac{k e^{i k r}}{k^2 + \tilde{\gamma}^2},
\end{align}
where $k := |\bm{k}|$, $\gamma := m c / \hbar$, $\tilde{\gamma}^2 := \gamma^2 - k^2_0$ and $r := |\bm{x} - \bm{y}|$. The integral has poles at $k = \pm i \tilde{\gamma}$, such that:
\begin{align} \nonumber
     &\frac{1}{4 \pi^2 i r} \int^{\infty}_{-\infty} dk\frac{k e^{i k r}}{k^2 + \tilde{\gamma}^2} = f(k_0) \\ \nonumber &:= \frac{1}{4 \pi r}  \Bigg(\theta(k^2_0 - \gamma^2) \left(e^{-r \sqrt{\gamma^2 - k^2_0}}-1\right) \\\label{eq:expdecay}&+ \theta(\gamma^2 -k^2_0) \left(e^{i r \sqrt{k^2_0 - \gamma^2}} - 1\right) + 1\Bigg).
\end{align}
Note that when $\gamma = 0$, as in the previous section, we obtain \eqref{eq:intkQG}.

We now follow the second method used in the above quantum gravity section in obtaining the expression for $\gamma^{(2)}_{ij}$:  we perform the time integrals and take a delta function approximation such that
\begin{align} \label{eq:virtialdecay2}
    &\int^{ c t}_0 d x^0 \int^{c t }_0 d z^0 e^{ i \gamma (z^0 - x^0)} \int \frac{d k^0}{2\pi} F(k_0) e^{i k_0 (x^0 - z^0)}\\ &=   \int^{c t }_0 d z^0  \int \frac{d k^0}{2\pi} F(k_0) e^{i z^0 (\gamma - k_0)} e^{i c t (k_0 - \gamma) / 2} \frac{\sin( c t (k_0 - \gamma)/2)}{k_0 - \gamma}\\
    &= 2 \int \frac{d k_0}{2 \pi} F(k_0) \frac{\sin^2(c t (k_0 - \gamma)/2)}{(k_0 - \gamma)^2}\\ \label{eq:intkandk0}
    &\approx  F(\gamma) \, ct
\end{align}
where we used $\lim_{x \rightarrow \infty} \sin^2(x s)/(x s^2) = \pi \delta(s)$, $F(k_0) := \int d^3 \bm{x}\, d^3 \bm{z} \Phi(\bm{x}) \Phi(\bm{z}) \tilde{\phi}_{2j} (\bm{x}) \tilde{\phi}_{1j} (\bm{z})  f(k_0)$, and note that $f(\gamma) = 1 / (4 \pi r)$ \cite{Peskin:1995ev}. Often interaction processes involving massive virtual particles come with exponential decay factors over space \cite{Peskin:1995ev}, as in the first term of \eqref{eq:expdecay}. However, this is not the case for the process considered here. This is because the contraction of the matter field with the position-like states -- see \eqref{eq:WickN} --  provides  a factor $e^{i \gamma (z^0 - x^0)}$ in \eqref{eq:virtialdecay2}. Without this factor, the time integral sets $k_0 = 0$, resulting in the evaluation of \eqref{eq:expdecay} but with $k_0 = 0$, and thus exponential decay with space. In contrast, with the factor $e^{i \gamma (z^0 - x^0)}$, we get  $k^0 = \gamma$ rather than $k^0 = 0$, resulting in  no exponential decay factors over space. However, despite no exponential decay with space, the corresponding particles are still virtual particles as they are in general off-shell, which is further discussed in Appendix \ref{sec:CGDiscussion} from a physical perspective. Note that by taking the above delta function approximation, we are essentially performing a non-relativistic approximation similar to as we did through equations \eqref{eq:QGdelta}-\eqref{eq:alphaTheta} in the quantum gravity calculation of the previous section.  However, just as for the quantum gravity case, this is just an approximation taken for computational ease, and the actual physical process is fundamentally relativistic, which is considered further in Appendix \ref{app:ct}. 

For the contractions where the  momentum conjugate is also used in the virtual propagator, i.e.\ $\wick{\c{\hat{\pi}}(x) \c {\hat{\phi}}(y)}$ and $\wick{\c {\hat{\pi}}(x) \c {\hat{\pi}}^{\dagger}(y)}$, the time derivatives result in an extra factor of $i \gamma$ and $\gamma^2$ respectively. Using  \eqref{eq:intkandk0} in \eqref{eq:beta4Comp}, we then obtain \cite{fn3}:
\begin{align}\label{eq:beta4ijFull}
    \beta^{(4)}_{ij} \approx \frac{M^2 t^2 m^4 }{4 \pi^2 \hbar^6 V^2 } \left(i \int d^3 \bm{x} \int d^3 \bm{y} \frac{\Phi(\bm{x}) \, \Phi(\bm{y}) \, \theta_{1i}(\bm{x}) \, \theta_{2j}(\bm{y})}{|\bm{x} - \bm{y}|}\right)^2.
\end{align}
As stated above,  the gravitational potential of the objects, $\Phi(\bm{x})$, is the \emph{same} irrespective of $i$ and $j$ - it is the same for each superposition branch since it is a classical potential. Despite this, and in contrast to the classical gravity amplitude \eqref{eq:alphaCG}, $\beta^{(4)}_{ij}$  will, in general, be different for each superposition branch because the object functions $\theta_{1i} (\bm{x})$  and $\theta_{2j} (\bm{y})$ are connected through the term $|\bm{x} - \bm{y}|$ in the denominator. This is analogous to the quantum gravity expression for  $\alpha^{(2)}_{ij}$ in \eqref{eq:alphaQG} - there the linking denominator  came from the virtual graviton, whereas here it comes from  virtual matter. That is, although $\Phi(\bm{x})$ does not quantum superpose, the virtual matter particles do and the distance they must travel in each superposition branch  is different just as is the case for virtual gravitons in $\alpha^{(2)}_{ij}$ in quantum gravity. This then leads to a different amplitude for each branch.

Since $\Phi(\bm{x})$  is coming from a superposition of matter in Eq.\ \eqref{eq:beta4ijFull}, we must consider  how exactly gravity  is  sourced by quantum matter in a fundamental theory of classical gravity. The two leading suggestions for how this occurs in a fundamental theory of classical gravity  are: (i) gravity is sourced by the mean expectation of matter $\nabla^2 \Phi(\bm{x}) = \xi   \langle \hat{T}_{00}\rangle$ \cite{ROSENFELD1963353,moller1962theories}, where  $\xi = 4 \pi G / c^2$, and the expectation is over the standard quantum state of matter or a generalization, such as a local description \cite{Kent_2018,kent2005nonlinearity,Giulini2023}; and (ii) gravity is sourced by stochastic fluctuations around the mean expectation \cite{DIOSI1987377,TilloyDiosi,layton2023weak,carney2024classical}:  $\nabla^2 \Phi(\bm{x}) = \xi  [ \langle \hat{T}_{00}\rangle + \delta T_{00}]$, where $\delta T_{00}$ is a stochastic quantity. The former has historically been studied more than the latter and is thus the option considered here, while  a discussion on (ii) is provided in Appendix \ref{sec:fundDec}. The theoretical consistency of both cases has been  debated, as  discussed in Appendix \ref{sec:consistent}, but neither has been ruled out experimentally. With option (i), $\Phi(\bm{x})$ in \eqref{eq:beta4ijFull} is the  sum of the average potentials of each mass distribution over their left and right states. For example, in  semi-classical Einstein gravity  \cite{moller1962theories,ROSENFELD1963353}, $\Phi(\bm{x})$ is sourced by the expectation of the quantum matter objects:
\begin{align}
    \Phi(\bm{x}) = - \frac{G}{c^2} \int d^3 \bm{y} \frac{ \left\langle \psi \left| \hat{T}_{00}(\bm{y}) \right|\psi\right\rangle}{|\bm{x} - \bm{y}|},
\end{align}
where $|\psi\rangle$ is the joint quantum state of the matter objects. This results in $\Phi(\bm{x})$ being the sum of the average potentials of each mass over their left and right states:
\begin{align} \label{eq:semiclassicalPhi}
    \Phi(\bm{x}) = \Phi_{C1}(\bm{x}) + \Phi_{C2}(\bm{x}),
\end{align}
with
\begin{align} 
    \Phi_{C\kappa}(\bm{x}) &:= \frac{1}{2} \left(\Phi_{\kappa L}(\bm{x}) + \Phi_{\kappa R}(\bm{x})\right),
\end{align}
and
\begin{align}\nonumber
    \Phi_{\kappa i }(\bm{x}) &:= - G M \bigg[ \left(\frac{3}{2 R} - \frac{|\bm{x} - \bm{X}_{\kappa i }|^2}{2R^3}\right) \theta(R - |\bm{x}- \bm{X}_{\kappa i }|) \\&+ \frac{\theta(|\bm{x}- \bm{X}_{\kappa i}| - R)}{|\bm{x}- \bm{X}_{\kappa i}|} \bigg],
\end{align}
such that $\Phi_{\kappa i} (\bm{x})$ is the gravitational potential of a spherical mass distribution of total mass $M$ at position $\bm{X}_{\kappa i}$, and  $\Phi_{C \kappa} (\bm{x})$ is the average gravitational potential of  spherical mass distributions  each of mass $M$  located at $\bm{X}_{\kappa L}$ and $\bm{X}_{\kappa R}$. Note that the same $\Phi(\bm{x})$ \eqref{eq:semiclassicalPhi} results if we chose to perform the expectation of $\hat{T}_{00}$ with the `local' state of matter \cite{kent2005nonlinearity,Kent_2018} or chose a relativistic collapse mechanism \cite{Helou_2017}, as discussed further in Appendix \ref{sec:consistent}.  

Plugging \eqref{eq:semiclassicalPhi} into \eqref{eq:beta4ijFull}, we are integrating all the different gravitational potentials over the different  superposition branches. We solve these integrals by integrating first over $\bm{y}$ and then over $\bm{x}$ using a well-known technique for finding the gravitational potential of an axially symmetric mass distribution: 
\begin{align} \label{eq:varPhixtheta}
    \varPhi(x',\theta_x) :=  \int d^3 \bm{y}'  \, \frac{\rho(y',\theta_y)}{|\bm{y}' - \bm{x'}|},
\end{align}
which can be written as
\begin{align}
   \varPhi(x',\theta_x) =  \sum^{\infty}_{n=0} \varPhi_n(x') P_n(\cos \theta_x),
\end{align}
where
\begin{align} \nonumber
    \varPhi_n(x') &= - \frac{2 \pi }{(n+ 1/2) x^{'(n+1)}} \int^{x'}_0 d y'\,  y^{'(n+2)} \rho_n(y') \\ \label{eq:varPhin} &- \frac{2 \pi x^{' n}}{n + 1/2} \int^{\infty}_{x'} d y'\, y^{' (1-n)} \rho_n (y').
\end{align}
For example, using the above, we can solve integrals such as
\begin{align}\nonumber
    I &= \int d^3 \bm{x} \int d^3 \bm{y} \,\frac{1}{|\bm{x} - \bm{X}_{1L}|\,|\bm{y} - \bm{x}| \, |\bm{y} - \bm{X}_{1R}|} \\&\times \theta(R - |\bm{x}-\bm{X}_{1R}|) \theta(R - |\bm{y}-\bm{X}_{2R}|)\\\nonumber
    &=  \int d^3 \bm{x} \frac{\theta(R - |\bm{x'} - \bm{\Delta_{1}}|)}{x'} \\&\times \int d^3 \bm{y'} \frac{1}{|\bm{y'} - \bm{d_{RR}}|} \frac{1}{|\bm{y'}-\bm{x'}|} \theta(R - y'),
\end{align}
where $\bm{y'} := \bm{y} - \bm{X}_{1R}$, $\bm{x'} = \bm{x} - \bm{X}_{1R}$, $\bm{d_{RR}} = \bm{X}_{1R} - \bm{X}_{2R}$ and $\bm{\Delta_{1}} = \bm{X}_{1R} - \bm{X}_{1L}$. We then choose the coordinate system $\bm{y}'$ such that its z-direction is along $\bm{d_{RR}}$. In this case
\begin{align}
    \frac{1}{|\bm{y}' - \bm{d_{RR}}|} &\equiv \frac{1}{d_{RR} \sqrt{1 - 2 \frac{y'}{d_{RR}} + \frac{y^{'2}}{d_{RR}}}} \\&= \frac{1}{d_{RR}} \sum^{\infty}_{m=0} P_m (\cos \theta_y) \left(\frac{y'}{d_{RR}}\right)^m,
\end{align}
where $d_{RR} = |\bm{d_{RR}}|$, $P_m(x)$ is the Legendre polynomials, and $\theta_y$ is the polar angle of the $\bm{y}'$ coordinate system. We can then write $I$ as
\begin{align} \label{eq:IB}
    I =  \int d^3 \bm{x}' \frac{\theta(R - |\bm{x}' - \bm{\Delta_1}| )}{x'} \int d^3 \bm{y}'  \frac{\rho(y',\theta_y)}{|\bm{y}' - \bm{x'}|},
\end{align}
where
\begin{align}
    \rho(y',\theta_y) := \frac{\theta(R - y') }{d_{RR}} \sum^{\infty}_{m=0} P_m (\cos \theta_y) \left(\frac{y'}{d_{RR}}\right)^m.
\end{align}
We can again now use the above solution  of the gravitational potential of an axially symmetric mass distribution. In this case, $\rho_n(y') := (n + 1/2) \int^{\pi}_0 \rho(y',\theta_y) P_n (\cos \theta_y) \sin \theta_y d \theta_y = \theta(R - x') x^{' n} / d^{n+1}_{RR}$ using the orthogonal property of the Legendre polynomials: $\int^1_{-1} P_n (x) P_m(x) dx = \delta_{nm} / (n+1/2)$. Plugging our $\rho_n(y')$ into \eqref{eq:varPhin}, we find
\begin{align}\nonumber
    \varPhi_n(x') &= \frac{2\pi}{(n+1/2)d^{n+1}_{RR}} \\\nonumber&\times\Bigg[ \left(\frac{R^2 x^{' n}}{2} - \frac{x^{' (n+2)} (n+1/2)}{2 n + 3}\right) \theta(R - x') \\&+ \left(\frac{R^{2 n + 3}}{x^{' (n+1)} (2 n + 3)} \right) \theta(x' - R) \Bigg].
\end{align}
Inserting this into $I$, only the second term survives in the assumption that the sphere states do not overlap ($\Delta x > 2 R$), leaving us with
\begin{align}\nonumber
     I &=  \sum_{n=0}^{\infty} \frac{2\pi  R^{2 n + 3}}{(n+1/2) (2 n + 3) d^{n+1}_{RR}}  \\&\times\int d^3 \bm{x}' \frac{P_n(\cos \theta_x)}{x^{' (n+2)}} 
 \theta(R - |\bm{x}' - \bm{\Delta_1}| ). 
\end{align}
We now assume $d_{RR} \gg R$. In this case,   we only need to consider the $n=0$ term:
\begin{align}
    I \approx \frac{4\pi  R^{3}}{3 d_{RR}}  \int d^3 \bm{x}' \frac{1}{x^{'2}}  \theta(R - |\bm{x'} - \bm{\Delta_1}|).
\end{align}
This can then be solved by integrating over the sphere at $\bm{\Delta_1}$ from the origin of the $\bm{x'}$ coordinate system with surfaces of constant radius \cite{Howl_2019}:
\begin{align}\nonumber
    I &\approx \frac{4 \pi R^{3}}{3 d_{RR}} \\&\times\int^{2 \pi}_0 \int^{\Delta x + R}_{\Delta x - R}  \int_0^{cos^{-1} ((r^2 + (\Delta x)^2 - R^2)/(2 r \Delta x)}  \sin\theta \, dr \, d \theta \, d\phi   \\
    &= \frac{8\pi^2  R^{3} }{3 d_{RR}} \left(R + \frac{1}{2 \Delta x} \left(R^2 - (\Delta x)^2\right)  \ln( \frac{\Delta x + R}{\Delta x - R} ) \right). 
\end{align}

 Applying this integration method to  \eqref{eq:beta4ijFull} with $\Phi(\bm{x})$ given by \eqref{eq:semiclassicalPhi}, it is possible to solve all the spatial integrals. Then, in the  approximations $\Delta x \gg R$ and $d_{ij} \gg R \, \, \forall i,j$ (which match the approximations used in deriving the final quantum gravity result \eqref{eq:alphaTheta}), $\beta^{(4)}_{ij}$ is found to be:
 \begin{align}\label{eq:betaCG}
    \beta^{(4)}_{ij} &\approx
      \left(\frac{6 }{25} \frac{i G^2 m^2 M^3 R t }{\hbar^3 d_{ij}}\right)^2.
\end{align}
Just as with the quantum gravity amplitude $\alpha^{(2)}_{ij}$, there is an inverse dependence on $d_{ij}$. Therefore, with $d_{RL} \ll \Delta x$, we have $\beta^{(4)}_{RL} =: \vartheta$, where 
\begin{align} \label{eq:vartheta}
    \sqrt{\vartheta} = \frac{6 }{25} \frac{ G^2 m^2 M^3 R \, t }{\hbar^3 \,d_{RL}}.
\end{align}
With $d_{RL} \ll \Delta x$, this amplitude dominates over all other $\beta^{(4)}_{ij}$ amplitudes. The  state \eqref{eq:stateAfterG} is  then  entangled  since, just as in  quantum gravity,  $\alpha_{RL}$ contains a contribution, $\vartheta$, that is not in any of the other amplitudes $\alpha_{ij}$ \cite{fn4}. With $\Delta x \ll d_{RL}$, just as for quantum gravity, the amplitudes $\beta_{ij}^{(4)}$ become similar and the parameter relevant to entanglement  becomes $\beta^{(4)}_{RL} \,\Delta x^2 / d^2_{RL}$. As with quantum gravity, this can be derived by considering the overlap of the superposition states \cite{Aspelmeyer2022}, or considering, for example, the negativity of the final state \cite{EntNegativity}.

Note that in addition to $\beta^{(4)}_{ij}$ contributing to $\alpha_{ij}$ in the final state \eqref{eq:psiAfterSG}, there will also be contributions, up to fourth order, from the processes considered above as well as additional processes discussed in  Appendix \ref{app:AdditionalCG}. However, as shown above and in Appendix \ref{app:AdditionalCG}, these contributions can be written as local relative phases between the objects (see \eqref{eq:PsiRelativeClassicalPhase} and \eqref{eq:relativePhase2}) and can thus be derived by considering local unitaries acting on the initial state (even though, for example, they will depend, like $\beta^{(4)}_{RL}$, explicitly on $d_{RL}$). Therefore, although these relative phases between the objects contribute to the amplitudes $\alpha_{ij}$, they do not contribute to the entanglement of the final state. In deriving the amplitudes, we have assumed a linear theory, whereas semi-classical Einstein gravity is known to be non-linear. However, since we are working perturbatively, up to at least fourth order,  the theory is approximately linear for the experiment, as  discussed in detail in Appendix \ref{sec:linear}.

\begin{figure}
    \centering
\includegraphics[width=0.35\textwidth]{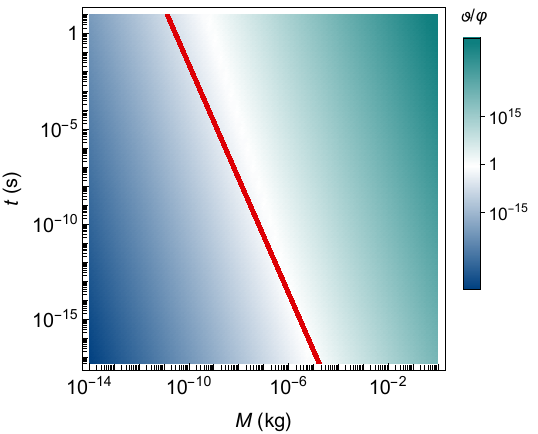}    
\caption{Comparison of   classical and quantum gravity effects. The relative strength of the considered effects,    $\vartheta / \varphi$, is shown for a range of   masses $M$ and times $t$ in the experiment described in  Section \ref{sec:exp}.   The red line and region to the right of the line  are where  the classical gravity effect and its associated entanglement would be significant ($\vartheta \geq 0.1$).  To evidence quantum gravity, this experiment must therefore operate to the left of this  line.  A minimum separation of $d_{RL} = 10 R$ is assumed, with  $R$ set by the total mass and  density (note  $\vartheta$ is independent of the density in this case), and  $m$ the mass of  ytterbium. The ratio $\overline{\vartheta}/\overline{\varphi}$, which characterizes the relative strength of the effects when  $d_{RL} \gg \Delta x$, is identical to $\vartheta/\varphi$. }
\label{fig:Fig4}
\end{figure}

In Section \ref{sec:QG}, we discussed how the quantum gravity effect depends strongly on the mass and time of the experiment. Similarly, the comparative strength of the   classical and quantum gravity effects calculated above depends strongly on the   mass and time of the experiment.  Figure \ref{fig:Fig4} compares  the classical and quantum gravity effects, $\varphi$ and $\vartheta$, for various  times and masses for the experiment described in Section \ref{sec:exp}  with ytterbium  as the   material. For relatively small masses $M \approx 10^{-14}\,\mathrm{kg}$ and large times $t \approx 2\,\mathrm{s}$ \cite{bose2017spin}, $\vartheta$ is  substantially smaller than $\varphi$. However, with masses approaching the Planck mass and beyond, $\vartheta$ becomes large enough ($\vartheta \approx 0.1$) such that entanglement due to the classical gravity process is significant, even at short times. Thus, the mere observation of entanglement at these values in the experiment described above could not be taken as evidence of quantum gravity. Although not  an issue for near-future experiments like \cite{bose2017spin}, this illustrates that  the  gravitational interaction  can in principle entangle quantum matter systems when gravity is either quantum or classical. 

\section{Summary}

We have argued above that, when treating matter at the QFT level, a classical gravity interaction can create entanglement between two matter systems through virtual matter processes - see Sections \ref{sec:PertCG} and \ref{sec:CG}.   As discussed in Section \ref{sec:CG}, with the matter systems in superpositions of locations, an entanglement process occurs that can be viewed as arising from the different distances the virtual  matter particles  have to travel between the systems in the different superposition branches (for a further discussion, see Appendix \ref{sec:CGDiscussion}). The considered  effect, see Section \ref{sec:CG}, was calculated for  a version of Feynman's experiment described in Section \ref{sec:exp}, and compared to the standard virtual graviton effect from perturbative quantum gravity in the experiment, Section \ref{sec:QG}.   For certain values of mass and time, it was found  that, in principle, the classical gravity interaction effect can be large such that the mere observation of entanglement in that  model experiment could not be used to evidence quantum gravity. However, in near-future experiments like \cite{bose2017spin} this is unlikely to be issue. Here, semi-classical Einstein gravity was used to characterize the classical gravity effect, but other models are possible. In Appendix \ref{sec:fundDec}, a version of stochastic classical theory is considered, which suppresses the effect due to fundamental decoherence from the gravitational field.

As discussed in Section \ref{sec:exp}, in the considered experiment,  the matter objects are described by  N00N states of complex scalar atoms with  wavepackets that are localized in space, and with the matter quantum field permeating all of space, as is  possible in QFT. Within the approximations and regime of the experiment, the wavepackets were found to stay localized  under the free evolution of the system.  The classical gravity process that is generating the entanglement  is then a virtual quantum matter process associated with the gravitational interaction that acts locally between the localized wavepackets of the experiment (the matter quantum field interacts with the localized wavepackets through the gravitational interaction). A promising implementation of the  experiment could be, for example, a cold atoms experiment with the atoms interacting during  free-fall atom fountains and potentially utilizing Feshbach resonances \cite{marletto2017gravitationallyinduced,Haine_2021,carney2021using,howl2023gravitationallyinducedentanglementcoldatoms}. For experiments with  strong trapping potentials and electromagnetic interactions,  N00N states of effective complex matter fields may not  be the best way to describe the matter objects and  some  matter fields may be  suppressed (and potentially enhanced) in certain places in space.  Future work will consider what connected processes occur in such experiments and how suppressed the processes can be, as well as any effects  associated with electromagnetism. Further screening effects could also potentially be used to remove the classical gravity effect.  Note that, from a purely theoretical perspective, even in the unphysical case of  infinite hard-wall boundaries for all matter fields, there can  still in principle be entangling processes through the  classical gravity interaction   due to the propagation of virtual matter particles of the interaction between or from the  boundaries (see for example, Appendix \ref{app:bound}). Thus, despite there being no virtual graviton propagators in the classical gravity process, the gravitational interaction can still, in principle, create entanglement between matter objects. To see further that the gravitational interaction is responsible for the entanglement effect in Section \ref{sec:CG},  we can  consider what happens if the interaction Hamiltonian $\hat{H}^{CG}_{int}$ for gravity is turned off, i.e.\  $\hat{H}^{CG}_{int} = 0$. In this case, we just have the free evolution, which approximately just contributes  a global phase as detailed in Section \ref{sec:QG} and below  \eqref{eq:globalphase}. It is then only  once the interaction Hamiltonian for classical gravity $\hat{H}^{CG}_{int}$, and with its spatially varying $\Phi(x)$ from the objects,  is turned on that we get entanglement. That the size of entanglement is proportional to $G$, Equation \eqref{eq:betaCG}, also illustrates that gravity is required for the considered entangling process.

Note that entanglement here is  not arising from just degrees of freedom that could be purely associated with the classical  gravitational field \cite{bose2017spin,marletto2017gravitationallyinduced,marletto2020witnessing,Galley2022nogotheoremnatureof,oppenheim2025covariantpathintegralsquantum,ludescher2025gravity}. As argued in Section \ref{sec:PertCG}, the very fact that gravity couples to matter  implies, when quantum matter is treated within QFT, the possibility of having a matter propagator that can generate entanglement through the gravitational interaction, and  regardless of the specific form of the classical gravity model.  Here, since the  gravitational field is taken to be classical, it cannot enter into a quantum superposition branch, and thus, in the non-relativistic gravity limit, $\Phi(\bm{x})$ is the same in each superposition branch.  This contrasts with previous works that have considered classical gravity creating entanglement through non-local processes where $\Phi(\bm{x})$ is essentially  implicitly or explicitly taken to act as a quantum operator and thus has different values in different superposition branches, as described further in Appendix \ref{sec:nonlocalCG}. In contrast, in the process described here, the virtual matter particles have to travel different distances in each superposition branch and thus go into a quantum superposition, resulting in  entanglement, as discussed in Section \ref{sec:CG}. Note that despite taking the non-relativistic gravity limit, the process that is being described can be considered local in the sense that   it involves the local interaction of virtual particles with the localized real particle wavepackets. Since it is a quantum mediator that is entangling the objects, this process does not break previous theorems on how LOCC cannot create entanglement.  Instead, we have taken a more general view of the gravitational interaction where it is not just the mediation of the gravitational field.

\bmhead{Acknowledgements}

We acknowledge the support of 
grant ID 62312  from the John Templeton Foundation, as part of the QISS project. RH also acknowledges grant ID 62420 from the John Templeton Foundation. The opinions expressed in this publication are those of the authors and do not necessarily reflect the views of the John Templeton Foundation. RH thanks members of the workshop ``A look at the interface between gravity and quantum theory 2024''  for insightful  discussions. Note that this  preprint has not undergone peer review. The Version of Record of this article is published in \emph{Nature}, and is available online at \url{http://doi.org/10.1038/s41586-025-09595-7}.  We sincerely thank the referees of the published article for their insightful and constructive comments, which helped to significantly improve the manuscript.

\bibliography{sn-bibliography}


\begin{thebibliography}{115}
\ifx \bisbn   \undefined \def \bisbn  #1{ISBN #1}\fi
\ifx \binits  \undefined \def \binits#1{#1}\fi
\ifx \bauthor  \undefined \def \bauthor#1{#1}\fi
\ifx \batitle  \undefined \def \batitle#1{#1}\fi
\ifx \bjtitle  \undefined \def \bjtitle#1{#1}\fi
\ifx \bvolume  \undefined \def \bvolume#1{\textbf{#1}}\fi
\ifx \byear  \undefined \def \byear#1{#1}\fi
\ifx \bissue  \undefined \def \bissue#1{#1}\fi
\ifx \bfpage  \undefined \def \bfpage#1{#1}\fi
\ifx \blpage  \undefined \def \blpage #1{#1}\fi
\ifx \burl  \undefined \def \burl#1{\textsf{#1}}\fi
\ifx \doiurl  \undefined \def \doiurl#1{\url{https://doi.org/#1}}\fi
\ifx \betal  \undefined \def \betal{\textit{et al.}}\fi
\ifx \binstitute  \undefined \def \binstitute#1{#1}\fi
\ifx \binstitutionaled  \undefined \def \binstitutionaled#1{#1}\fi
\ifx \bctitle  \undefined \def \bctitle#1{#1}\fi
\ifx \beditor  \undefined \def \beditor#1{#1}\fi
\ifx \bpublisher  \undefined \def \bpublisher#1{#1}\fi
\ifx \bbtitle  \undefined \def \bbtitle#1{#1}\fi
\ifx \bedition  \undefined \def \bedition#1{#1}\fi
\ifx \bseriesno  \undefined \def \bseriesno#1{#1}\fi
\ifx \blocation  \undefined \def \blocation#1{#1}\fi
\ifx \bsertitle  \undefined \def \bsertitle#1{#1}\fi
\ifx \bsnm \undefined \def \bsnm#1{#1}\fi
\ifx \bsuffix \undefined \def \bsuffix#1{#1}\fi
\ifx \bparticle \undefined \def \bparticle#1{#1}\fi
\ifx \barticle \undefined \def \barticle#1{#1}\fi
\bibcommenthead
\ifx \bconfdate \undefined \def \bconfdate #1{#1}\fi
\ifx \botherref \undefined \def \botherref #1{#1}\fi
\ifx \url \undefined \def \url#1{\textsf{#1}}\fi
\ifx \bchapter \undefined \def \bchapter#1{#1}\fi
\ifx \bbook \undefined \def \bbook#1{#1}\fi
\ifx \bcomment \undefined \def \bcomment#1{#1}\fi
\ifx \oauthor \undefined \def \oauthor#1{#1}\fi
\ifx \citeauthoryear \undefined \def \citeauthoryear#1{#1}\fi
\ifx \endbibitem  \undefined \def \endbibitem {}\fi
\ifx \bconflocation  \undefined \def \bconflocation#1{#1}\fi
\ifx \arxivurl  \undefined \def \arxivurl#1{\textsf{#1}}\fi
\csname PreBibitemsHook\endcsname

\bibitem[\protect\citeauthoryear{Rovelli}{2001}]{rovelli2001notesbriefhistoryquantum}
\begin{bchapter}
\bauthor{\bsnm{Rovelli}, \binits{C.}}:
\bctitle{Notes for a brief history of quantum gravity}.
In: \beditor{\bsnm{Rovelli}, \binits{C.}} (ed.)
\bbtitle{Quantum Gravity}.
\bpublisher{World Scientific},
\blocation{Singapore}
(\byear{2001}).
\doiurl{10.1142/9789812777386\_0059} .
\burl{https://www.worldscientific.com/doi/10.1142/9789812777386\_0059}
\end{bchapter}
\endbibitem

\bibitem[\protect\citeauthoryear{Carlip}{2008}]{Carlip_2008}
\begin{barticle}
\bauthor{\bsnm{Carlip}, \binits{S.}}:
\batitle{Is quantum gravity necessary?}
\bjtitle{Classical and Quantum Gravity}
\bvolume{25}(\bissue{15}),
\bfpage{154010}
(\byear{2008})
\doiurl{10.1088/0264-9381/25/15/154010}
\end{barticle}
\endbibitem

\bibitem[\protect\citeauthoryear{{Feynman}}{1957}]{FeynmanQG}
\begin{bchapter}
\bauthor{\bsnm{{Feynman}}, \binits{R.}}:
\bctitle{{The role of gravitation in physics}}.
In: \beditor{\bsnm{DeWitt}, \binits{C.M.}},
\beditor{\bsnm{Rickles}, \binits{D.}} (eds.)
\bbtitle{{Chapel Hill Conference Proceedings}},
pp. \bfpage{250}--\blpage{256}.
\bpublisher{Edition Open Access},
\blocation{North Carolina}
(\byear{1957}).
\doiurl{10.34663/9783945561294-00} .
\burl{http://www.edition-open-sources.org/sources/5/index.html}
\end{bchapter}
\endbibitem

\bibitem[\protect\citeauthoryear{Bose et~al.}{2017}]{bose2017spin}
\begin{barticle}
\bauthor{\bsnm{Bose}, \binits{S.}},
\bauthor{\bsnm{Mazumdar}, \binits{A.}},
\bauthor{\bsnm{Morley}, \binits{G.W.}},
\bauthor{\bsnm{Ulbricht}, \binits{H.}},
\bauthor{\bsnm{Toro\v{s}}, \binits{M.}},
\bauthor{\bsnm{Paternostro}, \binits{M.}},
\bauthor{\bsnm{Geraci}, \binits{A.A.}},
\bauthor{\bsnm{Barker}, \binits{P.F.}},
\bauthor{\bsnm{Kim}, \binits{M.S.}},
\bauthor{\bsnm{Milburn}, \binits{G.}}:
\batitle{{Spin Entanglement Witness for Quantum Gravity}}.
\bjtitle{Phys. Rev. Lett.}
\bvolume{119},
\bfpage{240401}
(\byear{2017})
\doiurl{10.1103/PhysRevLett.119.240401}
\end{barticle}
\endbibitem

\bibitem[\protect\citeauthoryear{Marletto and Vedral}{2017}]{marletto2017gravitationallyinduced}
\begin{barticle}
\bauthor{\bsnm{Marletto}, \binits{C.}},
\bauthor{\bsnm{Vedral}, \binits{V.}}:
\batitle{{Gravitationally Induced Entanglement between Two Massive Particles is Sufficient Evidence of Quantum Effects in Gravity}}.
\bjtitle{Phys. Rev. Lett.}
\bvolume{119},
\bfpage{240402}
(\byear{2017})
\doiurl{10.1103/PhysRevLett.119.240402}
\end{barticle}
\endbibitem

\bibitem[\protect\citeauthoryear{Kafri and Taylor}{2013}]{kafri2013noise}
\begin{barticle}
\bauthor{\bsnm{Kafri}, \binits{D.}},
\bauthor{\bsnm{Taylor}, \binits{J.}}:
\batitle{A noise inequality for classical forces}.
\bjtitle{arXiv preprint arXiv:1311.4558}
(\byear{2013})
\doiurl{https://arxiv.org/abs/1311.4558}
{[quant-ph]}
\end{barticle}
\endbibitem

\bibitem[\protect\citeauthoryear{Kafri et~al.}{2014}]{kafri2014classical}
\begin{barticle}
\bauthor{\bsnm{Kafri}, \binits{D.}},
\bauthor{\bsnm{Taylor}, \binits{J.M.}},
\bauthor{\bsnm{Milburn}, \binits{G.J.}}:
\batitle{{A Classical Channel Model for Gravitational Decoherence}}.
\bjtitle{New J. Phys.}
\bvolume{16}(\bissue{6}),
\bfpage{065020}
(\byear{2014})
\doiurl{10/gknst9}
\end{barticle}
\endbibitem

\bibitem[\protect\citeauthoryear{Krisnanda et~al.}{2017}]{krisnanda2017revealing}
\begin{barticle}
\bauthor{\bsnm{Krisnanda}, \binits{T.}},
\bauthor{\bsnm{Zuppardo}, \binits{M.}},
\bauthor{\bsnm{Paternostro}, \binits{M.}},
\bauthor{\bsnm{Paterek}, \binits{T.}}:
\batitle{{Revealing Non-Classicality of Inaccessible Objects}}.
\bjtitle{Phys. Rev. Lett.}
\bvolume{119}(\bissue{12}),
\bfpage{120402}
(\byear{2017})
\doiurl{10.1103/physrevlett.119.120402}
{\href{https://arxiv.org/abs/1607.01140}{{arXiv:1607.01140}}}
\end{barticle}
\endbibitem

\bibitem[\protect\citeauthoryear{Marletto and Vedral}{2020}]{marletto2020witnessing}
\begin{barticle}
\bauthor{\bsnm{Marletto}, \binits{C.}},
\bauthor{\bsnm{Vedral}, \binits{V.}}:
\batitle{Witnessing nonclassicality beyond quantum theory}.
\bjtitle{Phys. Rev. D}
\bvolume{102},
\bfpage{086012}
(\byear{2020})
\doiurl{10.1103/PhysRevD.102.086012}
\end{barticle}
\endbibitem

\bibitem[\protect\citeauthoryear{Galley et~al.}{2022}]{Galley2022nogotheoremnatureof}
\begin{barticle}
\bauthor{\bsnm{Galley}, \binits{T.D.}},
\bauthor{\bsnm{Giacomini}, \binits{F.}},
\bauthor{\bsnm{Selby}, \binits{J.H.}}:
\batitle{{A no-go theorem on the nature of the gravitational field beyond quantum theory}}.
\bjtitle{{Quantum}}
\bvolume{6},
\bfpage{779}
(\byear{2022})
\doiurl{10.22331/q-2022-08-17-779}
\end{barticle}
\endbibitem

\bibitem[\protect\citeauthoryear{Ludescher et~al.}{2025}]{ludescher2025gravity}
\begin{barticle}
\bauthor{\bsnm{Ludescher}, \binits{S.L.}},
\bauthor{\bsnm{Loveridge}, \binits{L.D.}},
\bauthor{\bsnm{Galley}, \binits{T.D.}},
\bauthor{\bsnm{M{\"u}ller}, \binits{M.P.}}:
\batitle{Gravity-mediated entanglement via infinite-dimensional systems}.
\bjtitle{arXiv preprint arXiv:2507.13201}
(\byear{2025})
\doiurl{10.48550/arXiv.2507.13201}
\end{barticle}
\endbibitem

\bibitem[\protect\citeauthoryear{Bose et~al.}{2025}]{bose2023massivequantumsystemsinterfaces}
\begin{barticle}
\bauthor{\bsnm{Bose}, \binits{S.}},
\bauthor{\bsnm{Fuentes}, \binits{I.}},
\bauthor{\bsnm{Geraci}, \binits{A.A.}},
\bauthor{\bsnm{Khan}, \binits{S.M.}},
\bauthor{\bsnm{Qvarfort}, \binits{S.}},
\bauthor{\bsnm{Rademacher}, \binits{M.}},
\bauthor{\bsnm{Rashid}, \binits{M.}},
\bauthor{\bsnm{Toro\v{s}}, \binits{M.}},
\bauthor{\bsnm{Ulbricht}, \binits{H.}},
\bauthor{\bsnm{Wanjura}, \binits{C.C.}}:
\batitle{Massive quantum systems as interfaces of quantum mechanics and gravity}.
\bjtitle{Rev. Mod. Phys.}
\bvolume{97},
\bfpage{015003}
(\byear{2025})
\doiurl{10.1103/RevModPhys.97.015003}
\end{barticle}
\endbibitem

\bibitem[\protect\citeauthoryear{Huggett et~al.}{2023}]{huggett2022quantum}
\begin{bbook}
\bauthor{\bsnm{Huggett}, \binits{N.}},
\bauthor{\bsnm{Linnemann}, \binits{N.}},
\bauthor{\bsnm{Schneider}, \binits{M.D.}}:
\bbtitle{{Quantum Gravity in a Laboratory?}}
\bsertitle{Elements in the Foundations of Contemporary Physics}.
\bpublisher{Cambridge University Press},
\blocation{Cambridge}
(\byear{2023}).
\doiurl{10.1017/9781009327541}
\end{bbook}
\endbibitem

\bibitem[\protect\citeauthoryear{Carney et~al.}{2019}]{Carney_2019}
\begin{barticle}
\bauthor{\bsnm{Carney}, \binits{D.}},
\bauthor{\bsnm{Stamp}, \binits{P.C.E.}},
\bauthor{\bsnm{Taylor}, \binits{J.M.}}:
\batitle{Tabletop experiments for quantum gravity: a user’s manual}.
\bjtitle{Classical and Quantum Gravity}
\bvolume{36}(\bissue{3}),
\bfpage{034001}
(\byear{2019})
\doiurl{10.1088/1361-6382/aaf9ca}
\end{barticle}
\endbibitem

\bibitem[\protect\citeauthoryear{Delić et~al.}{2020}]{CoolingMicrosolid}
\begin{barticle}
\bauthor{\bsnm{Delić}, \binits{U.}},
\bauthor{\bsnm{Reisenbauer}, \binits{M.}},
\bauthor{\bsnm{Dare}, \binits{K.}},
\bauthor{\bsnm{Grass}, \binits{D.}},
\bauthor{\bsnm{Vuletić}, \binits{V.}},
\bauthor{\bsnm{Kiesel}, \binits{N.}},
\bauthor{\bsnm{Aspelmeyer}, \binits{M.}}:
\batitle{Cooling of a levitated nanoparticle to the motional quantum ground state}.
\bjtitle{Science}
\bvolume{367}(\bissue{6480}),
\bfpage{892}--\blpage{895}
(\byear{2020})
\doiurl{10.1126/science.aba3993}
\end{barticle}
\endbibitem

\bibitem[\protect\citeauthoryear{Westphal et~al.}{2021}]{westphal2021measurement}
\begin{barticle}
\bauthor{\bsnm{Westphal}, \binits{T.}},
\bauthor{\bsnm{Hepach}, \binits{H.}},
\bauthor{\bsnm{Pfaff}, \binits{J.}},
\bauthor{\bsnm{Aspelmeyer}, \binits{M.}}:
\batitle{Measurement of gravitational coupling between millimetre-sized masses}.
\bjtitle{Nature}
\bvolume{591}(\bissue{7849}),
\bfpage{225}--\blpage{228}
(\byear{2021})
\doiurl{10.1038/s41586-021-03250-7}
\end{barticle}
\endbibitem

\bibitem[\protect\citeauthoryear{Panda et~al.}{2024}]{panda2024measuring}
\begin{barticle}
\bauthor{\bsnm{Panda}, \binits{C.D.}},
\bauthor{\bsnm{Tao}, \binits{M.J.}},
\bauthor{\bsnm{Ceja}, \binits{M.}},
\bauthor{\bsnm{Khoury}, \binits{J.}},
\bauthor{\bsnm{Tino}, \binits{G.M.}},
\bauthor{\bsnm{M{\"u}ller}, \binits{H.}}:
\batitle{Measuring gravitational attraction with a lattice atom interferometer}.
\bjtitle{Nature}
\bvolume{631},
\bfpage{1}--\blpage{6}
(\byear{2024})
\doiurl{10.1038/s41586-024-07561-3}
\end{barticle}
\endbibitem

\bibitem[\protect\citeauthoryear{Margalit et~al.}{2021}]{Folman}
\begin{barticle}
\bauthor{\bsnm{Margalit}, \binits{Y.}},
\bauthor{\bsnm{Dobkowski}, \binits{O.}},
\bauthor{\bsnm{Zhou}, \binits{Z.}},
\bauthor{\bsnm{Amit}, \binits{O.}},
\bauthor{\bsnm{Japha}, \binits{Y.}},
\bauthor{\bsnm{Moukouri}, \binits{S.}},
\bauthor{\bsnm{Rohrlich}, \binits{D.}},
\bauthor{\bsnm{Mazumdar}, \binits{A.}},
\bauthor{\bsnm{Bose}, \binits{S.}},
\bauthor{\bsnm{Henkel}, \binits{C.}},
\bauthor{\bsnm{Folman}, \binits{R.}}:
\batitle{{Realization of a complete Stern-Gerlach interferometer: Toward a test of quantum gravity}}.
\bjtitle{Science Advances}
\bvolume{7}(\bissue{22}),
\bfpage{2879}
(\byear{2021})
\doiurl{10.1126/sciadv.abg2879}
\end{barticle}
\endbibitem

\bibitem[\protect\citeauthoryear{Christodoulou et~al.}{2023}]{christodoulou2022gravity}
\begin{barticle}
\bauthor{\bsnm{Christodoulou}, \binits{M.}},
\bauthor{\bsnm{Biagio}, \binits{A.D.}},
\bauthor{\bsnm{Howl}, \binits{R.}},
\bauthor{\bsnm{Rovelli}, \binits{C.}}:
\batitle{Gravity entanglement, quantum reference systems, degrees of freedom}.
\bjtitle{Classical and Quantum Gravity}
\bvolume{40}(\bissue{4}),
\bfpage{047001}
(\byear{2023})
\doiurl{10.1088/1361-6382/acb0aa}
\end{barticle}
\endbibitem

\bibitem[\protect\citeauthoryear{Marletto and Vedral}{2019}]{marletto2019answers}
\begin{barticle}
\bauthor{\bsnm{Marletto}, \binits{C.}},
\bauthor{\bsnm{Vedral}, \binits{V.}}:
\batitle{{Answers to a few questions regarding the BMV experiment}}.
\bjtitle{arXiv:1907.08994}
(\byear{2019})
\doiurl{10.48550/arXiv.1907.08994}
\end{barticle}
\endbibitem

\bibitem[\protect\citeauthoryear{Bose et~al.}{2022}]{Bose2022}
\begin{barticle}
\bauthor{\bsnm{Bose}, \binits{S.}},
\bauthor{\bsnm{Mazumdar}, \binits{A.}},
\bauthor{\bsnm{Schut}, \binits{M.}},
\bauthor{\bsnm{Toro\v{s}}, \binits{M.}}:
\batitle{Mechanism for the quantum natured gravitons to entangle masses}.
\bjtitle{Phys. Rev. D}
\bvolume{105},
\bfpage{106028}
(\byear{2022})
\doiurl{10.1103/PhysRevD.105.106028}
\end{barticle}
\endbibitem

\bibitem[\protect\citeauthoryear{Cao}{1999}]{Cao_1999}
\begin{bbook}
\beditor{\bsnm{Cao}, \binits{T.}} (ed.):
\bbtitle{Conceptual Foundations of Quantum Field Theory}.
\bpublisher{Cambridge University Press},
\blocation{Cambridge}
(\byear{1999}).
\doiurl{10.1017/CBO9780511470813}
\end{bbook}
\endbibitem

\bibitem[\protect\citeauthoryear{Howl et~al.}{2021}]{howl2021nongaussianity}
\begin{barticle}
\bauthor{\bsnm{Howl}, \binits{R.}},
\bauthor{\bsnm{Vedral}, \binits{V.}},
\bauthor{\bsnm{Naik}, \binits{D.}},
\bauthor{\bsnm{Christodoulou}, \binits{M.}},
\bauthor{\bsnm{Rovelli}, \binits{C.}},
\bauthor{\bsnm{Iyer}, \binits{A.}}:
\batitle{Non-{{Gaussianity}} as a signature of a quantum theory of gravity}.
\bjtitle{PRX Quantum}
\bvolume{2}(\bissue{1}),
\bfpage{010325}
(\byear{2021})
\doiurl{10/gkq6wg}
{\href{https://arxiv.org/abs/2004.01189}{{arXiv:2004.01189}}}
\end{barticle}
\endbibitem

\bibitem[\protect\citeauthoryear{Christodoulou et~al.}{2023}]{LocallyHowl}
\begin{barticle}
\bauthor{\bsnm{Christodoulou}, \binits{M.}},
\bauthor{\bsnm{Di~Biagio}, \binits{A.}},
\bauthor{\bsnm{Aspelmeyer}, \binits{M.}},
\bauthor{\bsnm{Brukner}, \binits{{\v{C}}.}},
\bauthor{\bsnm{Rovelli}, \binits{C.}},
\bauthor{\bsnm{Howl}, \binits{R.}}:
\batitle{Locally mediated entanglement in linearized quantum gravity}.
\bjtitle{Phys. Rev. Lett.}
\bvolume{130},
\bfpage{100202}
(\byear{2023})
\doiurl{10.1103/PhysRevLett.130.100202}
\end{barticle}
\endbibitem

\bibitem[\protect\citeauthoryear{Wallace}{2022}]{wallace2022quantum}
\begin{barticle}
\bauthor{\bsnm{Wallace}, \binits{D.}}:
\batitle{Quantum gravity at low energies}.
\bjtitle{Studies in History and Philosophy of Science}
\bvolume{94},
\bfpage{31}--\blpage{46}
(\byear{2022})
\doiurl{10.1016/j.shpsa.2022.04.003}
\end{barticle}
\endbibitem

\bibitem[\protect\citeauthoryear{Donoghue}{2017}]{Donoghue:2017ovt}
\begin{barticle}
\bauthor{\bsnm{Donoghue}, \binits{J.}}:
\batitle{{Quantum gravity as a low energy effective field theory}}.
\bjtitle{Scholarpedia}
\bvolume{12}(\bissue{4}),
\bfpage{32997}
(\byear{2017})
\doiurl{10.4249/scholarpedia.32997}
\end{barticle}
\endbibitem

\bibitem[\protect\citeauthoryear{Peskin and Schroeder}{1995}]{Peskin:1995ev}
\begin{bbook}
\bauthor{\bsnm{Peskin}, \binits{M.E.}},
\bauthor{\bsnm{Schroeder}, \binits{D.V.}}:
\bbtitle{{An Introduction to Quantum Field Theory}}.
\bpublisher{Addison-Wesley},
\blocation{Reading, USA}
(\byear{1995}).
\doiurl{10.1201/9780429503559}
\end{bbook}
\endbibitem

\bibitem[\protect\citeauthoryear{Di~Biagio et~al.}{2023}]{dibiagio2023relativisticlocalityimplysubsystem}
\begin{botherref}
\oauthor{\bsnm{Di~Biagio}, \binits{A.}},
\oauthor{\bsnm{Howl}, \binits{R.}},
\oauthor{\bsnm{Brukner}, \binits{{\v{C}}.}},
\oauthor{\bsnm{Rovelli}, \binits{C.}},
\oauthor{\bsnm{Christodoulou}, \binits{M.}}:
Circuit locality from relativistic locality in scalar field mediated entanglement.
arXiv preprint arXiv:1311.4558,
(2023)
\doiurl{10.48550/arXiv.2305.05645}
\end{botherref}
\endbibitem

\bibitem[\protect\citeauthoryear{Fewster and Verch}{2015}]{fewster2019algebraicquantumfieldtheory}
\begin{bchapter}
\bauthor{\bsnm{Fewster}, \binits{C.J.}},
\bauthor{\bsnm{Verch}, \binits{R.}}:
\bctitle{Algebraic quantum field theory in curved spacetimes}.
In: \beditor{\bsnm{Brunetti}, \binits{R.}},
\beditor{\bsnm{Dappiaggi}, \binits{C.}},
\beditor{\bsnm{Fredenhagen}, \binits{K.}},
\beditor{\bsnm{Yngvason}, \binits{J.}} (eds.)
\bbtitle{Advances in Algebraic Quantum Field Theory}.
\bsertitle{Mathematical Physics Studies},
pp. \bfpage{125}--\blpage{189}.
\bpublisher{Springer},
\blocation{Cham, Switzerland}
(\byear{2015}).
\doiurl{10.1007/978-3-319-21353-8_4}
\end{bchapter}
\endbibitem

\bibitem[\protect\citeauthoryear{Weinberg}{1995}]{WeinbergVol1}
\begin{bbook}
\bauthor{\bsnm{Weinberg}, \binits{S.}}:
\bbtitle{The Quantum Theory of Fields, Volume 1: Foundations},
\bedition{1st} edn.
\bpublisher{Cambridge University Press},
\blocation{Cambridge, UK}
(\byear{1995}).
\doiurl{10.1017/CBO9781139644167}
\end{bbook}
\endbibitem

\bibitem[\protect\citeauthoryear{Mandl and Shaw}{2010}]{mandl2013quantum}
\begin{bbook}
\bauthor{\bsnm{Mandl}, \binits{F.}},
\bauthor{\bsnm{Shaw}, \binits{G.}}:
\bbtitle{Quantum Field Theory},
\bedition{2nd} edn.
\bpublisher{Wiley},
\blocation{Hoboken, NJ}
(\byear{2010})
\end{bbook}
\endbibitem

\bibitem[\protect\citeauthoryear{Dereziński}{2014}]{QFTClassSources}
\begin{barticle}
\bauthor{\bsnm{Dereziński}, \binits{J.}}:
\batitle{{Quantum fields with classical perturbations}}.
\bjtitle{Journal of Mathematical Physics}
\bvolume{55}(\bissue{7}),
\bfpage{075201}
(\byear{2014})
\doiurl{10.1063/1.4878920}
\end{barticle}
\endbibitem

\bibitem[\protect\citeauthoryear{Howl et~al.}{2019}]{Howl_2019}
\begin{barticle}
\bauthor{\bsnm{Howl}, \binits{R.}},
\bauthor{\bsnm{Penrose}, \binits{R.}},
\bauthor{\bsnm{Fuentes}, \binits{I.}}:
\batitle{{Exploring the unification of quantum theory and general relativity with a Bose–Einstein condensate}}.
\bjtitle{New J. Phys.}
\bvolume{21}(\bissue{4}),
\bfpage{043047}
(\byear{2019})
\doiurl{10.1088/1367-2630/ab104a}
\end{barticle}
\endbibitem

\bibitem[\protect\citeauthoryear{Aspelmeyer}{2022}]{Aspelmeyer2022}
\begin{bbook}
\bauthor{\bsnm{Aspelmeyer}, \binits{M.}}:
In: \beditor{\bsnm{Kiefer}, \binits{C.}} (ed.)
\bbtitle{{When Zeh meets Feynman: How to avoid the appearance of a classical World in gravity experiments}},
pp. \bfpage{85}--\blpage{95}.
\bpublisher{Springer},
\blocation{Cham}
(\byear{2022}).
\doiurl{10.1007/978-3-030-88781-0_5}
\end{bbook}
\endbibitem

\bibitem[\protect\citeauthoryear{Lapponi et~al.}{2025}]{lapponi2025gravitationalredshiftquantizedlinear}
\begin{botherref}
\oauthor{\bsnm{Lapponi}, \binits{A.}},
\oauthor{\bsnm{Ferreri}, \binits{A.}},
\oauthor{\bsnm{Bruschi}, \binits{D.E.}}:
Gravitational redshift via quantized linear gravity
(2025).
\url{https://arxiv.org/abs/2504.03956}
\end{botherref}
\endbibitem

\bibitem[\protect\citeauthoryear{Gupta}{1952}]{QGLorentzGupta1}
\begin{barticle}
\bauthor{\bsnm{Gupta}, \binits{S.N.}}:
\batitle{{Quantization of Einstein's Gravitational Field: Linear Approximation}}.
\bjtitle{Proceedings of the Physical Society. Section A}
\bvolume{65}(\bissue{3}),
\bfpage{161}
(\byear{1952})
\doiurl{10.1088/0370-1298/65/3/301}
\end{barticle}
\endbibitem

\bibitem[\protect\citeauthoryear{Gupta}{1968}]{QGLorentzGupta2}
\begin{barticle}
\bauthor{\bsnm{Gupta}, \binits{S.N.}}:
\batitle{Supplementary conditions in the quantized gravitational theory}.
\bjtitle{Phys. Rev.}
\bvolume{172},
\bfpage{1303}--\blpage{1307}
(\byear{1968})
\doiurl{10.1103/PhysRev.172.1303}
\end{barticle}
\endbibitem

\bibitem[\protect\citeauthoryear{Pav{\v{s}}i{\v{c}}}{2018}]{pavvsivc2018localized}
\begin{barticle}
\bauthor{\bsnm{Pav{\v{s}}i{\v{c}}}, \binits{M.}}:
\batitle{Localized states in quantum field theory}.
\bjtitle{Advances in Applied Clifford Algebras}
\bvolume{28}(\bissue{5}),
\bfpage{1}--\blpage{29}
(\byear{2018})
\doiurl{10.1007/s00006-018-0904-5}
\end{barticle}
\endbibitem

\bibitem[\protect\citeauthoryear{}{}]{fn6}
\begin{botherref}
In the non-relativistic limit, $\hat{H}_I$ would in principle also allow states of the form $|N-k\rangle_{1i}|N+k\rangle_{2j} $ and $|N+k\rangle_{1i}|N-k\rangle_{2j} $, where $k$ is some integer. These states are considered not seen in the experiment, for example, due to their different interactions with a magnetic field of a Stern-Gerlach device compared to $|N\rangle_{1i}|N\rangle_{2j} $. These states would only contribute to the entanglement of the final state since there can be no states of the form $|N-k\rangle_{1i}|N-k\rangle_{2j} $ or $|N+k\rangle_{1i}|N+k\rangle_{2j} $, as the total particle number is conserved in the non-relativistic limit.
\end{botherref}
\endbibitem

\bibitem[\protect\citeauthoryear{Maggiore}{2008}]{GWBook}
\begin{bbook}
\bauthor{\bsnm{Maggiore}, \binits{M.}}:
\bbtitle{Gravitational Waves {{Volume}} 1: {{Theory}} and {{Experiments}}}.
\bpublisher{{Oxford University Press}},
\blocation{{Oxford}}
(\byear{2008})
\end{bbook}
\endbibitem

\bibitem[\protect\citeauthoryear{DeWitt}{1967}]{PhysRev.162.1195}
\begin{barticle}
\bauthor{\bsnm{DeWitt}, \binits{B.S.}}:
\batitle{{Quantum Theory of Gravity. II. The Manifestly Covariant Theory}}.
\bjtitle{Phys. Rev.}
\bvolume{162},
\bfpage{1195}--\blpage{1239}
(\byear{1967})
\doiurl{10.1103/PhysRev.162.1195}
\end{barticle}
\endbibitem

\bibitem[\protect\citeauthoryear{}{}]{fn10}
\begin{botherref}
Note that for the phase factor to be $\approx e^{ - i m c x^0/\hbar}$ we generally also require $t \lesssim 2 m R^2 / \hbar$ unless further constraints are applied. Alternatively, higher order terms can be considered.
\end{botherref}
\endbibitem

\bibitem[\protect\citeauthoryear{}{}]{fn15}
\begin{botherref}
The approximation is good for $ R \gg \hbar / (mc)$ and $t \lesssim 2 m R^2 / \hbar$ as required also for \eqref{eq:WickN} as detailed above.
\end{botherref}
\endbibitem

\bibitem[\protect\citeauthoryear{Ogievetsky and Polubarinov}{1965}]{OGIEVETSKY1965167}
\begin{barticle}
\bauthor{\bsnm{Ogievetsky}, \binits{V.I.}},
\bauthor{\bsnm{Polubarinov}, \binits{I.V.}}:
\batitle{Interacting field of spin 2 and the einstein equations}.
\bjtitle{Annals of Physics}
\bvolume{35}(\bissue{2}),
\bfpage{167}--\blpage{208}
(\byear{1965})
\doiurl{10.1016/0003-4916(65)90077-1}
\end{barticle}
\endbibitem

\bibitem[\protect\citeauthoryear{de~Rham and Gabadadze}{2010}]{PhysRevD.82.044020}
\begin{barticle}
\bauthor{\bsnm{Rham}, \binits{C.}},
\bauthor{\bsnm{Gabadadze}, \binits{G.}}:
\batitle{Generalization of the fierz-pauli action}.
\bjtitle{Phys. Rev. D}
\bvolume{82},
\bfpage{044020}
(\byear{2010})
\doiurl{10.1103/PhysRevD.82.044020}
\end{barticle}
\endbibitem

\bibitem[\protect\citeauthoryear{}{}]{fn1}
\begin{botherref}
Such as gravitational self-interactions of the matter objects.
\end{botherref}
\endbibitem

\bibitem[\protect\citeauthoryear{Vidal and Werner}{2002}]{EntNegativity}
\begin{barticle}
\bauthor{\bsnm{Vidal}, \binits{G.}},
\bauthor{\bsnm{Werner}, \binits{R.F.}}:
\batitle{Computable measure of entanglement}.
\bjtitle{Phys. Rev. A}
\bvolume{65},
\bfpage{032314}
(\byear{2002})
\doiurl{10.1103/PhysRevA.65.032314}
\end{barticle}
\endbibitem

\bibitem[\protect\citeauthoryear{Krisnanda et~al.}{2020}]{krisnanda2020observable}
\begin{barticle}
\bauthor{\bsnm{Krisnanda}, \binits{T.}},
\bauthor{\bsnm{Tham}, \binits{G.Y.}},
\bauthor{\bsnm{Paternostro}, \binits{M.}},
\bauthor{\bsnm{Paterek}, \binits{T.}}:
\batitle{{Observable Quantum Entanglement Due to Gravity}}.
\bjtitle{npj Quantum Information}
\bvolume{6}(\bissue{1}),
\bfpage{12}
(\byear{2020})
\doiurl{10/ggz5q7}
{\href{https://arxiv.org/abs/1906.08808}{{arXiv:1906.08808}}}
\end{barticle}
\endbibitem

\bibitem[\protect\citeauthoryear{Howl et~al.}{2023}]{howl2023gravitationallyinducedentanglementcoldatoms}
\begin{barticle}
\bauthor{\bsnm{Howl}, \binits{R.}},
\bauthor{\bsnm{Cooper}, \binits{N.}},
\bauthor{\bsnm{Hackermüller}, \binits{L.}}:
\batitle{Gravitationally-induced entanglement in cold atoms}.
\bjtitle{arXiv:2304.00734}
(\byear{2023})
\doiurl{10.48550/arXiv.2304.00734}
{[quant-ph]}
\end{barticle}
\endbibitem

\bibitem[\protect\citeauthoryear{Christodoulou and Rovelli}{2019}]{christodoulou2019possibility}
\begin{barticle}
\bauthor{\bsnm{Christodoulou}, \binits{M.}},
\bauthor{\bsnm{Rovelli}, \binits{C.}}:
\batitle{{On the Possibility of Laboratory Evidence for Quantum Superposition of Geometries}}.
\bjtitle{Physics Letters B}
\bvolume{792},
\bfpage{64}--\blpage{68}
(\byear{2019})
\doiurl{10/gj6ssc}
\end{barticle}
\endbibitem

\bibitem[\protect\citeauthoryear{{van de Kamp, Thomas W. and Marshman, Ryan J. and Bose, Sougato and Mazumdar, Anupam}}{2020}]{PhysRevA.102.062807}
\begin{barticle}
\bauthor{\bsnm{{van de Kamp, Thomas W. and Marshman, Ryan J. and Bose, Sougato and Mazumdar, Anupam}}}:
\batitle{{Quantum gravity witness via entanglement of masses: Casimir screening}}.
\bjtitle{Phys. Rev. A}
\bvolume{102},
\bfpage{062807}
(\byear{2020})
\doiurl{10.1103/PhysRevA.102.062807}
\end{barticle}
\endbibitem

\bibitem[\protect\citeauthoryear{Pedernales and Plenio}{2023}]{EquivPlenio}
\begin{barticle}
\bauthor{\bsnm{Pedernales}, \binits{J.S.}},
\bauthor{\bsnm{Plenio}, \binits{M.B.}}:
\batitle{On the origin of force sensitivity in tests of quantum gravity with delocalised mechanical systems}.
\bjtitle{Contemporary Physics}
\bvolume{64}(\bissue{2}),
\bfpage{147}--\blpage{163}
(\byear{2023})
\doiurl{10.1080/00107514.2023.2286074}
{\href{https://arxiv.org/abs/https://doi.org/10.1080/00107514.2023.2286074}{{https://doi.org/10.1080/00107514.2023.2286074}}}
\end{barticle}
\endbibitem

\bibitem[\protect\citeauthoryear{Bengyat et~al.}{2024}]{MariosEquiv}
\begin{barticle}
\bauthor{\bsnm{Bengyat}, \binits{O.}},
\bauthor{\bsnm{Di~Biagio}, \binits{A.}},
\bauthor{\bsnm{Aspelmeyer}, \binits{M.}},
\bauthor{\bsnm{Christodoulou}, \binits{M.}}:
\batitle{Gravity-mediated entanglement between oscillators as quantum superposition of geometries}.
\bjtitle{Phys. Rev. D}
\bvolume{110},
\bfpage{056046}
(\byear{2024})
\doiurl{10.1103/PhysRevD.110.056046}
\end{barticle}
\endbibitem

\bibitem[\protect\citeauthoryear{Anastopoulos and Hu}{2018}]{anastopoulos2018commentaspinentanglement}
\begin{botherref}
\oauthor{\bsnm{Anastopoulos}, \binits{C.}},
\oauthor{\bsnm{Hu}, \binits{B.-L.}}:
Comment on ``A Spin Entanglement Witness for Quantum Gravity" and on ``Gravitationally Induced Entanglement Between Two Massive Particles Is Sufficient Evidence of Quantum Effects in Gravity''.
\doiurl{10.48550/arXiv.1804.11315}
\end{botherref}
\endbibitem

\bibitem[\protect\citeauthoryear{Anastopoulos et~al.}{2021}]{Anastopoulos_2021}
\begin{barticle}
\bauthor{\bsnm{Anastopoulos}, \binits{C.}},
\bauthor{\bsnm{Lagouvardos}, \binits{M.}},
\bauthor{\bsnm{Savvidou}, \binits{K.}}:
\batitle{Gravitational effects in macroscopic quantum systems: a first-principles analysis}.
\bjtitle{Classical and Quantum Gravity}
\bvolume{38}(\bissue{15}),
\bfpage{155012}
(\byear{2021})
\doiurl{10.1088/1361-6382/ac0bf9}
\end{barticle}
\endbibitem

\bibitem[\protect\citeauthoryear{Fragkos et~al.}{2022}]{Fragkos2022}
\begin{barticle}
\bauthor{\bsnm{Fragkos}, \binits{V.}},
\bauthor{\bsnm{Kopp}, \binits{M.}},
\bauthor{\bsnm{Pikovski}, \binits{I.}}:
\batitle{{On inference of quantization from gravitationally induced entanglement}}.
\bjtitle{AVS Quantum Science}
\bvolume{4}(\bissue{4}),
\bfpage{045601}
(\byear{2022})
\doiurl{10.1116/5.0101334}
\end{barticle}
\endbibitem

\bibitem[\protect\citeauthoryear{}{}]{fn3}
\begin{botherref}
See Appendix \ref{app:AtlDer} for a derivation of this result in momentum space.
\end{botherref}
\endbibitem

\bibitem[\protect\citeauthoryear{Rosenfeld}{1963}]{ROSENFELD1963353}
\begin{barticle}
\bauthor{\bsnm{Rosenfeld}, \binits{L.}}:
\batitle{On quantization of fields}.
\bjtitle{Nuclear Physics}
\bvolume{40},
\bfpage{353}--\blpage{356}
(\byear{1963})
\doiurl{10.1016/0029-5582(63)90279-7}
\end{barticle}
\endbibitem

\bibitem[\protect\citeauthoryear{M{\o}ller}{1962}]{moller1962theories}
\begin{barticle}
\bauthor{\bsnm{M{\o}ller}, \binits{C.}}:
\batitle{Les th{\'e}ories relativistes de la gravitation}.
\bjtitle{Colloques Internationaux CNRS}
\bvolume{91}(\bissue{1}),
\bfpage{15}--\blpage{29}
(\byear{1962})
\end{barticle}
\endbibitem

\bibitem[\protect\citeauthoryear{Kent}{2018}]{Kent_2018}
\begin{barticle}
\bauthor{\bsnm{Kent}, \binits{A.}}:
\batitle{{Simple refutation of the Eppley–Hannah argument}}.
\bjtitle{Classical and Quantum Gravity}
\bvolume{35}(\bissue{24}),
\bfpage{245008}
(\byear{2018})
\doiurl{10.1088/1361-6382/aaea20}
\end{barticle}
\endbibitem

\bibitem[\protect\citeauthoryear{Kent}{2005}]{kent2005nonlinearity}
\begin{barticle}
\bauthor{\bsnm{Kent}, \binits{A.}}:
\batitle{Nonlinearity without superluminality}.
\bjtitle{Physical Review A}
\bvolume{72}(\bissue{1}),
\bfpage{012108}
(\byear{2005})
\doiurl{10.1103/PhysRevA.72.012108}
\end{barticle}
\endbibitem

\bibitem[\protect\citeauthoryear{Giulini et~al.}{2023}]{Giulini2023}
\begin{bbook}
\bauthor{\bsnm{Giulini}, \binits{D.}},
\bauthor{\bsnm{Gro{\ss}ardt}, \binits{A.}},
\bauthor{\bsnm{Schwartz}, \binits{P.K.}}:
In: \beditor{\bsnm{Pfeifer}, \binits{C.}},
\beditor{\bsnm{L{\"a}mmerzahl}, \binits{C.}} (eds.)
\bbtitle{Coupling Quantum Matter and Gravity},
pp. \bfpage{491}--\blpage{550}.
\bpublisher{Springer},
\blocation{Cham}
(\byear{2023}).
\doiurl{10.1007/978-3-031-31520-6_16}
\end{bbook}
\endbibitem

\bibitem[\protect\citeauthoryear{Diósi}{1987}]{DIOSI1987377}
\begin{barticle}
\bauthor{\bsnm{Diósi}, \binits{L.}}:
\batitle{A universal master equation for the gravitational violation of quantum mechanics}.
\bjtitle{Physics Letters A}
\bvolume{120}(\bissue{8}),
\bfpage{377}--\blpage{381}
(\byear{1987})
\doiurl{10.1016/0375-9601(87)90681-5}
\end{barticle}
\endbibitem

\bibitem[\protect\citeauthoryear{Tilloy and Di\'osi}{2016}]{TilloyDiosi}
\begin{barticle}
\bauthor{\bsnm{Tilloy}, \binits{A.}},
\bauthor{\bsnm{Di\'osi}, \binits{L.}}:
\batitle{Sourcing semiclassical gravity from spontaneously localized quantum matter}.
\bjtitle{Phys. Rev. D}
\bvolume{93},
\bfpage{024026}
(\byear{2016})
\doiurl{10.1103/PhysRevD.93.024026}
\end{barticle}
\endbibitem

\bibitem[\protect\citeauthoryear{Layton et~al.}{2023}]{layton2023weak}
\begin{barticle}
\bauthor{\bsnm{Layton}, \binits{I.}},
\bauthor{\bsnm{Oppenheim}, \binits{J.}},
\bauthor{\bsnm{Russo}, \binits{A.}},
\bauthor{\bsnm{Weller-Davies}, \binits{Z.}}:
\batitle{The weak field limit of quantum matter back-reacting on classical spacetime}.
\bjtitle{Journal of High Energy Physics}
\bvolume{2023}(\bissue{8}),
\bfpage{1}--\blpage{43}
(\byear{2023})
\doiurl{10.1038/s41467-023-43348-2}
\end{barticle}
\endbibitem

\bibitem[\protect\citeauthoryear{Carney and Matsumura}{2025}]{carney2024classical}
\begin{barticle}
\bauthor{\bsnm{Carney}, \binits{D.}},
\bauthor{\bsnm{Matsumura}, \binits{A.}}:
\batitle{Classical-quantum scattering}.
\bjtitle{Classical and Quantum Gravity}
\bvolume{42}(\bissue{13}),
\bfpage{135010}
(\byear{2025})
\doiurl{10.1088/1361-6382/ade589}
\end{barticle}
\endbibitem

\bibitem[\protect\citeauthoryear{Helou and Chen}{2017}]{Helou_2017}
\begin{barticle}
\bauthor{\bsnm{Helou}, \binits{B.}},
\bauthor{\bsnm{Chen}, \binits{Y.}}:
\batitle{{Extensions of Born’s rule to non-linear quantum mechanics, some of which do not imply superluminal communication}}.
\bjtitle{Journal of Physics: Conference Series}
\bvolume{880}(\bissue{1}),
\bfpage{012021}
(\byear{2017})
\doiurl{10.1088/1742-6596/880/1/012021}
\end{barticle}
\endbibitem

\bibitem[\protect\citeauthoryear{}{}]{fn4}
\begin{botherref}
Note that unlike with the classical gravity amplitude $\beta^{(2)}_{ij}$ above, there are no other Feynman diagram amplitudes at fourth order or lower for $\beta^{(4)}_{RL}$ to combine with to make a separable state.
\end{botherref}
\endbibitem

\bibitem[\protect\citeauthoryear{Haine}{2021}]{Haine_2021}
\begin{barticle}
\bauthor{\bsnm{Haine}, \binits{S.A.}}:
\batitle{{Searching for signatures of quantum gravity in quantum gases}}.
\bjtitle{New J. Phys.}
\bvolume{23}(\bissue{3}),
\bfpage{033020}
(\byear{2021})
\doiurl{10.1088/1367-2630/abd97d}
\end{barticle}
\endbibitem

\bibitem[\protect\citeauthoryear{Carney et~al.}{2021}]{carney2021using}
\begin{barticle}
\bauthor{\bsnm{Carney}, \binits{D.}},
\bauthor{\bsnm{M{\"u}ller}, \binits{H.}},
\bauthor{\bsnm{Taylor}, \binits{J.M.}}:
\batitle{{Using an Atom Interferometer to Infer Gravitational Entanglement Generation}}.
\bjtitle{PRX Quantum}
\bvolume{2}(\bissue{3}),
\bfpage{030330}
(\byear{2021})
\doiurl{10/gmvfjc}
{\href{https://arxiv.org/abs/2101.11629}{{arXiv:2101.11629}}}
\end{barticle}
\endbibitem

\bibitem[\protect\citeauthoryear{Oppenheim and Weller-Davies}{2023}]{oppenheim2025covariantpathintegralsquantum}
\begin{barticle}
\bauthor{\bsnm{Oppenheim}, \binits{J.}},
\bauthor{\bsnm{Weller-Davies}, \binits{Z.}}:
\batitle{Covariant path integrals for quantum fields back-reacting on classical space-time}.
\bjtitle{arXiv preprint arXiv:2302.07283}
(\byear{2023})
\doiurl{10.48550/arXiv.2302.07283}
{[gr-qc]}
\end{barticle}
\endbibitem

\bibitem[\protect\citeauthoryear{Gradshteyn and Ryzhik}{2007}]{gradshteyn2014table}
\begin{bbook}
\bauthor{\bsnm{Gradshteyn}, \binits{I.S.}},
\bauthor{\bsnm{Ryzhik}, \binits{I.M.}}:
\bbtitle{{Table of Integrals, Series, and Products}}.
\bpublisher{Academic press},
\blocation{San Diego}
(\byear{2007})
\end{bbook}
\endbibitem

\bibitem[\protect\citeauthoryear{}{}]{fn12}
\begin{botherref}
With the last assumption dropped, we would ultimately obtain: \begin{align}\nonumber \beta &\propto \frac{ M m^2 t }{ \hbar^3} \int d^3 \bm{x}\, d^3 \bm{y} \, d^3 \bm{z} \frac{f(\bm{x}) f^{' \ast}(\bm{y}) \Phi(\bm{z}) \Phi(\bm{y})}{r_{zx} \, r_{zy}} \\\nonumber &\times \left(\frac{1}{(r_{zy} - r_{zx} + i \epsilon)^2} - \frac{1}{(r_{zy} + r_{zx})^2} \right), \end{align} where $r_{zy} := |\bm{z} - \bm{y}|$ and $r_{zx} := |\bm{z} - \bm{x}|$. Assumimng long times and that $\Phi(\bm{k})$ does not support extremely high momentum modes, as well as $d \gg R$, we would then have: \begin{align}\nonumber \beta^{(4)}_{ij} \sim \frac{ M^2 m^4 t^2 }{ \hbar^6} \left(\int d^3 \bm{x}\, d^3 \bm{y} \frac{ \Phi(\bm{x}) \Phi(\bm{y}) \theta_{1i}(\bm{x}) \theta_{2j}(\bm{y})}{|\bm{x} - \bm{y}|^4} \right)^2, \end{align} in this alternative approximation.
\end{botherref}
\endbibitem

\bibitem[\protect\citeauthoryear{Bose et~al.}{2017}]{bose2017spinSupp}
\begin{barticle}
\bauthor{\bsnm{Bose}, \binits{S.}},
\bauthor{\bsnm{Mazumdar}, \binits{A.}},
\bauthor{\bsnm{Morley}, \binits{G.W.}},
\bauthor{\bsnm{Ulbricht}, \binits{H.}},
\bauthor{\bsnm{Toro\v{s}}, \binits{M.}},
\bauthor{\bsnm{Paternostro}, \binits{M.}},
\bauthor{\bsnm{Geraci}, \binits{A.A.}},
\bauthor{\bsnm{Barker}, \binits{P.F.}},
\bauthor{\bsnm{Kim}, \binits{M.S.}},
\bauthor{\bsnm{Milburn}, \binits{G.}}:
\batitle{{Supplementray Material for ``Spin Entanglement Witness for Quantum Gravity'''}}.
\bjtitle{Phys. Rev. Lett.}
\bvolume{119},
\bfpage{240401}
(\byear{2017})
\doiurl{10.1103/PhysRevLett.119.240401}
\end{barticle}
\endbibitem

\bibitem[\protect\citeauthoryear{Dobkowski et~al.}{2025}]{dobkowski2025observationquantumfreefall}
\begin{botherref}
\oauthor{\bsnm{Dobkowski}, \binits{O.}},
\oauthor{\bsnm{Trok}, \binits{B.}},
\oauthor{\bsnm{Skakunenko}, \binits{P.}},
\oauthor{\bsnm{Japha}, \binits{Y.}},
\oauthor{\bsnm{Groswasser}, \binits{D.}},
\oauthor{\bsnm{Efremov}, \binits{M.}},
\oauthor{\bsnm{Marletto}, \binits{C.}},
\oauthor{\bsnm{Fuentes}, \binits{I.}},
\oauthor{\bsnm{Penrose}, \binits{R.}},
\oauthor{\bsnm{Vedral}, \binits{V.}},
\oauthor{\bsnm{Schleich}, \binits{W.P.}},
\oauthor{\bsnm{Folman}, \binits{R.}}:
Observation of quantum free fall and the consistency with the equivalence principle
(2025).
\url{https://arxiv.org/abs/2502.14535}
\end{botherref}
\endbibitem

\bibitem[\protect\citeauthoryear{}{}]{fn9}
\begin{botherref}
Note that since $\Phi(\bm{x})$ is assumed constant with time, no energy flows into the vertices. We are also assuming sufficiently long times that energy-momentum is respected at the vertices to a very good approximation.
\end{botherref}
\endbibitem

\bibitem[\protect\citeauthoryear{}{}]{fn8}
\begin{botherref}
Note that here $\hat{\phi}(x)$ need not be the effective quantum field of the particles of the matter objects.
\end{botherref}
\endbibitem

\bibitem[\protect\citeauthoryear{Page and Geilker}{1981}]{PageExp}
\begin{barticle}
\bauthor{\bsnm{Page}, \binits{D.N.}},
\bauthor{\bsnm{Geilker}, \binits{C.D.}}:
\batitle{{Indirect Evidence for Quantum Gravity}}.
\bjtitle{Phys. Rev. Lett.}
\bvolume{47},
\bfpage{979}--\blpage{982}
(\byear{1981})
\doiurl{10.1103/PhysRevLett.47.979}
\end{barticle}
\endbibitem

\bibitem[\protect\citeauthoryear{Oppenheim}{2023}]{oppenheim2018postquantum}
\begin{barticle}
\bauthor{\bsnm{Oppenheim}, \binits{J.}}:
\batitle{{A Postquantum Theory of Classical Gravity?}}
\bjtitle{Phys. Rev. X}
\bvolume{13},
\bfpage{041040}
(\byear{2023})
\doiurl{10.1103/PhysRevX.13.041040}
\end{barticle}
\endbibitem

\bibitem[\protect\citeauthoryear{Bassi et~al.}{2017}]{Bassi_2017}
\begin{barticle}
\bauthor{\bsnm{Bassi}, \binits{A.}},
\bauthor{\bsnm{Großardt}, \binits{A.}},
\bauthor{\bsnm{Ulbricht}, \binits{H.}}:
\batitle{Gravitational decoherence}.
\bjtitle{Classical and Quantum Gravity}
\bvolume{34}(\bissue{19}),
\bfpage{193002}
(\byear{2017})
\doiurl{10.1088/1361-6382/aa864f}
\end{barticle}
\endbibitem

\bibitem[\protect\citeauthoryear{Tilloy and Di\'osi}{2017}]{LeastDecoh}
\begin{barticle}
\bauthor{\bsnm{Tilloy}, \binits{A.}},
\bauthor{\bsnm{Di\'osi}, \binits{L.}}:
\batitle{{Principle of least decoherence for Newtonian semiclassical gravity}}.
\bjtitle{Phys. Rev. D}
\bvolume{96},
\bfpage{104045}
(\byear{2017})
\doiurl{10.1103/PhysRevD.96.104045}
\end{barticle}
\endbibitem

\bibitem[\protect\citeauthoryear{Di{\'o}si}{1989}]{diosi1989models}
\begin{barticle}
\bauthor{\bsnm{Di{\'o}si}, \binits{L.}}:
\batitle{{Models for Universal Reduction of Macroscopic Quantum Fluctuations}}.
\bjtitle{Phys. Rev. A}
\bvolume{40}(\bissue{3}),
\bfpage{1165}--\blpage{1174}
(\byear{1989})
\doiurl{10/fv2d9m}
\end{barticle}
\endbibitem

\bibitem[\protect\citeauthoryear{Penrose}{1996}]{penrose1996gravity}
\begin{barticle}
\bauthor{\bsnm{Penrose}, \binits{R.}}:
\batitle{{On Gravity's Role in Quantum State Reduction}}.
\bjtitle{General Relativity and Gravitation}
\bvolume{28},
\bfpage{581}--\blpage{600}
(\byear{1996})
\doiurl{10/d52jm5}
\end{barticle}
\endbibitem

\bibitem[\protect\citeauthoryear{Angeli and Carlesso}{2025}]{jzht-fbwt}
\begin{barticle}
\bauthor{\bsnm{Angeli}, \binits{O.}},
\bauthor{\bsnm{Carlesso}, \binits{M.}}:
\batitle{Entanglement in markovian hybrid classical-quantum theories of gravity}.
\bjtitle{Phys. Rev. D}
\bvolume{112},
\bfpage{024047}
(\byear{2025})
\doiurl{10.1103/jzht-fbwt}
\end{barticle}
\endbibitem

\bibitem[\protect\citeauthoryear{Trillo and Navascu\'es}{2025}]{trillo2024di}
\begin{barticle}
\bauthor{\bsnm{Trillo}, \binits{D.}},
\bauthor{\bsnm{Navascu\'es}, \binits{M.}}:
\batitle{{Di\'osi-Penrose model of classical gravity predicts gravitationally induced entanglement}}.
\bjtitle{Phys. Rev. D}
\bvolume{111},
\bfpage{121101}
(\byear{2025})
\doiurl{10.1103/PhysRevD.111.L121101}
\end{barticle}
\endbibitem

\bibitem[\protect\citeauthoryear{}{}]{fn11}
\begin{botherref}
Assuming that at $t=0$ we are able to start in a quantum superposition state.
\end{botherref}
\endbibitem

\bibitem[\protect\citeauthoryear{Andersen}{2019}]{Andersen_2019}
\begin{barticle}
\bauthor{\bsnm{Andersen}, \binits{T.C.}}:
\batitle{{Quantum statistics in Bohmian trajectory gravity}}.
\bjtitle{Journal of Physics: Conference Series}
\bvolume{1275}(\bissue{1}),
\bfpage{012038}
(\byear{2019})
\doiurl{10.1088/1742-6596/1275/1/012038}
\end{barticle}
\endbibitem

\bibitem[\protect\citeauthoryear{D{\"o}ner and Gro{\ss}ardt}{2022}]{doner2022gravitational}
\begin{barticle}
\bauthor{\bsnm{D{\"o}ner}, \binits{M.K.}},
\bauthor{\bsnm{Gro{\ss}ardt}, \binits{A.}}:
\batitle{Is gravitational entanglement evidence for the quantization of spacetime?}
\bjtitle{Foundations of Physics}
\bvolume{52}(\bissue{5}),
\bfpage{101}
(\byear{2022})
\doiurl{10.1007/s10701-022-00619-0}
\end{barticle}
\endbibitem

\bibitem[\protect\citeauthoryear{Mart\'{\i}n-Mart\'{\i}nez and Perche}{2023}]{Eduardo2023}
\begin{barticle}
\bauthor{\bsnm{Mart\'{\i}n-Mart\'{\i}nez}, \binits{E.}},
\bauthor{\bsnm{Perche}, \binits{T.R.}}:
\batitle{What gravity mediated entanglement can really tell us about quantum gravity}.
\bjtitle{Phys. Rev. D}
\bvolume{108},
\bfpage{101702}
(\byear{2023})
\doiurl{10.1103/PhysRevD.108.L101702}
\end{barticle}
\endbibitem

\bibitem[\protect\citeauthoryear{Kent}{2024}]{kent2024necessarilytreatmasseslocalized}
\begin{botherref}
\oauthor{\bsnm{Kent}, \binits{A.}}:
Should We Necessarily Treat Masses as Localized When Analysing Tests of Quantum Gravity?
\doiurl{10.48550/arXiv.2405.20514}
\end{botherref}
\endbibitem

\bibitem[\protect\citeauthoryear{Franzmann}{2024}]{franzmann2024bewhere}
\begin{botherref}
\oauthor{\bsnm{Franzmann}, \binits{G.}}:
To Be or Not to Be, but Where?
\doiurl{10.48550/arXiv.2405.21031}
\end{botherref}
\endbibitem

\bibitem[\protect\citeauthoryear{Marchese et~al.}{2025}]{marchese2024newtonslawsmotiongenerate}
\begin{barticle}
\bauthor{\bsnm{Marchese}, \binits{M.M.}},
\bauthor{\bsnm{Pl\'avala}, \binits{M.}},
\bauthor{\bsnm{Kleinmann}, \binits{M.}},
\bauthor{\bsnm{Nimmrichter}, \binits{S.}}:
\batitle{Newton's laws of motion generating gravity-mediated entanglement}.
\bjtitle{Phys. Rev. A}
\bvolume{111},
\bfpage{042202}
(\byear{2025})
\doiurl{10.1103/PhysRevA.111.042202}
\end{barticle}
\endbibitem

\bibitem[\protect\citeauthoryear{Telali et~al.}{2025}]{PhysRevD.111.085005}
\begin{barticle}
\bauthor{\bsnm{Telali}, \binits{E.}},
\bauthor{\bsnm{Perche}, \binits{T.R.}},
\bauthor{\bsnm{Mart\'{\i}n-Mart\'{\i}nez}, \binits{E.}}:
\batitle{Causality in relativistic quantum interactions without mediators}.
\bjtitle{Phys. Rev. D}
\bvolume{111},
\bfpage{085005}
(\byear{2025})
\doiurl{10.1103/PhysRevD.111.085005}
\end{barticle}
\endbibitem

\bibitem[\protect\citeauthoryear{Eppley and Hannah}{1977}]{eppley1977necessity}
\begin{barticle}
\bauthor{\bsnm{Eppley}, \binits{K.}},
\bauthor{\bsnm{Hannah}, \binits{E.}}:
\batitle{The necessity of quantizing the gravitational field}.
\bjtitle{Foundations of Physics}
\bvolume{7}(\bissue{1}),
\bfpage{51}--\blpage{68}
(\byear{1977})
\doiurl{10.1007/BF00715241}
\end{barticle}
\endbibitem

\bibitem[\protect\citeauthoryear{Gisin}{1990}]{GISIN19901}
\begin{barticle}
\bauthor{\bsnm{Gisin}, \binits{N.}}:
\batitle{Weinberg's non-linear quantum mechanics and supraluminal communications}.
\bjtitle{Physics Letters A}
\bvolume{143}(\bissue{1}),
\bfpage{1}--\blpage{2}
(\byear{1990})
\doiurl{10.1016/0375-9601(90)90786-N}
\end{barticle}
\endbibitem

\bibitem[\protect\citeauthoryear{Huggett and Callender}{2001}]{Huggett_Callender_2001}
\begin{barticle}
\bauthor{\bsnm{Huggett}, \binits{N.}},
\bauthor{\bsnm{Callender}, \binits{C.}}:
\batitle{Why quantize gravity (or any other field for that matter)?}
\bjtitle{Philosophy of Science}
\bvolume{68}(\bissue{S3}),
\bfpage{382}--\blpage{394}
(\byear{2001})
\doiurl{10.1086/392923}
\end{barticle}
\endbibitem

\bibitem[\protect\citeauthoryear{Mattingly}{2006}]{mattingly2006eppley}
\begin{barticle}
\bauthor{\bsnm{Mattingly}, \binits{J.}}:
\batitle{{Why Eppley and Hannah’s thought experiment fails}}.
\bjtitle{Physical Review D—Particles, Fields, Gravitation, and Cosmology}
\bvolume{73}(\bissue{6}),
\bfpage{064025}
(\byear{2006})
\doiurl{10.1103/PhysRevD.73.064025}
\end{barticle}
\endbibitem

\bibitem[\protect\citeauthoryear{Ballentine}{1970}]{RevModPhys.42.358}
\begin{barticle}
\bauthor{\bsnm{Ballentine}, \binits{L.E.}}:
\batitle{{The Statistical Interpretation of Quantum Mechanics}}.
\bjtitle{Rev. Mod. Phys.}
\bvolume{42},
\bfpage{358}--\blpage{381}
(\byear{1970})
\doiurl{10.1103/RevModPhys.42.358}
\end{barticle}
\endbibitem

\bibitem[\protect\citeauthoryear{Rozema et~al.}{2012}]{PhysRevLett.109.100404}
\begin{barticle}
\bauthor{\bsnm{Rozema}, \binits{L.A.}},
\bauthor{\bsnm{Darabi}, \binits{A.}},
\bauthor{\bsnm{Mahler}, \binits{D.H.}},
\bauthor{\bsnm{Hayat}, \binits{A.}},
\bauthor{\bsnm{Soudagar}, \binits{Y.}},
\bauthor{\bsnm{Steinberg}, \binits{A.M.}}:
\batitle{Violation of heisenberg's measurement-disturbance relationship by weak measurements}.
\bjtitle{Phys. Rev. Lett.}
\bvolume{109},
\bfpage{100404}
(\byear{2012})
\doiurl{10.1103/PhysRevLett.109.100404}
\end{barticle}
\endbibitem

\bibitem[\protect\citeauthoryear{Bahrami et~al.}{2014}]{Bahrami_2014}
\begin{barticle}
\bauthor{\bsnm{Bahrami}, \binits{M.}},
\bauthor{\bsnm{Großardt}, \binits{A.}},
\bauthor{\bsnm{Donadi}, \binits{S.}},
\bauthor{\bsnm{Bassi}, \binits{A.}}:
\batitle{{The Schrödinger–Newton equation and its foundations}}.
\bjtitle{New Journal of Physics}
\bvolume{16}(\bissue{11}),
\bfpage{115007}
(\byear{2014})
\doiurl{10.1088/1367-2630/16/11/115007}
\end{barticle}
\endbibitem

\bibitem[\protect\citeauthoryear{Kent}{1998}]{kent1998causality}
\begin{barticle}
\bauthor{\bsnm{Kent}, \binits{A.}}:
\batitle{Causality in time-neutral cosmologies}.
\bjtitle{Physical Review D}
\bvolume{59}(\bissue{4}),
\bfpage{043505}
(\byear{1998})
\doiurl{10.48550/arXiv.gr-qc/9703041}
\end{barticle}
\endbibitem

\bibitem[\protect\citeauthoryear{Mielnik}{2000}]{mielnik2000comments}
\begin{barticle}
\bauthor{\bsnm{Mielnik}, \binits{B.}}:
\batitle{{Comments on: ``Weinberg's Nonlinear Quantum Mechanics and Einstein-Podolsky-Rosen paradox'', by Joseph Polchinski}}.
\bjtitle{arXiv preprint quant-ph/0012041}
(\byear{2000})
\doiurl{10.48550/arXiv.quant-ph/0012041}
\end{barticle}
\endbibitem

\bibitem[\protect\citeauthoryear{Mead}{1964}]{mead1964possible}
\begin{barticle}
\bauthor{\bsnm{Mead}, \binits{C.A.}}:
\batitle{Possible connection between gravitation and fundamental length}.
\bjtitle{Physical Review}
\bvolume{135}(\bissue{3B}),
\bfpage{849}
(\byear{1964})
\doiurl{10.1103/PhysRev.135.B849}
\end{barticle}
\endbibitem

\bibitem[\protect\citeauthoryear{Kent}{2018}]{kent2018testing}
\begin{barticle}
\bauthor{\bsnm{Kent}, \binits{A.}}:
\batitle{Testing causal quantum theory}.
\bjtitle{Proceedings of the Royal Society A}
\bvolume{474}(\bissue{2220}),
\bfpage{20180501}
(\byear{2018})
\doiurl{10.1098/rspa.2018.0501}
\end{barticle}
\endbibitem

\bibitem[\protect\citeauthoryear{Donadi et~al.}{2021}]{donadi2021underground}
\begin{barticle}
\bauthor{\bsnm{Donadi}, \binits{S.}},
\bauthor{\bsnm{Piscicchia}, \binits{K.}},
\bauthor{\bsnm{Curceanu}, \binits{C.}},
\bauthor{\bsnm{Di{\'o}si}, \binits{L.}},
\bauthor{\bsnm{Laubenstein}, \binits{M.}},
\bauthor{\bsnm{Bassi}, \binits{A.}}:
\batitle{Underground test of gravity-related wave function collapse}.
\bjtitle{Nature Physics}
\bvolume{17}(\bissue{1}),
\bfpage{74}--\blpage{78}
(\byear{2021})
\doiurl{10.1038/s41567-020-1008-4}
\end{barticle}
\endbibitem

\bibitem[\protect\citeauthoryear{Oppenheim et~al.}{2023}]{oppenheim2023gravitationally}
\begin{barticle}
\bauthor{\bsnm{Oppenheim}, \binits{J.}},
\bauthor{\bsnm{Sparaciari}, \binits{C.}},
\bauthor{\bsnm{{\v{S}}oda}, \binits{B.}},
\bauthor{\bsnm{Weller-Davies}, \binits{Z.}}:
\batitle{Gravitationally induced decoherence vs space-time diffusion: testing the quantum nature of gravity}.
\bjtitle{Nature Communications}
\bvolume{14}(\bissue{1}),
\bfpage{7910}
(\byear{2023})
\doiurl{10.1038/s41467-023-43348-2}
\end{barticle}
\endbibitem

\bibitem[\protect\citeauthoryear{Di{\'o}si}{2024}]{diosi2024classicalquantumhybridcanonicaldynamics}
\begin{barticle}
\bauthor{\bsnm{Di{\'o}si}, \binits{L.}}:
\batitle{Classical-quantum hybrid canonical dynamics and its difficulties with special and general relativity}.
\bjtitle{Physical Review D}
\bvolume{110}(\bissue{8}),
\bfpage{084052}
(\byear{2024})
\end{barticle}
\endbibitem

\bibitem[\protect\citeauthoryear{Di\'osi}{2022}]{PhysRevD.106.L051901}
\begin{barticle}
\bauthor{\bsnm{Di\'osi}, \binits{L.}}:
\batitle{{Is there a relativistic Gorini-Kossakowski-Lindblad-Sudarshan master equation?}}
\bjtitle{Phys. Rev. D}
\bvolume{106},
\bfpage{051901}
(\byear{2022})
\doiurl{10.1103/PhysRevD.106.L051901}
\end{barticle}
\endbibitem

\bibitem[\protect\citeauthoryear{Grudka et~al.}{2024}]{grudka2024renormalisation}
\begin{barticle}
\bauthor{\bsnm{Grudka}, \binits{A.}},
\bauthor{\bsnm{Morris}, \binits{T.R.}},
\bauthor{\bsnm{Oppenheim}, \binits{J.}},
\bauthor{\bsnm{Russo}, \binits{A.}},
\bauthor{\bsnm{Sajjad}, \binits{M.}}:
\batitle{Renormalisation of postquantum-classical gravity}.
\bjtitle{arXiv:2402.17844}
(\byear{2024})
\doiurl{10.48550/arXiv.2402.17844}
\end{barticle}
\endbibitem

\bibitem[\protect\citeauthoryear{Penrose}{2016}]{PenroseFashion}
\begin{bbook}
\bauthor{\bsnm{Penrose}, \binits{R.}}:
\bbtitle{Fashion, Faith, and Fantasy in the New Physics of the Universe}.
\bpublisher{Princeton University Press},
\blocation{Princeton, NJ, USA}
(\byear{2016}).
\doiurl{10.1515/9781400880287} .
\burl{http://www.jstor.org/stable/j.ctvc775bn}
\end{bbook}
\endbibitem

\bibitem[\protect\citeauthoryear{Smolin}{2007}]{smolin2007trouble}
\begin{bbook}
\bauthor{\bsnm{Smolin}, \binits{L.}}:
\bbtitle{The Trouble with Physics: The Rise of String Theory, the Fall of a Science, and What Comes Next}.
\bpublisher{Houghton Mifflin Harcourt},
\blocation{Boston, MA}
(\byear{2007})
\end{bbook}
\endbibitem

\bibitem[\protect\citeauthoryear{Woit}{2006}]{woit2006not}
\begin{bbook}
\bauthor{\bsnm{Woit}, \binits{P.}}:
\bbtitle{Not Even Wrong: The Failure of String Theory and the Search for Unity in Physical Law}.
\bpublisher{Basic Books},
\blocation{New York, NY, USA}
(\byear{2006}).
\burl{https://www.basicbooks.com/titles/peter-woit/not-even-wrong/9780465092765/}
\end{bbook}
\endbibitem

\bibitem[\protect\citeauthoryear{Schwarz}{1997}]{schwarz1997status}
\begin{barticle}
\bauthor{\bsnm{Schwarz}, \binits{J.H.}}:
\batitle{The status of string theory}.
\bjtitle{arXiv preprint hep-th/9711029}
(\byear{1997})
\doiurl{10.48550/arXiv.hep-th/9711029}
\end{barticle}
\endbibitem

\bibitem[\protect\citeauthoryear{Nicolai et~al.}{2005}]{Nicolai_2005}
\begin{barticle}
\bauthor{\bsnm{Nicolai}, \binits{H.}},
\bauthor{\bsnm{Peeters}, \binits{K.}},
\bauthor{\bsnm{Zamaklar}, \binits{M.}}:
\batitle{Loop quantum gravity: an outside view}.
\bjtitle{Classical and Quantum Gravity}
\bvolume{22}(\bissue{19}),
\bfpage{193}
(\byear{2005})
\doiurl{10.1088/0264-9381/22/19/R01}
\end{barticle}
\endbibitem

\bibitem[\protect\citeauthoryear{Lami et~al.}{2024}]{Lami:2023gmz}
\begin{barticle}
\bauthor{\bsnm{Lami}, \binits{L.}},
\bauthor{\bsnm{Pedernales}, \binits{J.S.}},
\bauthor{\bsnm{Plenio}, \binits{M.B.}}:
\batitle{Testing the quantumness of gravity without entanglement}.
\bjtitle{Phys. Rev. X}
\bvolume{14},
\bfpage{021022}
(\byear{2024})
\doiurl{10.1103/PhysRevX.14.021022}
\end{barticle}
\endbibitem

\end{thebibliography}

\onecolumn

 \section*{Appendix}

\appendix 

\section{\large Further detail on entanglement from quantum gravity in QFT} \label{app:MoreQG}

Here, we further discuss the first-order and other second-order Dyson series terms in quantum gravity. At first order, there is no way to internally Wick contract the gravitational field, so this order does not contribute to the entangling process we are interested in. At second order, in addition to that discussed in the main text, there is also the Feynman diagram \ref{fig:CrissCross}\textcolor{blue}{a}. The  corresponding contraction is:
\begin{align}\label{eq:WickCrissCross}
   -\frac{1}{4 \hbar^2 c^2}   \int_t d^4 x \int_t d^4  y \,_{1i}{\langle} \wick{\c1 N| \, \,_{2j}{\langle} \c2 N|  \hat{\mathcal{T}}_{\mu \nu}[\c1{\hat{\phi}}^{\dagger} (x) \c3{\hat{\phi}}(x) ]          \c1{\hat{h}}^{\mu \nu}(x)  \c1{\hat{h}}^{\rho \sigma}(y)  \hat{\mathcal{T}}_{\rho \sigma}[\c2{\hat{\phi}}^{\dagger} (y)  \c2{\hat{\phi}}(y)] |\c2 N \rangle_{1i} \,| \c3N \rangle_{2j}}.
\end{align}
From \eqref{eq:WickN}, this amplitude contains the integral:
\begin{align} \label{eq:zeroDiff}
    \int d^3 \bm{x}\,\int d^3 \bm{y}\, \tilde{\phi}_{1j}(\bm{x}) \tilde{\phi}_{2j}(\bm{x}) \tilde{\phi}_{1j}(\bm{y}) \tilde{\phi}_{2j}(\bm{y}). 
\end{align}
Since the wavefunctions do not overlap, as described in the main text, this integral evaluates to zero. The Feynman diagram thus provides a vanishing amplitude. This is because there is no on-shell process such that an  atom from one object can propagate  to the other object,  given the initial and final states. Therefore, before second order we have a separable state, and find a non-separable state \eqref{eq:finalStateQG} at second order. Note that separability cannot be `repaired' by going to a higher (and thus numerically weaker) order in the  expansion - the fact that we see non-separability of the state at second order   is enough to demonstrate that the full state must be entangled, and as discussed above, this is just due to the second order expansion of the quantum phase $e^{i \varphi}$.

\section{\large Further detail on entanglement from classical gravity in QFT}

\subsection{Linearity} \label{sec:linear}

Up to Equation \eqref{eq:beta4ijFull}, we assumed that gravity is sourced by the quantum matter of the experiment but did not specify exactly how except that this must satisfy the assumptions of the experiment, such as    $\beta^{(4)}_{ij}$ is small enough for the perturbative approach in deriving it to be valid and that the potential is sufficiently spatially varying. After this, we specialized to semi-classical Einstein gravity \cite{ROSENFELD1963353,moller1962theories,Kent_2018,Helou_2017}. In this theory, the gravitational field is sourced by the  expectation of the energy-momentum tensor of matter, which results in $\Phi(\bm{x})$ following \eqref{eq:semiclassicalPhi} in the Newtonian regime. In the experiment, the largest value $\Phi(\bm{x})$ takes is of order $ G M / R$. We thus parametrize the evolution of the system in terms of the dimensionless parameter $\tilde{G} := G M / (R c^2)$, which is a very small number for the values we are interested in. For example, we can write $\Phi(\bm{x}) =: \tilde{G} c^2 \tilde{\Phi}[\psi(t)] (\bm{x})$ where:
\begin{align} \label{eq:tildePhi}
    \tilde{\Phi}[\psi(t)](\bm{x}):= -  \int d^3 \bm{\tilde{x}}' \frac{  \langle \tilde{\psi}(t) | \hat{\tilde{T}}_{00} (\bm{\tilde{x}'})  |\tilde{\psi}(t) \rangle }{|\bm{\tilde{x}} - \bm{\tilde{x}}'|},
\end{align}
with $\bm{\tilde{x}} := \bm{x} / R$, $\hat{\tilde{T}}_{00} := \hat{T}_{00} R^3 / (m c^2)$ and $|\tilde{\psi}\rangle := |\psi\rangle / \sqrt{N}$. We can also write 
the interaction Hamiltonian we are interested in for the regime of the experiment  \eqref{eq:HintNonRel} as $\hat{H}_{int} =: \tilde{G} \hat{\tilde{H}}_{int}[\tilde{\psi}(t)]$, where:
\begin{align}
\hat{\tilde{H}}_{int}[\psi(t)] := 4   \int d^3 \bm{x} \, \tilde{\Phi}[\psi(t)](\bm{x}) \left(\hat{\pi}(\bm{x}) \hat{\pi}^{\dagger}(\bm{x}) -  \frac{m^2 c^2}{2 \hbar^2} \hat{\phi}^{\dagger}(\bm{x}) \hat{\phi}(\bm{x})\right).
\end{align}
The quantum state of matter $|\psi\rangle$ then evolves according to the Schr\"{o}dinger equation:
\begin{align}
    i \hbar \frac{d |\psi\rangle}{dt} &= \hat{H} |\psi\rangle\\ \label{eq:nonlinear}
    &=\left(\hat{H}_0 + \tilde{G}\hat{\tilde{H}}_{int}[\psi(t)] \right) |\psi \rangle.
\end{align}
Since $\Phi(\bm{x})$  depends on the quantum state of matter, Equation \eqref{eq:nonlinear} is, in general, non-linear, which is thought to introduce  superluminal signalling unless the theory is modified further, as discussed  in Appendix \ref{sec:consistent}. However, because  we work to a perturbative order, and due to the experiment we are interested in, we operate in a regime where the theory is approximately linear as in standard quantum mechanics. Since $\tilde{G} \ll 1$, then for realistic times we can also expand $|\psi(t)\rangle$ in this parameter:
\begin{align}
    |\psi(t)\rangle  = |\psi^{(0)}(t)\rangle + \tilde{G} |\psi^{(1)}(t)\rangle + \tilde{G}^2 |\psi^{(2)}(t)\rangle + \cdots,
\end{align}
where $|\psi^{(n)}(0)\rangle  = 0$ for $ n > 0$. We can then plug this into \eqref{eq:tildePhi} and also expand $\tilde{\Phi}(\bm{x})$ in $\tilde{G}$:
\begin{align}
    \tilde{\Phi}[\psi(t)](\bm{x}) = \tilde{\Phi}_0[\psi(t)](\bm{x}) + \tilde{G} \tilde{\Phi}_1[\psi(t)](\bm{x}) + \cdots,
\end{align}
As discussed in the main text, $\hat{H}_0$ can be shown to approximately just act a global phase so that we can take $|\psi^{(0)}(t)\rangle \equiv |\psi(0)\rangle$. Therefore, $\tilde{G} c^2 \tilde{\Phi}_0[\psi(t)](\bm{x}) \equiv \Phi(\bm{x})$, with $\Phi(\bm{x})$ given by  \eqref{eq:semiclassicalPhi}, and is thus independent of time or the evolution of the quantum state of matter. To first order in $\tilde{G}$, we then have:
\begin{align} 
    i \hbar \frac{d |\psi (t) \rangle}{dt} 
    &=\left(\hat{H}_0 + \tilde{G} \hat{\tilde{H}}^0_{int} \right) |\psi (t)  \rangle, \\ \label{eq:firstorderU}
    &\equiv\left(\hat{H}_0 + \hat{H}_{int} \right) |\psi (t)  \rangle, 
\end{align}
with $\hat{H}_{int}$ and $\Phi(\bm{x})$ given by \eqref{eq:HintNonRel} and    \eqref{eq:semiclassicalPhi}, and we only keep solutions up to first order in $\tilde{G}$. We, therefore, to first order  have the linear equation of QFT in curved spacetime, but with $\Phi(\bm{x})$ given by \eqref{eq:semiclassicalPhi}. We know from calculating \eqref{eq:firstorderU} to first order that the gravitational potential of the objects is essentially unchanged compared to the zeroth order. That is, $\tilde{\Phi}[\psi(t)](\bm{x}) \approx \Phi(\bm{x})$ up to this order. This means that at second order, which depends in general on $\tilde{\Phi}_1[\psi(t)](\bm{x})$, we can again just apply \eqref{eq:firstorderU} and keep terms only up to second order. We can continue this procedure up to fourth order in $\tilde{G}$ where the considered entanglement effect occurs. As such, the state to fourth order still obeys the linear Schr\"{o}dinger equation \eqref{eq:firstorderU}  to a good approximation,  which can be solved through the Dyson series as discussed in the main text, as long as we only consider the result up to the  order we are interested in.

\subsection{Relativistic treatment} \label{app:ct}

Here we consider a derivation  of the classical gravity effect that keeps its  relativistic character more explicit. We start with:
\begin{align}\nonumber
  & I = \int^{c t}_0 d x^0 \int^{ct}_0 d  y^0 \frac{d^4 k}{(2 \pi)^4} \frac{1}{k^2 + m^2 c^2 / \hbar^2 + i \epsilon} e^{i k.(x-y)}   e^{i m c (x^0 - y^0)/ \hbar}.
    \end{align}
    In the main text we first performed the integral over $\bm{k}$. Here, we instead use the identity:
\begin{align}
    \frac{1}{A^n} = \frac{1}{(n-1)!} \int^{\infty}_0 du \, u^{n-1}\,e^{- u A}
\end{align}
to write:
\begin{align}\nonumber
  I =  \int^{c t}_0 d x^0 \int^{ct}_0 d  y^0 \int^{\infty}_0 du \,\int \frac{d^4 k}{(2 \pi)^4} e^{- u (k^2 + m^2 c^2 / \hbar^2 + i \epsilon)} e^{i k.(x-y)}   e^{i m c (x^0 - y^0)/ \hbar}.
    \end{align}
    We next perform the $\bm{k}$ integrals which are now Gaussian integrals:
\begin{align}\nonumber
  I =  \frac{1}{(4 \pi)^{3/2}} &\int^{c t}_0 d x^0 \int^{ct}_0 d  y^0 \int^{\infty}_0 du \, u^{-3/2} e^{- r^2 / ( 4 u)}\\
  &\times \int \frac{d k^0}{2 \pi} e^{- u (-(k^0)^2 + m^2 c^2 / \hbar^2 + i \epsilon)} e^{-i k^0 (x^0 -y^0)}   e^{i m c (x^0 - y^0)/ \hbar},
    \end{align}
    where $r := |\bm{x} - \bm{y}|$. Now we Wick rotate and perform the $k^0$ integral:
\begin{align}\nonumber
  I =   \frac{i}{16 \pi^2} \int^{c t}_0 d x^0 \int^{ct}_0 d  y^0 \int^{\infty}_0 du \, u^{-2} e^{- (r^2 - \tau^2) / ( 4 u)} e^{- u m^2 c^2 / \hbar^2}    e^{i m c \tau / \hbar},
  \end{align}
  where $\tau = x^0 - y^0$. Now we perform the $u$ integral using the identity:
  \begin{align}
      \int^{\infty}_0 du\, u^{ \alpha - 1} e^{ - \beta / u - \mu u} = 2 \left(\frac{\beta}{\mu}\right)^{\alpha / 2} K_{\alpha} (2 \sqrt{\beta \mu}),
  \end{align}
  where $K_{\alpha}(z)$ is the modified Bessel function of the second kind. This then gives:
\begin{align}\nonumber
  I =   \frac{i}{4 \pi^{2}} \int^{c t}_0 d x^0 \int^{ct}_0 d  y^0  \frac{\gamma \, e^{i \gamma  \tau}}{\sqrt{r^2 - \tau^2}}    K_1(\gamma \sqrt{r^2 - \tau^2}),
  \end{align}
  where $\gamma  = m c / \hbar$ and we have used $K_{-1} (z) = K_1 (z)$. We can write the full amplitude  as:
\begin{align}\nonumber
 \beta^{(4)}_{ij} \propto   \Big( \int^{c t}_{- ct} d \tau \int d^3 \bm{x}  d^3 \bm{y}\, A( c t,\tau) \theta(|\tau| - |\bm{x} - \bm{y}|)  &\frac{\Phi(\bm{x}) \Phi(\bm{y}) \theta_{1i}(\bm{x}) \theta_{2i}(\bm{x})}{  \sqrt{\tau^2 - |\bm{x} - \bm{y}|^2}}\\  \label{eq:betaRel}    &\times K_1(i \gamma \sqrt{\tau^2 - |\bm{x} - \bm{y}|^2}) \, e^{i \gamma \tau}\Big)^2,
  \end{align}
  where $A(t,\tau) := (c t - |\tau|) = ( ct - |x^0 - y^0|)$ with $ c t > d_{ij}$.   To this more relativistic expression we can then apply the long-time  approximations used in the main text to derive \eqref{eq:beta4ijFull}. In this case, we can approximate $ A(t,\tau) = (c t - |\tau|) \approx c t$,  extend the limits of the integration to infinity, change  variables to $ \tau = r \cosh s$ and make  use of  the following identity \cite{gradshteyn2014table}:
\begin{align}
\int^{\infty}_0 ds \, e^{ (\xi - z) \cosh s} J_{2 \nu}[2 \sqrt{ z \xi} \sinh s]  = I_{\nu} (z) K_{\nu} (\xi),
  \end{align}
  where  $I_{\nu}(z)$ is the modified Bessel function of the first kind and and $J_{\nu}(z)$ is the Bessel function of the first kind.

  Note that  the above more relativistic expression (keeping the mediating virtual particles of the quantum field relativistic), is of similar form to the quantum gravity amplitude in the relativistic setting (keeping gravity relativistic) - see \eqref{eq:gammarel}. In fact, it would almost be identical if the graviton were made slightly massive, as suggested in some beyond the Standard Model theories \cite{OGIEVETSKY1965167,PhysRevD.82.044020} - essentially, the only difference would be that there would be no phase factor $\exp(i \gamma \tau)$ or classical field. We could also move the effect to a fully relativistic setting by keeping the gravitational field explicitly relativistic. However, note that again the non-relativistic treatment was only a mathematical simplification, similar to how the quantum gravity interaction is normally treated when considering entanglement. In both cases, keeping it relativistic does not remove the effect.

\subsection{Momentum space} \label{app:AtlDer}

Here we derive \eqref{eq:virtialdecay2} using a slightly different methodology to that used in the main text. To start with we assume matter objects in some general wavepackets $|\psi\rangle$ and $|\psi'\rangle$ \eqref{eq:wavepacket} and only specialize to an N-particle position state $|N\rangle$ \eqref{eq:Nstate} towards the end of the derivation. Considering just the amplitude involving the field $\hat{\phi}$ for now, ignoring the momentum conjugate, we are interested in contractions of the following form for the entangling fourth order process:
\begin{align}\nonumber
        &\beta^{(4)}_{ij}  = \frac{16 m^8}{  \hbar^{12} c^4}  \int_t d^4 x \, \int_t d^4  y \,  \int_t d^4 z\, \int_t d^4 w\, \Phi(\bm{w})\, \Phi(\bm{z})\, \Phi(\bm{y})\, \Phi(\bm{x}) \,\times \\ \label{eq:beta4Alt} &\,_{1i}\langle \wick{\c1 \psi| \, _{2j}{\langle} \c2 \psi'|  \c2{\hat{\phi}}^{\dagger} (w) \c3{\hat{\phi}} (w)      \c1{\hat{\phi}}^{\dagger} (z) \c4{\hat{\phi}} (z) \c3{\hat{\phi}}^{\dagger}(y) \c5{\hat{\phi}} (y)
      \c4{\hat{\phi}}^{\dagger} (x) \c6{\hat{\phi}} (x)  |\c5 \psi \rangle_{1i} \,| \c6 \psi'} \rangle_{2j}.
\end{align}
We write the wavepackets $|\psi\rangle$ and  $|\psi'\rangle$ as
\begin{align}  \label{eq:psi1}
    |\psi\rangle &=  \int  \frac{d^3 \bm{p}}{(2 \pi)^3} \frac{1}{\sqrt{2 \omega_{\bm{p}}}} f(\bm{p}) |\bm{p}\rangle,\\ \label{eq:psi2}
    |\psi'\rangle &=  \int  \frac{d^3 \bm{p}}{(2 \pi)^3} \frac{1}{\sqrt{2 \omega_{\bm{p}}}} f'(\bm{p}) |\bm{p}\rangle,
\end{align}
with $\int \frac{d^3 \bm{p}}{(2 \pi)^3} |f(\bm{p})|^2 = 1 = \int \frac{d^3 \bm{p}}{(2 \pi)^3} |f'(\bm{p})|^2$. We can then use:
\begin{align}
    \wick{\c{\hat{\phi}} (x)
      |\c {\bm{p}} \rangle = c \sqrt{\hbar}\,e^{i p.x}},
\end{align}
such that:
\begin{align}
    \wick{\c{\hat{\phi}} (x)
      |\c \psi \rangle = c \sqrt{\hbar} \int \frac{d^3 \bm{p}}{(2 \pi)^3} \frac{1}{\sqrt{2\omega_{\bm{p}}}} f(\bm{p}) e^{i p.x}}.
\end{align}
We also take the Fourier transform of  $\Phi(\bm{x})$, i.e.\  $\Phi(\bm{x}) = \int \frac{d^3 \bm{p}}{(2 \pi)^3} \Phi(\bm{k}) e^{i \bm{k}.\bm{x}}$. Plugging this all into \eqref{eq:beta4Alt} gives $\beta^{(4)}_{ij} = \beta^2$, where:
\begin{align}\nonumber
        \beta
        &:= \frac{4 m^4 c}{\hbar^4}  \int_t  d^4  x\,  d^4 y  \,   \int \frac{d^3 \bm{k}}{(2 \pi)^3}  \frac{d^3 \bm{k'}}{(2 \pi)^3}  \frac{d^3 \bm{p'}}{(2 \pi)^3}\frac{d^3 \bm{p}}{(2 \pi)^3} \frac{d^4 q}{(2\pi)^4} \frac{1}{\sqrt{2\omega_{\bm{p'}}}}  \frac{1}{\sqrt{2\omega_{\bm{p}}}} f(\bm{p}) f^{' \ast}(\bm{p'}) \Phi(\bm{k}) \Phi(\bm{k}') \\ \label{eq:startingpoint}
        &\times e^{i\bm{k}'.\bm{x}} e^{i\bm{k}.\bm{y}} e^{ip.y} e^{-i p'.x} \frac{i}{q^2 + m^2 c^2/\hbar^2 + i \epsilon} e^{i q. (x - y)}.
\end{align}
We first integrate over $q^0$ so that we write the Feynman propagator in terms of Wightman functions: $\theta(x^0 - y^0) D(x - y)  + \theta(y^0 - x^0) D(y - x)$, where:
\begin{align}
    D(x - y) = \int \frac{d^3 \bm{q}}{(2 \pi)^3} \frac{1}{2 \omega_q} e^{iq.(x-y)},
\end{align}
with $\omega_q = c \sqrt{ \bm{q}^2 + m^2 c^2 / \hbar^2}$. Then  we integrate over space and subsequently $\bm{q}$ to get:
\begin{align}\nonumber
        \beta
        &= \frac{4 i m^4 c^2 \, (2 \pi)^3}{\hbar^4}      \int_0^{c t}  d  x^0\,  d y^0  \,   \int \frac{d^3 \bm{k}}{(2 \pi)^3}  \frac{d^3 \bm{k'}}{(2 \pi)^3}   \frac{d^3 \bm{p}}{(2 \pi)^3} \frac{d^3 \bm{p}'}{(2 \pi)^3}  \frac{1}{\sqrt{2\omega_{\bm{p'}}}}  \frac{1}{\sqrt{2\omega_{\bm{p}}}} f(\bm{p}) f^{' \ast}(\bm{p'}) \Phi(\bm{k}) \Phi(\bm{k}')   \\ \label{eq:BeforeTimeInts}
        &\times\frac{e^{-i \omega_{\bm{p}} y^0/c} e^{i \omega_{\bm{p'}} x^0 / c}}{2 \omega_q}  \left(\theta(x^0 - y^0) e^{-i \omega_q (x^0 - y^0)/c} + \theta(y^0 - x^0) e^{i \omega_q (x^0 - y^0)/c}\right) \delta^{(3)}( \bm{p} + \bm{k} + \bm{k}'-\bm{p}')
        \end{align}
where $\omega_q = c \sqrt{(\bm{p} + \bm{k})^2 + m^2 c^2 / \hbar^2}$.  Assuming   $\omega_{\bm{p}} \neq \omega_{q}$ and $\omega_q \neq \omega_{\bm{p'}}$, we integrate over time to find:
\begin{align}\nonumber
        \beta
        &= \frac{4 i  m^4 c^4 \, (2 \pi)^3}{\hbar^4}         \int \frac{d^3 \bm{k}}{(2 \pi)^3}  \frac{d^3 \bm{k'}}{(2 \pi)^3}   \int \frac{d^3 \bm{p}}{(2 \pi)^3} \frac{d^3 \bm{p}'}{(2 \pi)^3}  \frac{1}{\sqrt{2\omega_{\bm{p'}}}}  \frac{1}{\sqrt{2\omega_{\bm{p}}}} f(\bm{p}) f^{' \ast}(\bm{p'}) \\ \nonumber
        &\times\Phi(\bm{k}) \Phi(\bm{k}') \delta^{(3)}( \bm{p} + \bm{k} + \bm{k}'-\bm{p}')\\ \label{eq:timeInt}
&\times\left(\frac{g(\bm{k},\bm{p})}{ c^2 (\bm{k} + \bm{p})^2}  \, \left( \frac{e^{i (\omega_p - \omega_{p'}) t}-1}{\omega_p - \omega_{p'}}\right)    + \frac{e^{i (\omega_p - \omega_{p'}) t} - e^{i (\omega_p + \omega_{q}) t}}{2 \omega_q (\omega_q + \omega_p)(\omega_q + \omega_{p'})} - \frac{e^{i (\omega_q - \omega_{p'}) t} - 1}{2 \omega_q  (\omega_q - \omega_p)(\omega_q - \omega_{p'})}\right),
        \end{align}
where  $g(\bm{k},\bm{p}) = 1 / (1 - \bm{p}^2 / (\bm{k} + \bm{p})^2)$. 
Now we finally take $f(\bm{p})$ and $ f'(\bm{p})$ to be the Fourier transforms of spherical position wavefunctions, as assumed in the main text. This then allows us to take the non-relativistic approximation where we take the integrals to be dominated by $\bm{p}^2 \ll m^2 c^2 / \hbar^2$ and $\bm{p'}^2 \ll m^2 c^2 / \hbar^2$ since $f(\bm{p})$ and $f(\bm{p}')$ drop off quickly well before $\bm{p}^2$ approaches $m^2 c^2 / \hbar^2$, as discussed in the main text. 
Note that we do not need to make such an approximation for $\Phi(\bm{k})$. Although not necessary, as discussed below, it is possible for $\Phi(\bm{k})$ to support higher momentum modes than $f(\bm{p})$ due to the interference of all the different gravitational potentials. That is, the Fourier transform of \eqref{eq:semiclassicalPhi} can support  larger oscillatory peaks than $f(\bm{p})$.   This then leaves us with:
\begin{align}\nonumber
        \beta
        &= \frac{2 i M m^2  \, (2 \pi)^3}{c \hbar^3}          \int \frac{d^3 \bm{k}}{(2 \pi)^3}  \frac{d^3 \bm{k'}}{(2 \pi)^3}   \frac{d^3 \bm{p}}{(2 \pi)^3} \int \frac{d^3 \bm{p}'}{(2 \pi)^3}   f(\bm{p}) f^{' \ast}(\bm{p'}) \Phi(\bm{k}) \Phi(\bm{k}')\\ \label{eq:beforeApprox}
        &\times\left(\frac{- i c t }{(\bm{k} + \bm{p})^2}
        +\frac{1 }{ 2\Delta q} \left[ \frac{e^{i ( \Delta q - m c / \hbar) c t} - 1}{(\Delta q -m c / \hbar)^2} + \frac{e^{i (\Delta q + m c / \hbar)  c t} - 1}{(\Delta q + m c/ \hbar )^2}\right] \right)\delta^{(3)}( \bm{p} + \bm{k} + \bm{k}'-\bm{p}'),
        \end{align}
where $\Delta q :=  \sqrt{k^2 + 2\bm{k}.\bm{p} + m^2 c^2 / \hbar^2}$ \cite{fn12}. Here we have hit a resonance $\omega_p \approx \omega_{p'}$, which picks out the linear time dependence in the first term, but leaving, in general, oscillatory terms elsewhere. Note the same  expression results if we had taken the above approximations before performing the time integrals. In addition to $\bm{p}^2 \ll m^2 c^2 / \hbar^2$, since $f(\bm{p})$ are peaked at zero with typical width $1/R$, we also typically require $t \ll 2 m R^2  /\hbar$ as noted in the main text. Next, we take the long time limit $c t \gg 1$ of \eqref{eq:beforeApprox} as in Section \ref{sec:CG}, which includes $ t \gg \hbar / (m c^2)$:
\begin{align}\nonumber
        \beta
        &\approx \frac{2  M m^2 t }{ \hbar^3}          \int \frac{d^3 \bm{k}}{(2 \pi)^3}  \frac{d^3 \bm{k'}}{(2 \pi)^3}    \frac{d^3 \bm{p}}{(2 \pi)^3}\frac{d^3 \bm{p}'}{(2 \pi)^3}   f(\bm{p}) f^{' \ast}(\bm{p'}) \Phi(\bm{k}) \Phi(\bm{k}')\frac{1}{(\bm{k} + \bm{p})^2} 
        (2 \pi)^3 \delta^{(3)}( \bm{p} + \bm{k} + \bm{k}'-\bm{p}')\\\nonumber
         &=   \frac{2  M m^2 t }{ \hbar^3}    \,   \int \frac{d^3 \bm{k}}{(2 \pi)^3}  \frac{d^3 \bm{k'}}{(2 \pi)^3}    \frac{d^3 \bm{p}}{(2 \pi)^3}  \frac{d^3 \bm{p}'}{(2 \pi)^3}   f(\bm{p}) f^{' \ast}(\bm{p'}) \Phi(\bm{k}) \Phi(\bm{k}') \\  &\times \int \frac{d^3{\bm{q}}}{(2 \pi)^3} \frac{1}{\bm{q}^2} (2 \pi)^3  \delta^{(3)}(\bm{q} - \bm{p} - \bm{k}) (2 \pi)^3  \delta^{(3)}( \bm{q} + \bm{k}'-\bm{p}').
\end{align}
Assuming that $\Phi(\bm{k})$ does not support extremely high momentum modes, and that $f(\bm{p})$ supports momentum modes up to  $\sim 1/R$, the above approximation also, in general, requires support of  $\bm{k}$ modes greater than  $\sim (\sqrt{1 + 0.2 m R^2 / (\hbar \, t)} - 1)/R$. This, together with $ t \ll 2 m R^2 /\hbar$ would set a limit of $\bm{k}$ modes equal to or greater than $\sim 0.41 / R$, which is always satisfied by the Fourier transform of \eqref{eq:semiclassicalPhi}.    So if the potential were external and not sufficiently spatially varying, e.g.\ it were  constant over the matter objects, then this condition is not satisfied (equivalently, $\omega_{\bm{p}} = \omega_{\bm{q}}$ before \eqref{eq:timeInt}): in this case, in the approximation $\bm{p}^2 \ll m^2 c^2 / \hbar^2$ as specified above, $\beta$, and thus $\beta^{(4)}_{ij}$, would evaluate to zero as would be expected. 

Now we move back to position space by performing reverse Fourier transformations for $\Phi(\bm{k)}$,  $f(\bm{p})$ and $f'(\bm{p'})$. Then, after performing the integrals over $\bm{p}$, $\bm{p}'$, $\bm{k}$ and $\bm{k}'$ we end up with:
\begin{align}
    \beta
        &=  \frac{2  M m^2 t }{ \hbar^3}      \int d^3  \bm{x}\,  d^3 \bm{y}        f(\bm{x}) f^{' \ast}(\bm{y}) \Phi(\bm{x}) \Phi(\bm{y})   \int   \frac{d^3{\bm{q}}}{(2 \pi)^3} \frac{1}{\bm{q}^2}  e^{i\bm{q}.(\bm{x} - \bm{y}) }.
\end{align}
Integrating over $\bm{q}$  leads us to \eqref{eq:beta4ijFull} from the main text.

\subsection{Gaussians} \label{app:Gaussian}

In the main text, for simplicity, we considered that the wavepackets of the experiment are prepared in unit-step spherical wavefunctions: $\tilde{\phi}_{\kappa i}(\bm{x}) = \theta(R_{\kappa i} - |\bm{x} - \bm{X}_{\kappa i}|) / \sqrt{V_{\kappa i}}$, where we  allow  for the idea that the wavepackets have different radii. Although theoretically simple, in practice, such wavefunctions would be challenging to generate exactly and so here we  consider Gaussian distributions:  $\tilde{\phi}_{\kappa i}(\bm{x}) = e^{-|x - X_{\kappa i}|^2 /  2 \sigma^2_{x,\kappa i} - |y - Y_{\kappa i}|^2 /  2 \sigma^2_{y,\kappa i} - |z - Z_{\kappa i}|^2 /  2 \sigma^2_{z,\kappa i}} / \pi^{3/4} (\sigma_{x,\kappa i} \sigma_{y,\kappa i} \sigma_{z,\kappa i})^{1/2}$. For simplicity, we take $\sigma_{x,\kappa i} = \sigma_{y,\kappa i} = \delta \sigma_{\kappa i}$ and  $\sigma_{z,\kappa i} = \sigma_{\kappa i}$ where $\delta \in \mathbb{R}$, and consider the wavepackets to all be aligned along the $z$-axis.  In order to relate the two cases of unit-step functions and Gaussians, it is appropriate to match the densities  $|\tilde{\phi}_{\kappa i}(\bm{x})|^2$. The standard way to do this is to  consider the mean-square radius of the unit-step density, which is $\langle |x|^2 \rangle =   \int d^3 x |x|^2 \theta(R_{\kappa i} - x)/V_{\kappa i} = \sqrt{3/5} R_{\kappa i}$. The variance for the Gaussian density function is instead $\langle |x|^2 \rangle = 3 \sigma_{\kappa i}^2/2$. Matching these,  sets $\sigma_{\kappa i} = \sqrt{2/5}R_{\kappa i}$. 

In addition to being more realistic, another advantage in using Gaussians is that we can analytically solve the amplitudes of the processes of interest with fewer approximations.  As in the main text, we start with the field in a state of the form:
\begin{align} \label{eq:psi0}
   |\psi(0)\rangle = \frac{1}{2}  \left(|N\rangle_{1L}  + |N\rangle_{1R} \right) \left( |N\rangle_{2L}  +  |N\rangle_{2R} \right) |0\rangle,
\end{align}
where $|0\rangle $ indicates that the vast majority of field modes are essentially in the vacuum.   As before, we write $|N\rangle_{\kappa i}$ as:
\begin{align} \label{eq:x}
    |N\rangle_{\kappa i} &:= \frac{1}{\sqrt{N!}} \int \prod^N_{j=1} d^3 \bm{x}_j \tilde{\phi}_{\kappa i}(\bm{x}_j) |\bm{x}_j \rangle,
\end{align}
which is shorthand for:
\begin{align} 
   |N\rangle_{\kappa i} &:= \frac{1}{\sqrt{N!}}
\int d^3\bm{x}_1\cdots d^3\bm{x}_N\,
\prod_{j=1}^N \tilde{\phi}_{\kappa i}(\bm{x}_j)\,
|\bm{x}_1,\dots,\bm{x}_N\rangle_{\mathrm{sym}}.
\end{align} 
This can also be written as:
\begin{align}    
  |N\rangle_{\kappa i}  &= \frac{1}{\sqrt{N!}} \int \prod^N_{j=1} d^3 \bm{k}_j \frac{\tilde{\phi}_{\kappa i}(\bm{k}_j)}{\sqrt{2 \omega_{\bm{k}_j}}} |\bm{k}_j \rangle\\
    &= \frac{1}{\sqrt{N!}} \int \prod^N_{j=1} d^3 \bm{k}_j \tilde{\phi}_{\kappa i}(\bm{k}_j) \hat{a}^{\dagger}_{\bm{k}_j} |0 \rangle,
\end{align}
where $|\bm{x} \rangle = \int \frac{d^3 \bm{k}}{(2 \pi)^3} \frac{1}{\sqrt{2 \omega_{\bm{k}}}} e^{-i\bm{k}.\bm{x}} |\bm{k} \rangle$ and $\tilde{\phi}_{\kappa i} (\bm{k}) = \int d^3 \bm{x} \tilde{\phi}_{\kappa i} (\bm{x}) e^{-i \bm{k}.\bm{x}}$. 

The free Hamiltonian is the following:
\begin{align} \label{eq:H0}
    H_0 &= \int d^3 \bm{x} \left[  \hat{\pi} (\bm{x}) \hat{\pi}^{\dagger} (\bm{x}) + \nabla \hat{\phi}^{\dagger} (\bm{x}) \nabla \hat{\phi} (\bm{x}) + \frac{m^2 c^2}{\hbar^2} \hat{\phi}^{\dagger} (\bm{x}) \hat{\phi} (\bm{x})\right]. 
\end{align}
We can choose to write $U_0 = e^{- i H_0 t} $ when acting on $|N\rangle_{\kappa i}$ as  $|N\rangle_{0\kappa i} := U_0 |N\rangle_{\kappa i} $.When acting $U_0$ on the initial state $|\psi(0)\rangle $, we can then write:
\begin{align} \label{eq:psi0}
   |\psi_0(t)\rangle = U_0 |\psi(0)\rangle = \frac{1}{2}  \left(|N\rangle_{01L}  + |N\rangle_{01R} \right) \left( |N\rangle_{02L}  +  |N\rangle_{02R} \right) |0\rangle.
\end{align}
Here, we are tracking the wavepackets under free evolution - we define the time-dependent wavepacket modes by evolving the initial wavepacket profiles with the free Hamiltonian: $\tilde{\phi}_{0\kappa i} (\bm{k}) :=  \tilde{\phi}_{\kappa i} (\bm{k}) \exp\{ -i \omega_{\bm{k} } t\}$.

We now consider the classical gravity interaction. From the main text, in the non-relativistic limit of gravity, we can write:
\begin{align}
 H_{int}
= \frac{4}{c^{2}} \int d^{3}\bm{x}\,\Phi(\mathbf{x})
\!\left(\hat\pi(\mathbf{x})\hat\pi^{\dagger}(\mathbf{x})
- \frac{m^{2}c^{2}}{2\hbar^{2}}
\hat\phi^{\dagger}(\mathbf{x})\hat\phi(\mathbf{x})\right).
\end{align}
Here, $\Phi(\bm{x})$ is the classical gravitational potential generated by the wavepackets, which we assume to be time-independent for simplicity. Since it is unknown how gravity is sourced by matter in a fundamentally classical gravity, $\Phi(\bm{x})$ can be quite general. In the majority of this section we will consider the semi-classical Einstein version of classical gravity \cite{moller1962theories,ROSENFELD1963353}. For a single wavepacket of density $\rho(\bm{x}) = M |\tilde{\phi}(\bm{x})|^2$, in semi-classical gravity the potential would be defined, as standard, as $\nabla^2 \Phi(\bm{x}) = - 4 \pi G \rho(\bm{x})  $ such that $\Phi(\bm{k}) = - 4 \pi G \rho(\bm{k}) / k^2$. For a Gaussian wavepacket, $\tilde{\phi}(\bm{k})_{\kappa i} = (4 \pi \sigma^2)^{3/4} \delta e^{-\sigma^2(\delta^2 (k_x^2 + k_y^2) + k_z^2)/2} e^{-i\bm{k}.\bm{X}_{\kappa i}}$ we then have $\Phi(\bm{k}) = - 4 \pi G M e^{-\sigma_L^2(\delta^2_L (k_x^2 + k_y^2) + k_z^2)/4} e^{-i\bm{k}.\bm{X}_{\kappa i}} / k^2$, where  $\sigma_L = \sigma$ and $\delta_L = \delta$. Note we use the notation $\sigma_L$ for the potential since, as we show below, it is also possible for $\sigma_L \neq \sigma$ and $\delta_L \neq \delta$, with $\sigma$ and $\delta$ defined for the wavepackets.

Allowing  for the possibility that $H = H_0 + H_{int}$ generates entanglement in the chosen basis \eqref{eq:psi0}, and that the experiment is setup  in that basis  (for example, the magnets of the experiment are only able to direct the states $|N\rangle_{0\kappa i} |N\rangle_{0\lambda j}$ to the measurement location), we
are interested  in final states of the form: 
\begin{align}\nonumber
    |\psi(t)\rangle = \frac{1}{\mathcal{N}}  \Bigg(&\alpha_{LL} |N\rangle_{01L}  |N\rangle_{02L}  +  \alpha_{LR} |N\rangle_{01L} |N\rangle_{02R}  \\ \label{eq:stateAfterG}
    &+ \alpha_{RL} |N\rangle_{01R} |N\rangle_{02L}   + \alpha_{RR} |N\rangle_{01R} |N\rangle_{02R} \Bigg).
\end{align}
 We can calculate the amplitude $\beta_{ij}$ as discussed in the main text - we compute $_{0\kappa i}{\langle} N | \, _{0\lambda  j}{\langle} N| \psi(t)\rangle = \, _{0\kappa i}{\langle} N | \, _{0\lambda  j}{\langle} N| U_0 U_I |\psi(0)\rangle = \, _{\kappa i}{\langle} N | \, _{\lambda  j}{\langle} N|  U_I |\psi(0)\rangle = \, _{\kappa i}{\langle} N | \, _{\lambda  j}{\langle} N|  U_I |N\rangle_{\kappa i} |N\rangle_{\lambda j} $  (see below for a discussion on this when the wavepackets are Gaussian). This is essentially what was considered in the main text when approximately neglecting the free evolution of the wavepackets, whereas here we do not make this approximation. Note that in the case of step functions, as long as $|\bm{X}_{\kappa i } - \bm{X}_{\lambda j}| \geq 2 R$, the wavepacket modes are exactly orthonormal: $_{\kappa i}{\langle} N | N \rangle_{\lambda j} = \delta_{\kappa i,\lambda j}$ since $\int d^3 \bm{x} \tilde{\phi}^{\ast}(\bm{x})_{\kappa i} \tilde{\phi}(\bm{x})_{\lambda j} = 0$, which is preserved under $U_0$.  However, in the Gaussian case, this  orthonormallity is only up to the overlap of the Gaussian profiles. That is, $_{\kappa i}{\langle} N| N\rangle_{\lambda j} = e^{-N d^2 / (4 \sigma^2)}$ when $\kappa i \neq \lambda j$, which is preserved by $U_0$. Therefore, in the Gaussian case we need to be careful that $\beta_{ij}$ is coming from the dynamics and not the fact that the states are not exactly orthogonal. One method is to simply extract out the dynamic part by computing $ \, _{\kappa i}{\langle} N | \, _{\lambda  j}{\langle} N|  U_I - 1 |N\rangle_{\kappa i} |N\rangle_{\lambda j}$  or simply making sure that $\beta_{ij} = \, _{\kappa i}{\langle} N | \, _{\lambda  j}{\langle} N|  U_I  |N\rangle_{\kappa i} |N\rangle_{\lambda j} \gg  \, _{\kappa i}{\langle} N | \, _{\lambda  j}{\langle} N|  1 |N\rangle_{\kappa i} |N\rangle_{\lambda j}$. For the Gaussian wavepackets, $\, _{\kappa i}{\langle} N | \, _{\lambda  j}{\langle} N|  N\rangle_{\kappa i} |N\rangle_{\lambda j} \sim  1 + N^2 e^{-d^2 / (2 \sigma^2)}$ which can easily be made insignificant   - e.g.\ with $d = 15 R$ and $N = 10^{20}$, then  $N^2 e^{-d^2 / (2 \sigma^2)} \sim 10^{-83}$. In writing $\beta_{ij} = \, _{\kappa i}{\langle} N | \, _{\lambda  j}{\langle} N|  U_I |N\rangle_{\kappa i} |N\rangle_{\lambda j}$, we are also assuming orthogonality between other modes, such as $_{1R}{\langle} N | \, _{2L}{\langle} N|N\rangle_{1L} |N\rangle_{2R} = 0$. For the unit step wavefunctions these are exactly zero, and for Gaussians, since  $\Delta x \gg d >  2R$ and $N \gg 1$,  then $_{1R}{\langle} N | \, _{2L}{\langle} N|N\rangle_{1L} |N\rangle_{2R} \sim e^{-N (\Delta x)^2/(2 \sigma^2)}$ is vanishingly small.\footnote{Alternatively, we can re-define the wavepackets so that they are exactly orthogonal and compute the amplitude in that new basis. That is, we can define for example $\varphi_{1 R} (\bm{x}) = \tilde{\phi}_{1 R} (\bm{x})$ and $\varphi_{2 L} (\bm{x}) = [\tilde{\phi}_{2 L} (\bm{x}) - \epsilon \tilde{\phi}_{1 R} (\bm{x})]/\sqrt{1 - |\epsilon|^2}$ where $\epsilon = e^{-d^2 / (4 \sigma^2)}$ and then compute $\, _{\kappa i}{\langle} N | \, _{\lambda  j}{\langle} N|  U_I |N\rangle_{\kappa i} |N\rangle_{\lambda j}$ in this basis, with $\, _{\kappa i}{\langle} N  |N\rangle_{\kappa j\neq i} = 0$ and $\, _{\kappa i}{\langle} N | \, _{\lambda  j}{\langle} N|  U_I |N\rangle_{\kappa i} |N\rangle_{\lambda j} = 1$.}

Expanding $U_I$ as the Dyson series to fourth order, the amplitude $\beta_{ij}$ is made up of processes at second  and fourth order that occur in very different ways. The second order process is  depicted in Figure \ref{fig:CrissCross}\textcolor{blue}{a}, and the fourth order process in Figure \ref{fig:Fig2}\textcolor{blue}{e}, where, in this field-theory picture, we can think of virtual particles travelling between the wavepackets. The second order process can be evaluated from the following Wick contraction:
    \begin{align}\nonumber
        \beta^{(2)}_{ij}  = \frac{-4  m^4 }{   \hbar^{6} c^2}  \int^{ct}_0 d x^0 \, \int^{ct}_0 d y^0 \int d^3 \bm{x} \, \int d^3 \bm{y} \,  \,_{1i}{\langle} \wick{ \c3 N| \, _{2j}{\langle} \c1 N| \Phi(\bm{x}) \c1{\hat{\phi}}^{\dagger} (x) \c2{\hat{\phi}} (x) \Phi(\bm{y}) \c3{\hat{\phi}}^{\dagger} (y) \c4{\hat{\phi}} (y)    | \c2 N \rangle_{1i} \,| \c4 N \rangle_{2j}},
\end{align}
whereas the fourth order process that we are interested in is:
    \begin{align} \nonumber
        &\beta^{(4)}_{ij}  = \frac{16  m^8 }{  \hbar^{12} c^4}  \int d^4 x  \int d^4 y \, \int d^4 z  \int d^4 w \,  \times \\ \label{eq:beta2Wick} &\,_{1i}{\langle} \wick{ \c4 N| \, _{2j}{\langle} \c1 N| \Phi(\bm{x}) \Phi(\bm{y}) \Phi(\bm{z}) \Phi(\bm{w}) \c1{\hat{\phi}}^{\dagger} (x) \c3{\hat{\phi}} (x) \c3{\hat{\phi}}^{\dagger} (y) \c2{\hat{\phi}} (y)  \c4{\hat{\phi}}^{\dagger} (z) \c5{\hat{\phi}} (z) \c5{\hat{\phi}}^{\dagger} (w) \c6{\hat{\phi}} (w)  | \c2 N \rangle_{1i} \,|  \c6 N \rangle_{2j}}.
\end{align}
Although the full theory is  local, the second order process has the flavour of a ``direct'' interaction  between the two wavepackets (although the full theory is still assumed local). In contrast, the fourth order case  involves virtual particle processes between the wavepackets: each wavepacket interacts with the local modes of the full field, which mediate an interaction between the packets. Note that the fourth order process considered above is not the higher order version of the second order case, which is instead:
    \begin{align}\nonumber
        \beta^{(4a)}_{ij}  &= \frac{16  m^8 }{  \hbar^{12} c^4}  \int d^4 x  \int d^4 y \, \int d^4 w  \int d^4 z \times \\&\,  \,_{1i}{\langle} \wick{ \c1 N \c5| \, _{2j}{\langle} \c3 N \c7| \Phi(\bm{x}) \Phi(\bm{y})
        \Phi(\bm{z}) \Phi(\bm{w})\c3{\hat{\phi}}^{\dagger} (x) \c4{\hat{\phi}} (x) \c1{\hat{\phi}}^{\dagger} (y) \c2{\hat{\phi}} (y)
\c7{\hat{\phi}}^{\dagger} (x) \c8{\hat{\phi}} (x) \c5{\hat{\phi}}^{\dagger} (y) \c6{\hat{\phi}} (y)
        | \c2 N \c6 \rangle_{1i} \,|  \c4 N \c8 \rangle_{2j}}.
\end{align}
We initially consider the second order process. For notational convenience, we write the wavefunctions  $\tilde{\phi}_{1i}(\bm{x})$ and $\tilde{\phi}_{2j}(\bm{x})$ as $f(\bm{x})$ and $f'(\bm{x})$. We  use:
\begin{align}
    \wick{\c{\hat{\phi}} (x)
      |\c {\bm{p}} \rangle = c \sqrt{\hbar}\,e^{i p.x}},
\end{align}
such that, as shown in the main text:
\begin{align}
    \wick{  \c{\hat{\phi}}(x)  |\c N \rangle_{1 i}} 
  &= c \sqrt{\hbar} \sqrt{N}  \int \frac{d^3 \bm{k}}{(2 \pi)^3} \frac{1}{\sqrt{2 \omega_{\bm{k}}}} e^{i k. x} f(\bm{k}) |N-1 \rangle_{\kappa i}.
\end{align}
The second order process can then be written as $\beta^{(2)}_{ij} = \beta_2^2$ where
\begin{align}\nonumber
        \beta_2  
        &\approx\frac{-2 i M m c}{\hbar^2 } \int^{c t}_0 d x^0 \int d^3 \bm{x} \int \frac{d^3 \bm{p}}{(2 \pi)^3} \frac{d^3 \bm{p}'}{(2 \pi)^3}   e^{-i p'.x} e^{i p.x} \frac{f(\bm{p})}{\sqrt{2 \omega_p}} \frac{f^{' \ast}(\bm{p'})}{\sqrt{2} \omega_{p'}}   \Phi(\bm{x})\\
        &=\frac{-2 i M m c}{\hbar^2 } \int^{c t}_0 d x^0 \int d^3 \bm{x} \int \frac{d^3 \bm{k}}{(2 \pi)^3} \frac{d^3 \bm{p}}{(2 \pi)^3} \frac{d^3 \bm{p}'}{(2 \pi)^3}  e^{i \omega_{p'} x^0} e^{-i \omega_p x^0 } e^{-i \bm{p'}.\bm{x}} e^{i \bm{p}.\bm{x}} e^{i \bm{k}.\bm{x}} \frac{f(\bm{p})}{\sqrt{2 \omega_p}} \frac{f^{'\ast}(\bm{p'})}{\sqrt{2} \omega_{p'}}  \Phi(\bm{k}) \\
        &=\frac{-2 i M m c}{\hbar^2 } \int^{c t}_0 d x^0  \int \frac{d^3 \bm{k}}{(2 \pi)^3} \frac{d^3 \bm{p}}{(2 \pi)^3} \frac{d^3 \bm{p}'}{(2 \pi)^3} e^{i \omega_{p'} x^0} e^{-i \omega_p x^0 } \delta^{(3)}(\bm{p} + \bm{k} - \bm{p'}) \frac{f(\bm{p})}{\sqrt{2 \omega_p}} \frac{f^{'\ast}(\bm{p'})}{\sqrt{2} \omega_{p'}}  \Phi(\bm{k}) \\
        &\approx \frac{- i M  }{ c \hbar}   \int^{c t}_0 d x^0  \int \frac{d^3 \bm{k}}{(2 \pi)^3} \frac{d^3 \bm{p}}{(2 \pi)^3} \frac{d^3 \bm{p}'}{(2 \pi)^3}  e^{i p^{' 2} \tau/2} e^{-i p^{ 2} \tau/2 } \delta^{(3)}(\bm{p} + \bm{k} - \bm{p'})  f(\bm{p}) f^{'\ast}(\bm{p'}) \Phi(\bm{k})\\
        &= \frac{- i M  }{ c \hbar}   \int^{c t}_0 d x^0  \int \frac{d^3 \bm{k}}{(2 \pi)^3} \frac{d^3 \bm{p}}{(2 \pi)^3}  e^{i (\bm{p} + \bm{k})^2 \tau/2} e^{-i p^{ 2} \tau/2 }   f(\bm{p}) f^{'\ast}(\bm{p} + \bm{k}) \Phi(\bm{k})\\
        &= \frac{- i M  m }{  \hbar^2}   \int^{ \alpha}_0 d \tau  \int \frac{d^3 \bm{k}}{(2 \pi)^3} \frac{d^3 \bm{p}}{(2 \pi)^3}  e^{i (\bm{p} + \bm{k})^2 \tau/2} e^{-i p^{ 2} \tau/2 }   f^{'}(\bm{p}) f^{'}(\bm{p} + \bm{k}) \Phi(\bm{k}) e^{i \bm{p}.\bm{d}},
\end{align}
where the first line assumed $N \gg 1$, and we defined  $\tau := \hbar x^0 / (m c)$, $\alpha := \hbar t / m $, $\bm{d} = |\bm{X}_{1i} - \bm{X}_{2j}|$ and we assumed that  $\sigma \gg \hbar / (m c)$ so that the wavepackets behave approximately non-relativistically. We next assume that only the potential generated by the wavepacket states $1R$ and $2L$ is relevant given that the other states are far away in the assumption that $\Delta x \gg d$. We then write $\Phi(\bm{k}) \rightarrow \Phi(\bm{k})(1+ e^{i\bm{k}.\bm{d}})/2$, where $\Phi(\bm{k})$ is the potential generated by one of the wavepacket states and the factor of $1/2$ is in the assumption of a version of classical gravity similar to (or equivalent to) semi-classical Einstein gravity. Including the other potentials just introduces additional phase factors. We then would like to compute the following:
\begin{align}\label{eq:firstOrderStatPhase}
        \beta_2  &= \frac{- i M  m }{  2 \hbar^2}   \int^{ \alpha}_0 d \tau  \int \frac{d^3 \bm{k}}{(2 \pi)^3} \frac{d^3 \bm{p}}{(2 \pi)^3}  e^{i (\bm{p} + \bm{k})^2 \tau/2} e^{-i p^{ 2} \tau/2 }   f^{'}(\bm{p}) f^{'}(\bm{p} + \bm{k}) \Phi(\bm{k}) (1 + e^{i \bm{k}.\bm{d}}) e^{i \bm{p}.\bm{d}}.
\end{align}
Using the Gaussian distributions discussed above, and introducing a Schwinger parameter $s$, this can be evaluated to $\beta_2
= [\beta_2(0) + \beta_2(1)]/2$, where $\beta_2(c)$ is:
\begin{align} \label{eq:beta1c}
\beta_2(c)
= 
-\frac{i m GM^2}{2\sqrt{\pi}\,\hbar^2}\;
e^{-\frac{d^2}{4\sigma^2}}
\int_{0}^{\alpha} d\tau
\int_0^\infty ds\;
\frac{
e^{\frac{B(\tau,c)^2}{4(A_z(\tau)+s)}
}}{
(A_\perp(\tau)+s)\sqrt{A_z(\tau)+s}
}.
\end{align}
and:
\begin{align}
B(\tau,c)&:=\left(\frac{\tau}{2\sigma^2}-i\Big(c-\frac12\Big)\right)d,\\
A_\perp(\tau)&:=\frac{\delta^2 \sigma^2+\delta^2_L \sigma_L^2}{4}+\frac{\tau^2}{4\sigma^2\delta^2},\\
\qquad
A_z(\tau)&:=\frac{\sigma^2+\sigma_L^2}{4}+\frac{\tau^2}{4\sigma^2}.
\end{align}
With $\delta = \delta_L = 1$, this can be further simplified to:
\begin{align} \label{eq:beta1spherefull}
   \beta_2(c) 
   &=\frac{- 2 i m G M^2}{\hbar^2  } e^{-d^2/(4 \sigma^2)} \int^{\alpha}_0 d \tau \frac{1}{B} \mathrm{erfi}(B/(2 \sqrt{A})),
\end{align}
 where 
\begin{align}
    B &= d (\tau/\sigma^2 + i (1 - 2 c)),\\
    A &= \sigma^2 + \sigma^2_L+ \tau^2/\sigma^2.
\end{align}
The latter result used the identity:
\begin{align}
    \int^{\infty}_0 d k e^{- A k^2} \sinh(B k)/(Bk) = \pi \frac{\mathrm{erfi}(B / (2 \sqrt{A})}{2 B}.
\end{align}
 In the assumption that $ d \gg 2 \sigma$, we can  approximate $\mathrm{erfi}(x) \approx \frac{e^{x^2}}{\sqrt{\pi } x} (1 + \frac{1}{2x^2} + \cdots)$ when $\mathrm{Re}(x^2) > 0$.

When $\alpha \ll \sigma^2$, i.e.\ ignoring the free evolution entirely, and taking $\delta = 1$ for simplicity, we obtain:
\begin{align}
    \beta_2 &\approx - \frac{2i G M^2 t }{\hbar (d/2)} e^{-d^2/(4 \sigma^2)} \mathrm{erf}(d / (2 \sqrt{\sigma^2 + \sigma_L^2}))\\ \label{eq:beta1SmallTimes}
    &\approx - \frac{2i G M^2 t }{\hbar (d/2)} e^{-d^2/(4 \sigma^2)},
\end{align}
where in the last line we assumed $d \gg 2 \sigma$ and $\sigma > \sigma_L$. Note that this is just the quantum gravity phase $\varphi_{ij}$ multiplied by the Gaussian overlap of the wavepackets at $t = 0$. In the main text we found that this amplitude was approximately zero for unit-step wavefunctions when $\alpha \ll \sigma^2$, which matches the result here as there is no overlap of the unit-step wavefunctions at the initial time.  With $\delta \gg 1$, instead we obtain:
\begin{align}
    \beta_2 &\approx - \frac{2i G M^2 t }{\hbar S} e^{-d^2/(4 \sigma^2)}
\end{align}
where $S = d/2$ when $d \gg 2 \delta \sqrt{\sigma^2 + \sigma_L^2}$ and $S = \delta \sqrt{\sigma^2 + \sigma_L^2}/ \sqrt{\pi}$ when $d \ll 2 \delta \sqrt{\sigma^2 + \sigma_L^2}$.

In contrast, when $d \gg 2 \sigma$, but $\alpha \gg \sigma^2$ (and $\alpha \gg \delta^2 \sigma^2$) such that free evolution of the wavepackets is non-negligible, we can approximate $\beta^{(2)}(c)$ by taking an endpoint approximation for the $s$ integral around $s = 0$ and for the $\tau$ integral around $\tau = \alpha$ (since these are  the regions that  will be the least exponentially suppressed). This results in:
\begin{align} \label{eq:beta1Gaussian}
    \beta_2     &\approx- \frac{16 i m \sigma^3 \delta^2}{\sqrt{\pi} \hbar^2} G M^2  \frac{\alpha}{d^4}     e^{ - \frac{d^2 \sigma^2}{2\alpha^2}} e^{ - \frac{d^2 \sigma^2_L}{4\alpha^2}} \sin(d^2 / (2 \alpha)).
\end{align}
Taking $\sigma_L = \sigma$ then gives
\begin{align} \label{eq:beta1GaussianSame}
    \beta_2    &\approx- \frac{16 i m \sigma^3 \delta^2}{\sqrt{\pi} \hbar^2} G M^2  \frac{\alpha}{d^4}     e^{ - \frac{3d^2 \sigma^2}{4\alpha^2}}  \sin(d^2 / (2 \alpha)).
\end{align}
Numerical evaluation of \eqref{eq:beta1c} shows that \eqref{eq:beta1Gaussian} and \eqref{eq:beta1GaussianSame} are accurate approximations  as long as $d \gg 2 \sigma$. Note that it is now possible to avoid exponential suppression at late times in this regime. If $\sigma_L \ll \sigma$, for which we will consider a physical case below, then this depends essentially on just the mode overlap time  $\alpha \sim d \sigma / \sqrt{2}$, at which point gravity can induce non-negligible entanglement.  Note, however, that even when not exponentially suppressed, due to the sinusoidal factor, the amplitude still oscillates rapidly for times $ \alpha \ll d^2$ and is often zero for such times.  The result \eqref{eq:beta1Gaussian} can also be straightforwardly obtained by integrating over $\tau$ in \eqref{eq:firstOrderStatPhase} and then performing a stationary phase evaluation. Such an evaluation results in:
\begin{align}
    \beta_2 \approx  -\frac{ M m }{\hbar^2} \frac{1}{\alpha} \frac{1}{(4 \pi^3)}      
    f'(\bm{d}/\alpha) f'(0) \frac{1}{d^2} \Phi(-\bm{d}/\alpha ) \sin(d^2 / (2 \alpha))  .
  \end{align}
  Using $\bm{d} = (0,0,d)$ and plugging in the expressions for $f'(\bm{x})$ and $\Phi(\bm{x})$ results in \eqref{eq:beta1Gaussian}.

  If on the other hand $\alpha \gg \sigma^2$ but $\alpha \ll \delta^2 \sigma^2$ then we obtain either by stationary phase or from \eqref{eq:beta1c} that:
\begin{align} 
    \beta_2     &\approx- \frac{16 i m \sigma }{\sqrt{\pi} \hbar^2 (\delta^2 \sigma^2 + \delta^2_L \sigma^2_L) } G M^2  \frac{\alpha^3}{d^4}     e^{ - \frac{d^2 \sigma^2}{2\alpha^2}} e^{ - \frac{d^2 \sigma^2_L}{4\alpha^2}} \sin(d^2 / (2 \alpha)).
\end{align}
With $\sigma_L = \sigma$, we then have:
  \begin{align}
    \beta_2    &\approx- \frac{8 i m  }{\sqrt{\pi} \hbar^2 \delta^2 \sigma } G M^2  \frac{\alpha^3}{d^4}     e^{ -3 \frac{d^2 \sigma^2}{4\alpha^2}}  \sin(d^2 / (2 \alpha)).
\end{align}

  We now consider the fourth order process. The Wick contraction \eqref{eq:beta2Wick} gives $\beta^{(4)}_{ij} = \beta_4^2$ with:
\begin{align}\nonumber
        \beta_4
        &\approx \frac{4 i m^3 M c}{\hbar^4}  \int_t  d^4  x\,  d^4 y  \,   \int \frac{d^3 \bm{k}}{(2 \pi)^3}  \frac{d^3 \bm{k'}}{(2 \pi)^3}  \frac{d^3 \bm{p'}}{(2 \pi)^3}\frac{d^3 \bm{p}}{(2 \pi)^3} \frac{d^4 q}{(2\pi)^4} \frac{1}{\sqrt{2\omega_{\bm{p'}}}}  \frac{1}{\sqrt{2\omega_{\bm{p}}}} f(\bm{p}) f^{' \ast}(\bm{p'}) \Phi(\bm{k}) \Phi(\bm{k}') \\ \label{eq:startingpoint}
        &\times e^{i\bm{k}'.\bm{x}} e^{i\bm{k}.\bm{y}} e^{ip.y} e^{-i p'.x} \frac{i}{q^2 + m^2 c^2/\hbar^2 + i \epsilon} e^{i q. (x - y)}\\
        &:= \frac{4 i m^3 M c}{\hbar^4}  \int_t  d^4  x\,  d^4 y  \,   \int \frac{d^3 \bm{k}}{(2 \pi)^3}  \frac{d^3 \bm{k'}}{(2 \pi)^3}  \frac{d^3 \bm{p'}}{(2 \pi)^3}\frac{d^3 \bm{p}}{(2 \pi)^3} \frac{d^4 q}{(2\pi)^4} \frac{1}{\sqrt{2\omega_{\bm{p'}}}}  \frac{1}{\sqrt{2\omega_{\bm{p}}}} f'(\bm{p}) f^{' \ast}(\bm{p'}) e^{i \bm{p}.\bm{d}} \, \Phi(\bm{k}) \Phi(\bm{k}') \\ \label{eq:startingpoint}
        &\times e^{i\bm{k}'.\bm{x}} e^{i\bm{k}.\bm{y}} e^{ip.y} e^{-i p'.x} \frac{i}{q^2 + m^2 c^2/\hbar^2 + i \epsilon} e^{i q. (x - y)}.
\end{align}
We first integrate over $q^0$ so that we write the Feynman propagator in terms of Wightman functions: $\theta(x^0 - y^0) D(x - y)  + \theta(y^0 - x^0) D(y - x)$, where:
\begin{align}
    D(x - y) = \int \frac{d^3 \bm{q}}{(2 \pi)^3} \frac{1}{2 \omega_q} e^{iq.(x-y)},
\end{align}
with $\omega_q = c \sqrt{ \bm{q}^2 + m^2 c^2 / \hbar^2}$. Then  we integrate over space and subsequently $\bm{q}$ to get:
\begin{align}\nonumber
        \beta_4
        &= \frac{4 i m^3 M c^2 \, (2 \pi)^3}{\hbar^4}      \int_0^{c t}  d  x^0\,  d y^0  \,   \int \frac{d^3 \bm{k}}{(2 \pi)^3}  \frac{d^3 \bm{k'}}{(2 \pi)^3}   \frac{d^3 \bm{p}}{(2 \pi)^3} \frac{d^3 \bm{p}'}{(2 \pi)^3}  \frac{1}{\sqrt{2\omega_{\bm{p'}}}}  \frac{1}{\sqrt{2\omega_{\bm{p}}}} f(\bm{p}) f^{' \ast}(\bm{p'}) \Phi(\bm{k}) \Phi(\bm{k}')   \\ \label{eq:BeforeTimeInts}
        &\times\frac{e^{-i \omega_{\bm{p}} y^0/c} e^{i \omega_{\bm{p'}} x^0 / c}}{2 \omega_q}  \left(\theta(x^0 - y^0) e^{-i \omega_q (x^0 - y^0)/c} + \theta(y^0 - x^0) e^{i \omega_q (x^0 - y^0)/c}\right) \delta^{(3)}( \bm{p} + \bm{k} + \bm{k}'-\bm{p}'),
        \end{align}
where $\omega_q = c \sqrt{(\bm{p} + \bm{k})^2 + m^2 c^2 / \hbar^2}$. If we were to next perform the time integrals we would find that the $\theta(y^0 - x^0)$ component scales as $\omega_q + \omega_{p'}$ whereas the $\theta(x^0 - y^0)$ component scales as $\omega_q - \omega_p$. In the non-relativistic approximation of the wavepackets, the $\theta(y^0 - x^0)$ part is thus heavily suppressed for sufficient times as expected (as $m c^2 t / \hbar $). In this approximation (see the main text for when this approximation is not used), we can thus neglect this component and write:
  \begin{align}\nonumber
        \beta_4
        &\approx \frac{4 i m^3 M c^2 \, }{\hbar^4}      \int_0^{c t}  d  x^0\,  d y^0  \,   \int \frac{d^3 \bm{k}}{(2 \pi)^3}  \frac{d^3 \bm{k'}}{(2 \pi)^3}   \frac{d^3 \bm{p}}{(2 \pi)^3} \frac{d^3 \bm{p}'}{(2 \pi)^3}  \frac{1}{\sqrt{2\omega_{\bm{p'}}}}  \frac{1}{\sqrt{2\omega_{\bm{p}}}} f(\bm{p}) f^{' \ast}(\bm{p'}) \Phi(\bm{k}) \Phi(\bm{k}')   \\ \label{eq:BeforeTimeInts}
        &\times\frac{e^{-i \omega_{\bm{p}} y^0/c} e^{i \omega_{\bm{p'}} x^0 / c}}{2 \omega_q}  \theta(x^0 - y^0) e^{-i \omega_q (x^0 - y^0)/c}(2 \pi)^3 \delta^{(3)}( \bm{p} + \bm{k} + \bm{k}'-\bm{p}')\\\nonumber
         &\approx \frac{ i m^3 M  }{\hbar^4 }      \int_0^{\alpha}  d  \tau \,  d \tau'  \,   \int \frac{d^3 \bm{k}}{(2 \pi)^3}  \frac{d^3 \bm{k'}}{(2 \pi)^3}   \frac{d^3 \bm{p}}{(2 \pi)^3}  f'(\bm{p}) f'(\bm{p} + \bm{k} + \bm{k'} ) \Phi(\bm{k}) \Phi(\bm{k}')   \, e^{i\bm{p}.\bm{d}}\\  \label{eq:beta2BefStatPhase} &\times  e^{-i (p^2- (p+ k)^2) \tau'/2} e^{i  ((p + k + k')^2  - (p+k)^2) \tau/2}  \theta(\tau - \tau').
        \end{align}
  Using the Gaussian distributions we can perform Gaussian integrals to end up with $\beta_4 = \beta_4(1,0) + \beta_4(0,0) + \beta_4(1,1) + \beta_4(0,1)$, where:
\begin{align}
\beta_4(c,c') 
=
\frac{i G^2 m^3 M^3 }{4 \pi \hbar^4}\,
\int_{0}^{\alpha} d\tau\int_{0}^{\alpha} d\tau'\;
\theta(\tau-\tau')\;
\int_0^\infty ds\int_0^\infty ds'\;
\frac{1}{\sqrt{\det M}}\;
\exp\!\left(-\frac{d^2}{4\sigma^2}+\frac14\,\mathcal{J}^T \mathcal{M}^{-1}\mathcal{J}\right),
\end{align}
with:
\begin{align}
\mathcal{J}(\tau,\tau')&=
\binom{-\dfrac{\tau' }{2\sigma^2}+i\big(c-\frac12\big)}
{-\dfrac{\tau }{2\sigma^2}+i\big(c'-\frac12\big)}d,\\
\mathcal{M}(\tau,\tau',s,s')&=
\begin{pmatrix}
s+\dfrac{\sigma_L^2+\sigma^2}{4}+\dfrac{\tau'^2}{4\sigma^2},
&
\dfrac{\sigma^2}{4}+\dfrac{\tau\tau'}{4\sigma^2}-i\dfrac{\tau-\tau'}{4}
\\[6pt]
\dfrac{\sigma^2}{4}+\dfrac{\tau\tau'}{4\sigma^2}-i\dfrac{\tau-\tau'}{4},
&
s'+\dfrac{\sigma_L^2+\sigma^2}{4}+\dfrac{\tau^2}{4\sigma^2}
\end{pmatrix},\\
\sqrt{\det M}
&=
\bigl(A_\perp B_\perp - C_\perp^{\,2}\bigr)
\;
\sqrt{A_z B_z - C_z^{\,2}},\\
A_\perp &=
s + \frac{\delta^2 \sigma^2 + \delta_L^2 \sigma_L^2}{4}
+ \frac{\tau'^2}{4\sigma^2\delta^2},\\[4pt]
B_\perp &=
s' + \frac{\delta^2 \sigma^2 + \delta_L^2 \sigma_L^2}{4}
+ \frac{\tau^2}{4\sigma^2\delta^2},\\[4pt]
C_\perp &=
\frac{\sigma^2}{4}\,\delta^2
+ \frac{\tau\tau'}{4\sigma^2\delta^2}
- i\,\frac{\tau-\tau'}{4},\\
A_z &=
s + \frac{\sigma_L^2+\sigma^2}{4}
+ \frac{\tau'^2}{4\sigma^2},\\[4pt]
B_z &=
s' + \frac{\sigma_L^2+\sigma^2}{4}
+ \frac{\tau^2}{4\sigma^2},\\[4pt]
C_z &=
\frac{\sigma^2}{4}
+ \frac{\tau\tau'}{4\sigma^2}
- i\,\frac{\tau-\tau'}{4},
\end{align}
and $\bm{d} = (0,0,d)$.  In the case that $\delta = \delta_L = 1$, this can be simplified to:
\begin{align}\nonumber
     \beta_4(c,c')   &= \frac{ i m^3  (4 \pi)^2 G^2 M^3 }{\hbar^4 }    e^{- d^2/(4 \sigma^2)} \int_0^{\alpha}  d  \tau \,  d \tau' \, \int^1_0 da \, \frac{e^{d^2 a^2(i (2 c - 1) - \tau'/ \sigma^2)^2/(4 A)}}{2 \pi^{3/2} \sqrt{A}}     \\
        &\times \theta(\tau - \tau')      \frac{\mathrm{erfi}\left(\frac{ B(\tau,\tau',a)}{2 \sqrt{C(\tau,\tau',a)}}\right)}{2  \pi   B(\tau,\tau',a)},
\end{align}
with
\begin{align}
 B(\tau,\tau',a)  &=   d  [a^2((2c - 1)i - \tau'/ \sigma^2) (i (\tau - \tau') - \sigma^2 -  \tau \tau' / \sigma^2)  / A -  \tau/ \sigma^2 + (2c' - 1) i],\\
    C(\tau,\tau',a)  &=   \sigma^2_L  + \sigma^2 + \tau^2/( \sigma^2) - a^2 (i (\tau - \tau') - \sigma^2 -  \tau \tau' / \sigma^2)^2  /A,\\
     A &= \sigma_L^2  +\sigma^2 + \tau^{'2}/ \sigma^2.
\end{align}
If it were the case that  $\sigma_L \ll \sigma$, for which we consider a physical scenario below,  and $\alpha \ll \sigma^2 $ such that freee evolution can be entirely ignored, $\alpha \gg \sigma_L^2$ (and $\alpha \gg \delta_L \sigma_L \delta \sigma$) and $d \gg 2 \sigma$, then we can approximate the integral for $c=1$ and $c'=0$ with an endpoint approximation near $s=0 = s'$ (and $a = 1$ in the case of the $\delta =1$ expression), and $u = \alpha$, $v=0$ (where $u = \tau - \tau'$ and $v = \tau'$), resulting in:
\begin{align}\label{eq:beta210}
    \beta_4(1,0) &= \frac{  - 16 \sqrt{2} i m^3 G^2 M^3 \alpha^{13/2} e^{-3 i \pi / 4}}{\pi \hbar^4  \sigma^3  \delta^2 d^8 }  e^{-\frac{d^2 \sigma^2_L }{2 \alpha^2 }} e^{i \frac{d^2 }{2 \alpha}},
\end{align}
whereas $\beta_4(0,0)$, $\beta_4(1,1)$, and  $\beta_4(0,1)$ are always exponentially suppressed as $e^{-d^2 / (4 \sigma^2)}$, such that we can approximate $\beta_4\approx \beta_4(1,0)$. Equation \eqref{eq:beta210} can also be derived by starting from \eqref{eq:beta2BefStatPhase}, integrating over time, Taylor expanding with $\alpha \ll \sigma^2$ and then performing a stationary phase approximation to end up with: 
 \begin{align}
 \beta_4(1,0)  &=\frac{-4 M m^3}{\hbar^4}         \frac{1}{(2 \pi)^3} \frac{(2 \pi)^{3/2} e^{-3 i  \pi/4}}{\alpha^{3/2} } f'(\bm{x}=0) f^{' \ast}(\bm{x} =0) \Phi(\bm{d}/\alpha) \Phi(- \bm{d}/\alpha) \, e^{i d^2/(2\alpha)}    \frac{1}{(d/\alpha)^4}.
\end{align}
Plugging in the Gaussians and using $\sigma_L \ll \sigma$ and $\bm{d} = (0,0,d)$ results in \eqref{eq:beta210}. Note that in contrast to the first order case in this regime, \eqref{eq:beta1SmallTimes}, the second order process' amplitude can be non-exponentially suppressed at times towards $\alpha = \sigma_L d / \sqrt{2}$, while the first order process is always $e^{-d^2 / (4 \sigma^2)}$. This is due to the fourth order process involving local interactions between the wavepacket modes and higher momentum modes of the field, facilitating, from this point-of-view, virtual particle processes between the wavepackets and allowing the fourth order process to be significantly greater than the second order case, which has more of a direct  flavour as discussed above. Numerical analysis shows that \eqref{eq:beta210} is a good approximation to \eqref{eq:beta2BefStatPhase} in the appropriate regime. If instead $\alpha \gg \sigma^2_L$ but $\alpha \ll \delta_L \sigma_L \delta \sigma$, then we obtain:
\begin{align}
    \beta_4(1,0) &= \frac{  - 16 \sqrt{2} i m^3 G^2 M^3 \alpha^{15/2} e^{- i \pi / 4}}{\pi \hbar^4   d^8 \delta^2_L \sigma^2_L  \sigma  (2 \delta^2 \sigma^2 + \delta_L^2 \sigma_L^2 )}  e^{-\frac{d^2 \sigma^2_L }{2 \alpha^2 }} e^{i \frac{d^2 }{2 \alpha}}.
\end{align}
We now consider the alternative case where $\alpha \gg \sigma^2$ such that the free evolution of the wavepackets is non-negligible (but below we still work with $\alpha \ll \sigma d /\sqrt{2}$ such that wavepacket mode overlap  is insignificant). If also $\alpha \gg \delta^2 \sigma^2$, $\alpha \ll d^2$ and $d \gg 2 \sigma$, then for the  $c=1$ and $c'=0$ amplitude, an endpoint approximation near $s=0 = s'$, and $u = \alpha$, $v=0$ (where $u = \tau - \tau'$ and $v = \tau'$) results in:
\begin{align} \label{eq:beta210Long}
    \beta_4(1,0)   &=  \frac{ - 16 \sqrt{2} i m^3  G^2 M^3 \alpha^5}{\hbar^4 \pi  d^8}  e^{-\frac{d^2 \sigma^2_L}{2 \alpha^2}} e^{i \frac{d^2 }{2 \alpha}}.
\end{align}
Again this can be derived by first integrating over time in \eqref{eq:beta2BefStatPhase} and then performing a stationary phase approximation, resulting in 
\begin{align}
 \beta_4(1,0)
&=\frac{4 M m^3}{\hbar^4} 
\frac{1}{\alpha^3} \frac{1}{(2 \pi)^3}     
f'(\bm{x} = 0) f'(0) \Phi(\bm{d}/\alpha) \Phi(- \bm{d}/\alpha) \, e^{i d^2/(2\alpha)}    \frac{1}{(d/\alpha)^4}.
        \end{align}
In addition to the $\beta_4(1,0)$ amplitude, the amplitudes $\beta(0,0)$ and $\beta(1,1)$  can also be computed by endpoint  approximations at $s = 0 = s'$ and $u = 0$ and $v=\alpha$, resulting in:
\begin{align}\label{eq:beta00}
    \beta_4(0,0) &\approx \frac{1024 \sqrt{2} i m^4 (GM)^2}{ 3 \pi \hbar^4}  \frac{\alpha^5}{d^8} \frac{\sigma^3}{ \sigma^3_L}  e^{i d^2 / (2 \alpha)} e^{-\sigma^2 d^2 / (2 \alpha^2)} e^{-\sigma_L^2 d^2 / (8 \alpha^2)},\\\label{eq:beta11}
   \beta_4(1,1) &\approx \frac{1024 \sqrt{2} i m^4 (GM)^2}{ \pi \hbar^4}  \frac{\alpha^5}{d^8} \frac{\sigma^3}{ \sigma^3_L}  e^{-i d^2 / (2 \alpha)} e^{-\sigma^2 d^2 / (2 \alpha^2)} e^{-\sigma_L^2 d^2 / (8 \alpha^2)},
\end{align}
which can also be derived from stationary phase approximations. The $\beta_4(0,1)$ amplitude can also be shown to scale as $ e^{-\sigma^2 d^2 / (2 \alpha^2)} e^{-\sigma_L^2 d^2 / (4 \alpha^2)}$, i.e.\ it is more suppressed than the other amplitudes. Note that, while the amplitude  $\beta_4(1,0)$ goes as $e^{-\sigma^2_L d^2 / (2 \alpha^2)}$, the other amplitudes scale with   $e^{-\sigma^2 d^2 / (2 \alpha^2)}$ similar to the first order case, and so are significantly more suppressed given that $\sigma_L \ll \sigma$. We can then take $\beta_4 \approx \beta_4(1,0)$ as above. Also, just as with the case $\alpha \ll \sigma^2$, the fourth order process can be significantly larger than the second order process, particularly when $\alpha \ll \sigma d / \sqrt{2}$, which is essentially the mode overlap time, and $\alpha \sim \sigma_L d / \sqrt{2}$.  Numerical analysis shows that \eqref{eq:beta210}, \eqref{eq:beta00} and \eqref{eq:beta11} are good approximations to \eqref{eq:beta2BefStatPhase} in the appropriate regimes.

Other regimes of interest could be the following: with $\alpha \gg \sigma^2$ but $\alpha \ll \delta^2 \sigma^2$ and $\alpha \ll \delta^2_L \sigma_L^2$, we have:
\begin{align}
\beta_4(1,0)   &\approx    \frac{ - 32 \sqrt{2} i m^3  G^2 M^3 \alpha^7}{\hbar^4 \pi  d^8 \delta^2_L \sigma_L^2 (2 \delta^2 \sigma^2 + \delta_L^2 \sigma_L^2)}  e^{-\frac{d^2 \sigma^2_L}{2 \alpha^2}} e^{i \frac{d^2 }{2 \alpha}},
\end{align}
whereas if $\alpha \gg \sigma^2$ and $\alpha \gg \delta \sigma \delta_L \sigma_L$ but $\alpha \ll \delta^2_L \sigma_L^2$ then
\begin{align}
\beta_4(1,0)   &\approx    \frac{ - 32 \sqrt{2} i m^3  G^2 M^3 \alpha^5}{\hbar^4 \pi  d^8} \frac{\delta^2 \sigma^2}{\delta^2_L \sigma_L^2}   e^{-\frac{d^2 \sigma^2_L}{2 \alpha^2}} e^{i \frac{d^2 }{2 \alpha}},
\end{align}
and if $\alpha \ll \delta \sigma \delta_L \sigma_L$ then  
\begin{align}
\beta_4(1,0)   &\approx    \frac{ - 32 \sqrt{2} i m^3  G^2 M^3 \alpha^7}{\hbar^4 \pi  d^8 \delta_L^4 \sigma_L^4}    e^{-\frac{d^2 \sigma^2_L}{2 \alpha^2}} e^{i \frac{d^2 }{2 \alpha}},
\end{align}
and finally with $\alpha \ll \delta^2 \sigma^2$ and $\alpha \gg \delta_L^2 \sigma_L^2$, we obtain:
\begin{align}
\beta_4(1,0)   &\approx  \frac{ - 32 \sqrt{2} i m^3  G^2 M^3 \alpha^6}{\hbar^4 \pi  d^8 \delta^2 \sigma^2}    e^{-\frac{d^2 \sigma^2_L}{2 \alpha^2}} e^{i \frac{d^2 }{2 \alpha}}.
\end{align}

In the cases above, we assumed that $\sigma_L \ll \sigma$. If instead, we take $\sigma_L = \sigma$, then with $\alpha \gg \sigma^2$ (and $\alpha \gg \delta^2 \sigma^2)$, the fourth order processes can be approximated again using endpoint or stationary phase analysis, resulting in:
\begin{align}
\beta_4(10) 
&\approx  \frac{   i  16 \sqrt{3} m^3 G^2 M^3  \alpha^5 }{\pi \hbar^4 d^8}      e^{id^2/(2 \alpha)} e^{-5 \sigma^2 d^2/ (12 \alpha^2)},
\\ 
\beta_4(00) &\approx \frac{1024 \sqrt{2} i m^3 G^2 M^3}{3 \pi \hbar^4}  \frac{\alpha^5}{d^8}  e^{-i d^2 / (2 \alpha)} e^{-5 \sigma^2 d^2 / (8 \alpha^2)},\\
   \beta_4(11) &\approx \frac{1024 \sqrt{2} i m^3 G^2 M^3 }{ \pi \hbar^4}  \frac{\alpha^5}{d^8}  e^{i d^2 / (2 \alpha)} e^{-5 \sigma^2 d^2 / (8 \alpha^2)}, \\
   \beta_4(01) &\sim \frac{ i m^3 G^2 M^3}{ \sqrt{\pi} \hbar^4}  \frac{\alpha^2 \sigma^3 \delta^2}{d^5}  e^{-i d^2 / (2 \alpha)} e^{-3 \sigma^2 d^2 / (4 \alpha^2)}. 
\end{align}
Numerical analysis shows that these equations are good approximations to \eqref{eq:beta2BefStatPhase} in this regime. Note that the fourth order processes, particularly the $\beta_4(1,0)$ component, can be much less exponentially suppressed than the second order case \eqref{eq:beta1GaussianSame}. For example, with $\alpha = 0.1 \sqrt{5/12} \sigma d$, the exponential factor in $\beta_4(1,0)$ is 35 orders of magnitude larger than than the exponential factor in  $\beta_2$. With $\sigma_L = \sigma$ and $\alpha \gg \sigma^2$, but $\alpha \ll \delta^2 \sigma^2$, we alternatively obtain:  
\begin{align}
\beta_4(10) 
&\approx  \frac{   i  16 \sqrt{3} m^3 G^2 M^3  \alpha^7 }{\pi \hbar^4 d^8 \sigma^4 \delta^4}      e^{id^2/(2 \alpha)} e^{-5 \sigma^2 d^2/ (12 \alpha^2)}.
\end{align}

Future work will  consider relaxing $\Phi(\bm{x})$ as a static potential in the above calculations, consider the behaviour before the superpositions states (above, gravity is essentially turned on at $t = 0$ as in \cite{bose2017spin,marletto2017gravitationallyinduced}) and also relax the form of the experiment and  final state to allow for movement to other modes, such as higher momentum modes. With the assumption that the potential is static, the latter can be suppressed due to the final state being off-shell, but  continuous modes have been assumed.

Following \cite{bose2017spinSupp} that a phase of order $10^{-4}$ could be measurable and thus considered non-negligible, then in this way, the fourth order effect can be  an order of magnitude greater  than the second order case and also non-negligible, for example, with $\delta \sim 35 $ and times of a few seconds and masses $M \sim 10^{-4}\, \mathrm{kg}$, and $\delta \sim 1 $ and times of order $10\,\mathrm{s}$ and masses $M \sim 10^{-7}\, \mathrm{kg}$,   if densities of solids are theoretically assumed. Alternatively, $\delta \sim  350 $ can provide times of tens of seconds  with masses of order $10 \, \mathrm{g}$ and, for example, $\delta \sim 5$ can provide times of hundreds of seconds  and masses of order $10^{-7}\, \mathrm{kg}$, and  $\delta \sim 1$ provides masses of order $10^{-9}\, \mathrm{kg}$ with gas densities. Note that even with less demanding parameters, it will still be the case that at certain times entanglement in the state of interest is coming from the fourth order process and not the second order one since the first oscillates as $\sin(d^2/(2 \alpha))$ and is thus zero for times $\alpha = d^2 / (2  n \pi)$, where $n \in \mathbb{N}$. For early times, $\alpha \ll d^2$, the second-order amplitude is thus often zero.\footnote{Note that the $\sin(d^2 / 2 \alpha) $ scaling is not specific to Gaussian wavefunctions and occurs in general, for example, also for step function wavefunctions.}  In this case, for example, non-negligible entanglement for the state of interest would occur when $\delta \sim 10$, and a few seconds with $M \sim 10^{-6}\,\mathrm{kg}$ and  $\delta \sim 1$ and tens of seconds with  masses of order $M  \sim 10^{-9}\,\mathrm{kg}$ and solid densities, and for  $\delta \sim 15$ and  tens of  seconds with $M \sim 10^{-5}\,\mathrm{kg}$ and gas densities. 

Above, as well as taking $\sigma_L = \sigma$, for which we considered some example numbers, we also  considered that it could be the case that $\sigma_L \ll \sigma$. In theory-independent arguments for why classical gravity cannot create entanglement, the theory of gravity is of course not specified. Such arguments thus do not make any claims that we must have $\sigma_L = \sigma$ and cannot take $\sigma_L \ll \sigma$. Furthermore, since the way matter is sourced by gravity is not experimentally known for a fundamentally classical theory of gravity, it could be the case that the sourcing mechanism demands $\sigma_L \ll \sigma$ in such a theory. However, when taking a specific theory of classical gravity, such as semi-classical Einstein gravity, it is also possible for  $\sigma_L \ll \sigma$. For example, rather than, or in addition to, a mass distribution being placed in a superposition of positions, it could also be placed in a superposition of shapes. An example could be a cold atoms experiment where in one superposition branch the atoms are free and in a spherical configuration, whereas in another branch they are confined to a trap that squeezes them in the z-direction, such that the total mass and maximum density stay the same but the shapes are different in each branch. This would be very similar to a recently performed experiment \cite{dobkowski2025observationquantumfreefall} where a wavepacket is  placed in a superposition of spin states with one state sensitive to an external magnetic field, and the other not, such that one superposition involves the wavepacket in free fall and the other trapped. In our case, the magnetic field could also be used to  squeeze the trapped wavepacket's width, and we would need macroscopic superpositions.  Taking $\sigma$ to be the width of the distribution in the z-direction for the  branch with the untrapped wavepacket, then the gravitational potential from these objects would have $\sigma_L = \sigma$. But according to semi-classical Eisenstein gravity, this branch would also feel the gravitational potential from the other branch where the distribution is squeezed. In the squeezed branch, the potential will have $\sigma_L \ll \sigma$ and thus the free branch will be affected by potentials with  $\sigma_L \ll \sigma$. This then results in the first order amplitudes \eqref{eq:beta1SmallTimes} and \eqref{eq:beta1Gaussian},  and second order amplitudes \eqref{eq:beta210} and \eqref{eq:beta210Long} depending on whether we work at times for which free evolution can be ignored ($\alpha \ll \sigma^2$) or not ($\alpha \gg \sigma^2$). In this way it is easier for the considered fourth order effect to be much greater than the second order case. For example, when $\alpha \ll \sigma^2$, a significant second-order amplitude can be generated for times under $1\,\mathrm{s}$ and masses $M = 0.1\,\mathrm{mg}$ if $\delta \sim 1000$, or $M= 10^{-8}\,\mathrm{kg}$ after a few seconds if $\delta \sim 5 $, assuming trapped solid densities; and seconds  
 with $M = 0.1 \, \mathrm{mg}$ masses if $\delta$ is of order $100$, or with $M = 10^{-9}\,\mathrm{kg}$ after tens of seconds if $\delta \sim 1$, assuming that $\sigma/\sigma_L \sim 20 - 30$ and gas densities for both the trapped and untrapped distributions. If instead $\alpha \gg \sigma^2$, we can have significant entanglement in the state of interest for times  lower than $1\,\mathrm{s}$ with $M \sim 0.1{mg}$, a few seconds with $M \sim 10^{-8}\,\mathrm{kg}$  and order ten seconds with $M \sim 10^{-10}\,\mathrm{kg}$, when assuming trapped solid densities; and, for example, a few seconds with $M = 10^{-7}\,\mathrm{kg}$ and $\delta \sim 10$, or $M = 10^{-9} \,\mathrm{kg}$ after tens of seconds if we have gas densities and take $\sigma/\sigma_L \sim 5 - 7$.

Note that with just superpositions of radii, as long as $d \gg R$, the standard quantum gravity effect, considered in Section \ref{sec:QG}, will be utterly negligible. In fact, if we re-introduced spherical unit-step functions  but the radii of the trapped spherical distribution is reduced such that  we still have $\sigma_L \ll \sigma$, then the quantum gravity effect would be identically zero since there is no longer a superposition of positions, only radii, which will contribute the same phase. This is then an interesting situation where there would be identically zero entanglement from quantum gravity but significant entanglement in the state of interest due to classical gravity.  

\subsection{Additional second-order processes} \label{app:AdditionalCG}

In addition to the second-order process that is the classical analogue of the process considered in Section \ref{sec:QG} for entanglement in quantum gravity, there is also the classical analogue of the quantum gravity process Fig.\  \ref{fig:CrissCross}\textcolor{blue}{a}, which is provided in Fig.\  \ref{fig:CrissCross}\textcolor{blue}{b}. This was found to be vanishing in the quantum gravity section as discussed in Appendix \ref{app:MoreQG}, and for the same reasons (no overlap of the objects' wavefunctions), is also vanishing in classical gravity. 

Also at second order there are  contractions involving virtual matter propagators within each object (not between the objects):
\begin{align}\nonumber
         &\int_t d^4 x \, d^4  y \, \,_{1i}\langle \wick{\c1 N| \, _{2j}{\langle}  N|   \Phi(\bm{y})  \Phi(\bm{x})  \hat{\mathcal{T}}[\c1{\hat{\phi}}^{\dagger}(y) \c2{\hat{\phi}} (y)]
      \hat{\mathcal{T}}[\c2{\hat{\phi}}^{\dagger} (x) \c3{\hat{\phi}} (x) ] |\c3 N \rangle_{1i} \,|  N} \rangle_{2j}\\ \label{eq:contVirtWithin}
      &+\int_t d^4 x \, d^4  y \, \,_{1i}\langle \wick{ N| \, _{2j}{\langle} \c1 N|   \Phi(\bm{y})  \Phi(\bm{x})  \hat{\mathcal{T}}[\c1{\hat{\phi}}^{\dagger}(y) \c2{\hat{\phi}} (y)]
      \hat{\mathcal{T}}[\c2{\hat{\phi}}^{\dagger} (x) \c3{\hat{\phi}} (x) ] | N \rangle_{1i} \,| \c3 N} \rangle_{2j} =: \delta_{1j}+\delta_{2j}.
\end{align}
These also just contribute a local relative phase and thus no entanglement: for the different superposition branches $ij$, the overall amplitude due to this process up to second order is $\alpha^{(2)}_{ij} \approx 1 + \delta_{1i} + \delta_{2j}$, which is just that expected from a product state (up to order second order). That is:
\begin{align}
    |\psi\rangle  &\propto \alpha^{(2)}_{LL} |N\rangle_{1L} |N\rangle_{2L} + \alpha^{(2)}_{LR} |N\rangle_{1L} |N\rangle_{2R} +  \alpha^{(2)}_{RL} |N\rangle_{1R} |N\rangle_{2L} + \alpha^{(2)}_{RR} |N\rangle_{1R} |N\rangle_{2R}\\    
    &\propto (1 + \delta_{1L}+\delta_{2L})|N\rangle_{1L} |N\rangle_{2L}+ (1 + \delta_{1L}+\delta_{2R})|N\rangle_{1L} |N\rangle_{2R} \\&+  (1 + \delta_{1R}+\delta_{2L})|N\rangle_{1R} |N\rangle_{2L} + (1 + \delta_{1R}+\delta_{2R})|N\rangle_{1R} |N\rangle_{2R},
\end{align}
which we can write (to order $\delta_{\kappa i})$ as
\begin{align}
    |\psi\rangle \propto \left[(1 + \delta_{1L})|N\rangle_{1L} + (1 + \delta_{1R} )|N\rangle_{1R} \right] \otimes \left[(1 + \delta_{2L})|N\rangle_{2L} + (1 + \delta_{2R} |N\rangle_{2R}) \right],
\end{align}
where the amplitudes $\delta_{\kappa i}$ collect an $i$ from the virtual matter propagator  so that they are just first order contributions of $e^{i\delta_{\kappa i}}$. From geometry, $\delta_{1L} = \delta_{2R}$ and $\delta_{1R} = \delta_{2L}$, so this process just contributes 
\begin{align} \label{eq:relativePhase2}
    |\psi\rangle \propto \left( |N\rangle_{1L} + e^{i \theta} |N\rangle_{1R} \right) \otimes \left(|N\rangle_{2L} + e^{-i \theta} |N\rangle_{2R} \right),
\end{align}
where $i \theta := \delta_{2L} - \delta_{2R}$. This process then, by itself, does not contribute towards entanglement.

As discussed in the main text, if we took final states of the form $|N+k\rangle_{1i} \, |N-k\rangle_{2j}$ to be seen in the experiment, with $k$ an integer, then there would also be second order processes  contributing a relative phase that would combine with the processes at fourth order.

\subsection{Virtual matter in quantum gravity} \label{sec:QGVirtMatter}

Although in the main text  we have considered virtual matter processes from a classical gravity perspective, virtual matter processes will also generally exist  in a quantum theory of gravity. In this case,  the gravitational potential as well as the virtual matter will be in a quantum superposition. That is, Equation \eqref{eq:beta4ijFull} for $\beta^{(4)}_{ij}$ becomes:
\begin{align} \label{eq:alpha4}
    \kappa^{(4)}_{ij} := \frac{m^6 t^2 N^2 }{4 \pi^2 \hbar^6  } \left(i \int d^3 \bm{x} \int d^3 \bm{y} \frac{\Phi_{ij}(\bm{x}) \, \Phi_{ij}(\bm{y}) \, \theta_{1i}(\bm{x}) \, \theta_{2j}(\bm{y})}{|\bm{x} - \bm{y}|}\right)^2,
\end{align}
where:
\begin{align} \label{eq:PhiPotentialsQG}
    \Phi_{ij}(\bm{x}) := \Phi_{1i}(\bm{x}) + \Phi_{2j}(\bm{x}),
    \end{align}
with:
    \begin{align} \label{eq:PhiQG}
    \Phi_{\kappa i}(\bm{x}) := - G M \bigg[ \left(\frac{3}{2 R} - \frac{|\bm{x} - \bm{X}_{\kappa i}|^2}{2R^3}\right) \theta(R - |\bm{x}- \bm{X}_{\kappa i}|) + \frac{\theta(|\bm{x}- \bm{X}_{\kappa i}| - R)}{|\bm{x}- \bm{X}_{\kappa i}|} \bigg].
\end{align}
This is straightforwardly derived from the quantum gravity Hamiltonian in the full Newtonian regime - Equation \eqref{eq:HIonRel} with a hat added to $\Phi(x)$ to indicate that it can be superposed:
\begin{align} \label{eq:QGNewtLimit}
  \hat{H}_{I}  &= \frac{4}{c^2} \int d^3 \bm{x}  \, \hat{\Phi}(\bm{x})  \Big(   \hat{\pi}(x) \hat{\pi}^{\dagger}(x)   - \frac{m^2 c^2}{2\hbar^2} \hat{\phi}^{\dagger}(x) \hat{\phi}(x)\Big),
\end{align}
and with $\hat{\Phi}(\bm{x})$ now  written as:
\begin{align}
    \hat{\Phi}(\bm{x}) = -\frac{G}{c^2} \int d^3 \bm{y}\, \frac{\hat{T}_{00}(\bm{y})}{|\bm{x} - \bm{y}|}.
\end{align}
Using \eqref{eq:WickN}, we then have $\, _{1i}{\langle} N| \hat{\Phi} (\bm{x}) |N\rangle_{1i} = \Phi_{1i}(\bm{x}) \, _{1i}{\langle} N| N\rangle_{1i} = \Phi_{1i}(\bm{x})$ and $\, _{1i}{\langle} N| \, _{2j}{\langle} N|\hat{\Phi} (\bm{x}) |N\rangle_{1i} |N\rangle_{2j} =  \Phi_{1i}(\bm{x}) + \Phi_{2j} (\bm{x})$  in the approximation $R \gg \hbar / (mc)$ used to describe the experiment (and since the spheres do not overlap). That is, $\hat{\Phi} (\bm{x}) $ directly acts on the Hilbert space of matter \cite{christodoulou2022gravity}.

The relevant process then derives at fourth order as in the classical gravity case, with the amplitude $\kappa^{(4)}_{ij}$ deriving from:
\begin{align}\nonumber
        &\kappa^{(4)}_{ij}  =  \frac{1}{16 \, \hbar^4 c^4}   \int_t d^4 x \, d^4  y \,   d^4 z\,  d^4 w\,     \times \\ \nonumber &\,_{1i}\langle \wick{\c1 N| \, _{2j}{\langle} \c2 N| \hat{\Phi}(\bm{w})  
 \hat{\Phi}(\bm{z})  \hat{\Phi}(\bm{y})  \hat{\Phi}(\bm{x}) \hat{\mathcal{T}}_{\mu \nu}[\c2{\hat{\phi}}^{\dagger} (w) \c3{\hat{\phi}} (w)]      \hat{\mathcal{T}}_{\rho \sigma}[\c1{\hat{\phi}}^{\dagger} (z) \c4{\hat{\phi}} (z)] \hat{\mathcal{T}}_{\gamma \delta}[\c3{\hat{\phi}}^{\dagger}(y) \c5{\hat{\phi}} (y)]
      \hat{\mathcal{T}}_{\kappa \lambda}[\c4{\hat{\phi}}^{\dagger} (x) \c6{\hat{\phi}} (x) ] |\c5 N \rangle_{1i} \,| \c6 N} \rangle_{2j},
\end{align}
where one could also act symbolic Wick contractions between $ \hat{\Phi}(\bm{x})$ and the matter states, noting that $ \hat{\Phi}(\bm{x})$ leaves the states intact. Alternatively, the process can also be derived at higher order from the full relativistic Hamiltonian \eqref{eq:HIRel}, where there are virtual gravitons as well as virtual matter mediating between the masses. That is, although we can  have only virtual matter exchange in classical gravity, in quantum gravity this exchange is always accompanied by graviton exchange in the considered experiment, such that you could not strictly separate the two effects.

With \eqref{eq:PhiPotentialsQG} inserted into \eqref{eq:alpha4}, the spatial integrals can be solved using the same method as in Section \ref{sec:CG}. In the approximation that $R \ll \Delta x$ and $R \ll d_{ij}$, which were also assumed in Section \ref{sec:CG}, we find:
\begin{align}    
\kappa^{(4)}_{ij} = \left(\frac{24}{25} \frac{i G^2 m^2 M^3 R t }{\hbar^3 d_{ij}}\right)^2,
\end{align}
which has the same form as the semi-classical Einstein case \eqref{eq:betaCG} except for a slightly larger numerical factor due to the $1/2$ coming from the average of the potentials \eqref{eq:semiclassicalPhi} compared to \eqref{eq:PhiPotentialsQG}. The higher factor can also be considered as due to there being  a superposition of gravitons \emph{and} virtual matter, with both contributing to the entanglement in quantum gravity, whereas it is only the latter in a classical theory of gravity. With $d_{RL} \ll \Delta x$, the amplitude $\kappa^{(4)}_{RL}$ dominates over all others, with:
\begin{align}  \label{eq:alpha4RL}  
\kappa^{(4)}_{RL} = \left(\frac{24 }{25} \frac{i G^2 m^2 M^3 R t }{\hbar^3 d_{RL}}\right)^2.
\end{align}

\subsection{Further discussion on the processes} \label{sec:CGDiscussion}

It could be  thought that since there are no virtual graviton propagators, it is not the gravitational interaction that is creating the  entanglement considered in Section \ref{sec:CG}. We can see why it is the gravitational interaction that is creating the entanglement from the effective Feynman diagram for the process, Fig. \ref{fig:Fig2}\textcolor{blue}{e}: with $\hat{H}^{CG}_{int}$ turned off (or if we had a constant potential $\Phi(\bm{x})$), there would be no vertices, and the only way to create entanglement would be if a real particle from $1i$ freely propagates (diffuses) to $2j$. However, we have assumed  fixed position-like modes for the experiment - see Equations \eqref{eq:Nstate} and \eqref{eq:globalphase} - such that this contribution is vanishingly small, leaving approximately just  a global phase as discussed around \eqref{eq:globalphase}. With the interaction turned on, vertices can be reintroduced which provide additional momentum  to flow from gravity (as understood from the Fourier transform of $\Phi(\bm{x})$, with $\Phi(\bm{x})$ spatially varying as in \eqref{eq:semiclassicalPhi} - see Section \ref{sec:CG}). If we just kept one vertex, say where $1i$ connects to the potential, then we have  Figure \ref{fig:CrissCross}\textcolor{blue}{b}, which could still be thought of as involving diffusion of a real particle from $1i$ to $2j$. However, this diagram evaluates to zero as discussed around Equation \eqref{eq:zeroDiff}. One way to see this is  that the process is forbidden from energy-momentum conservation: while additional momentum  flows into the vertex  from gravity, it is not then possible to respect the energy-momentum relation for a real particle with zero momentum to propagate out of (as well as into) the vertex \cite{fn9}. This illustrates that the particle must be off-shell, it must be virtual, and, therefore,  there must be another vertex at $2j$. That is, only the diagram Fig.\ \ref{fig:Fig2}\textcolor{blue}{e} with interaction vertices and virtual (not real) particles propagating between the objects can create entanglement given the initial and final states, further illustrating that  $\hat{H}^{CG}_{int}$ is essential to the process - without it, there would be no observable entanglement. We see then that the classical gravity interaction is  responsible for creating the observed entanglement and it is necessary that there is a virtual particle that is  off shell. Just as the leading order process for quantum gravity can be described as atomic particles being created and annihilated with virtual gravitons between them (Section \ref{sec:QG}), similarly we see that there are also classical gravity effects that can be described as atomic particles being annihilated and created but with virtual matter particles between them.

Note that, in the non-relativistic gravity limit, the interaction Hamiltonian for classical gravity Equation \eqref{eq:HintNonRel} is not, by itself, a spatial entangling operator for the matter field. That is, since all the operators act on the same spatial position, and there are no spatial derivatives, the operator cannot by itself entangle different spatial regions of the field. Instead, it is the free Hamiltonian $\hat{H}_0$ that contains the spatial derivatives required, in principle, for spatial entanglement - although, as discussed above, $\hat{H}_0$ by itself is also not enough to create observable entanglement due to the experimental conditions. Therefore,  we need \emph{both} $\hat{H}_{int}$ and $\hat{H}_0$ to generate the observable entanglement. This can be seen from the fact that in the interaction picture $\hat{H}_{int}$ picks up time dependence through $\hat{H}_0$ resulting in the Dyson series, and equivalently from the Feynman diagrams: $\hat{H}_{int}$ provides the vertices and $\hat{H}_0$ the free propagation of the virtual particles in Fig.\ \ref{fig:Fig2}\textcolor{blue}{e}, with both effects required to generate the entanglement process as discussed above. This is also the case for the quantum gravity effect where the spatial matter field derivatives in the full relativistic interaction Hamiltonian \eqref{eq:HQG}  play no role in generating entanglement due to the assumed non-relativistic in and out matter states \eqref{eq:WickN} - see the discussion above \eqref{eq:alphaQG2}. This can be seen more clearly when taking the same simplifying non-relativistic limit assumed for the classical gravity interaction - the corresponding interaction Hamiltonian in this case for quantum gravity  is  \eqref{eq:HintNonRel} but with a hat added to $h(\bm{x})$ - see \eqref{eq:QGNewtLimit}.  This is also not a spatially entangling operator and, as with the classical gravity interaction, in order to generate the quantum-gravity entanglement, \emph{both} $\hat{H}_{int}$ and $\hat{H}_0$ (in this case $\hat{H}^G_0$) are required. As with classical gravity, we can see this from the relevant Feynman diagram (Fig.\ \ref{fig:Fig1}\textcolor{blue}{a} in the main text) where  $\hat{H}_{int}$ provides the vertices and $\hat{H}_{0}$ provides the free propagation of the virtual gravitons \cite{GWBook}.

In Section \ref{sec:CG}, we considered entangling processes via virtual matter processes of an effective quantum field for the particles of the matter objects. However, in theory, there will be further entangling processes via other virtual matter mechanisms associated with a classical gravity interaction that could also involve different quantum matter fields, including the electromagnetic field. Additionally, other entangling processes may occur once it is taken into account that the matter systems themselves will not just consist of pure excitations of the matter fields in a quantum field perspective.

\subsection{Process for infinite time}

In the main text we considered matter distributions that were formed of position-like wavepackets, and allowed them to interact  for a finite time.  Here, for comparison, we consider a theoretical case where we have general momentum wavepackets of matter interacting for infinite time. This is closer to typical calculations in QFT where particles are taken to be free and far apart in the asymptotic past and future, and  interact non-negligibly when they are close. 

We start from \eqref{eq:startingpoint} but do not take the Fourier transform of the potentials:
\begin{align}\nonumber
        \beta
        &= \frac{4 m^4 c}{\hbar^4}  \int_t  d^4  x\,  d^4 y  \, \frac{d^3 \bm{p'}}{(2 \pi)^3}\frac{d^3 \bm{p}}{(2 \pi)^3} \frac{d^4 q}{(2\pi)^4} \frac{1}{\sqrt{2\omega_{\bm{p'}}}}  \frac{1}{\sqrt{2\omega_{\bm{p}}}} f(\bm{p}) f^{' \ast}(\bm{p'}) \Phi(\bm{x}) \Phi(\bm{y}) \\ \label{eq:startingpoint}
        &\times  e^{ip.y} e^{-i p'.x} \frac{i}{q^2 + m^2 c^2/\hbar^2 + i \epsilon} e^{i q. (x - y)},
\end{align}
where $f(\bm{p})$ and $f(\bm{p})$ are momentum wavefunctions as defined in \eqref{eq:psi1}-\eqref{eq:psi2}. Here, we will not assume the Fourier transform of unit step functions and instead consider the corresponding wavepackets $|\psi\rangle$ and $|\psi'\rangle$ to be closer to momentum eigenstates.   We  take the Fourier transform of $f(\bm{p})$: $f(\bm{p}) = \int d^3 \bm{x} f(\bm{x}) e^{- i \bm{p}.\bm{x}}$, leaving us with:
\begin{align}\nonumber
        \beta
        &= \frac{4 m^4 c}{\hbar^4}  \int  d^4  x\,  d^4 y  \, \int d^3 \bm{z}\,d^3 \bm{w}  \int   \frac{d^3 \bm{p}}{(2 \pi)^3} \frac{d^3 \bm{p'}}{(2 \pi)^3}\frac{d^4 q}{(2\pi)^4} \frac{1}{\sqrt{2\omega_{\bm{p'}}}}  \frac{1}{\sqrt{2\omega_{\bm{p}}}}  \\
        &\times e^{- ip^0 x^0} e^{i p^{0'} y^0} e^{i\bm{p}.(\bm{x} - \bm{z})} e^{-i\bm{p}'.(\bm{y} - \bm{w})}  \frac{i}{q^2 + m^2 c^2/\hbar^2 + i \epsilon} e^{i q. (x - y)} f(\bm{z}) f^{' \ast}(\bm{w}) \Phi(\bm{x}) \Phi(\bm{y}).
\end{align}
We now integrate time from $-\infty$ to $+\infty$. Ignoring any time dependence on the potentials   $\Phi(x)$ due to the matter distributions now potentially significantly moving with time,  results in the delta functions $(2 \pi)^2 \delta(p^0 + q^0) \delta(p^{'0} + q^0)$. Integrating over $q^0$ and then using $p^0 = \sqrt{\bm{p}^2 + m^2 c^2 / \hbar^2}$, leaves us with:
\begin{align}\nonumber
        \beta
        &= \frac{4 m^4 c}{\hbar^4}  \int  d^3  \bm{x}\,  d^3 \bm{y}  \, \int d^3 \bm{z}\,d^3 \bm{w}  \int   \frac{d^3 \bm{p}}{(2 \pi)^3} \frac{d^3 \bm{p'}}{(2 \pi)^3}\frac{d^3 \bm{q}}{(2\pi)^3} \frac{1}{\sqrt{2\omega_{\bm{p'}}}}  \frac{1}{\sqrt{2\omega_{\bm{p}}}}  (2 \pi)  \delta(-p^0 +p^{0'}) \\
        &\times e^{i\bm{p}.(\bm{x} - \bm{z})} e^{-i\bm{p}'.(\bm{y} - \bm{w})}  \frac{i}{-\bm{p}^2 + \bm{q}^2  + i \epsilon} e^{i \bm{q}. (\bm{x} - \bm{y})}  f(\bm{z}) f^{' \ast}(\bm{w}) \Phi(\bm{x}) \Phi(\bm{y}).
\end{align} 
We next integrate over $\bm{p}'$ using the delta function identity:
\begin{align}
    \delta(f(|\bm{p'}|)) = \sum_i \frac{\delta(|\bm{p}| - |\bm{p'}_i|)}{f'(|\bm{p'}|)},
\end{align}
where $|\bm{p}'|_i$ are the roots of $f(|\bm{p'}|) := \sqrt{\bm{p}^{'2} + m^2 c^2 / \hbar^2}) - p^0$. This results in:
\begin{align}\nonumber
        \beta
        &= \frac{8 m^4 }{\hbar^4} \frac{1}{(2 \pi)} \int  d^3  \bm{x}\,  d^3 \bm{y}  \, \int d^3 \bm{z}\,d^3 \bm{w}  \int   \frac{d^3 \bm{p}}{(2 \pi)^3} \frac{d^3 \bm{q}}{(2\pi)^3}    \\
        &\times e^{i\bm{p}.(\bm{x} - \bm{z})} \frac{\sin(|\bm{p}| |\bm{y} - \bm{w}|)}{|\bm{y} - \bm{w}| }  \frac{i}{-\bm{p}^2 + \bm{q}^2  + i \epsilon} e^{i \bm{q}. (\bm{x} - \bm{y})}  f(\bm{z}) f^{' \ast}(\bm{w}) \Phi(\bm{x}) \Phi(\bm{y}).
\end{align}
Integrating over   $\bm{q}$ and then the angular variables of $\bm{p}$ provides:
\begin{align}\nonumber
        \beta
        &= \frac{8 m^4 }{\hbar^4}  \frac{1}{2 \pi}\int  d^3  \bm{x}\,  d^3 \bm{y}  \, \int d^3 \bm{z}\,d^3 \bm{w}  \int   \frac{d |\bm{p}| }{(2 \pi)^3}    |\bm{p}|\, \frac{\sin(|\bm{p}||\bm{x} - \bm{z}|)}{|\bm{x} - \bm{z}|} \frac{\sin(|\bm{p}|  |\bm{y} - \bm{w}|)}{|\bm{y} - \bm{w}| }  \frac{e^{i |\bm{p}| |\bm{x} - \bm{y}|}}{|\bm{x} - \bm{y}|} \\&\times f(\bm{z}) f^{' \ast}(\bm{w}) \Phi(\bm{x}) \Phi(\bm{y}).
\end{align}
Finally, integrating over $|\bm{p}|$ with a regulator $\epsilon$ results in:
 \begin{align}\nonumber
        \beta
        &:= \frac{2 m^4}{\hbar^4}  \frac{1}{(2 \pi)^4} \int  d^3  \bm{x}\,  d^3 \bm{y}  \, d^3 \bm{z}\,d^3 \bm{w}  \frac{f(\bm{z}) f^{' \ast}(\bm{w}) \Phi(\bm{x}) \Phi(\bm{y})}{|\bm{x} - \bm{z}| |\bm{x} - \bm{y}| |\bm{y} - \bm{w}|}  \\\nonumber
        &\times \Big(\frac{1}{(|\bm{x} - \bm{y}| + |\bm{x} - \bm{z}| + |\bm{w} - \bm{y}| + i \epsilon)^2} - \frac{1}{(|\bm{x} - \bm{y}| + |\bm{x} - \bm{z}| - |\bm{w} - \bm{y}|+ i \epsilon)^2} \\ &+ \frac{1}{(|\bm{x} - \bm{y}| - |\bm{x} - \bm{z}| + |\bm{w} - \bm{y}|+ i \epsilon)^2} - \frac{1}{(|\bm{x} - \bm{y}| - |\bm{x} - \bm{z}| - |\bm{w} - \bm{y}|+ i \epsilon)^2}\Big).
\end{align}
As stated above, this result is not applicable to the situation considered in the main text since it assumes infinite time, which for initial position-like states would mean they always interact and eventually entirely spread out. In the main text we had a finite time since we assumed a non-relativistic approximation (a roughly finite spread in momentum) that picks up a resonance. In contrast, here, the infinite time means the energy spread of the wavepackets smears out  delta functions, giving a finite result. For relevant times, inputting the step functions for $f(\bm{x})$ and $f'(\bm{x})$ that were used in the main text (and thus also including a factor of $N$) would be expected to lead to an overestimate of the effect. For example, taking $\bm{z}$ and $\bm{w}$ to be relatively fixed, $\Phi \sim G M / R$, $\int d^3 x f(x) \sim \sqrt{V}$ and $|\bm{x} - \bm{y}| \sim d_{RL} \approx 10 R$, results in a value for $\beta^2$ that is several orders of magnitude greater than that calculated in the main text for Planck masses $M \approx 10^{-8}\,\mathrm{kg}$ and times at which $\beta^{(4)}_{RL}$ was found to be significant.

\subsection{Boundaries} \label{app:bound}

Here we consider an example of how classical gravity can in principle lead to entanglement through virtual matter processes even in the theoretical case where there are, unphysical, infinite hard-wall boundary conditions on all matter fields. For simplicity, we consider a massive complex scalar field in $1+1$ dimensions interacting through classical gravity. In order to use a perturbative treatment, we first consider finite potentials, which can later be extended to infinite potentials. The Hamiltonian is then $\hat{H} = \hat{H}_0 + \hat{H}_{int}$, where $\hat{H}_0$ is the free Hamiltonian of the complex scalar field, and  $\hat{H}_{int}$ is:
\begin{align}
    \hat{H}_{int} &= 2 \frac{m}{\hbar} \int d \boldsymbol{x}\,   [\frac{m}{\hbar} \Phi (x) + \lambda c ( \,\delta(\bm{x} - \bm{X}_{1 i}) + \delta(\bm{x}-\bm{X}_{2 j})\, )] \, \hat{\phi}^{\dagger}(\bm{x}) \hat{\phi}(\bm{x})\\
    &= 2 \frac{m^2}{\hbar^2} \int d \boldsymbol{x}\,   \Phi (x) \hat{\phi}^{\dagger}(\bm{x}) \hat{\phi}(\bm{x})  + \lambda \frac{m c}{\hbar} 
    \hat{\phi}^{\dagger}_{1 i} \hat{\phi}_{1 i} + \lambda \frac{m c}{\hbar} 
    \hat{\phi}^{\dagger}_{2j} \hat{\phi}_{2j},
\end{align}
where $\hat{\phi}_{\kappa i} := \hat{\phi}(\bm{X}_{\kappa i})$. Here, for simplicity, we have taken a gauge where $h_{\mu \nu} = 2 \Phi \eta_{\mu \nu}$, with $\mu,\nu =\{0,1\}$ and the metric signature is $(-,+)$. The vectors $\bm{X}_{\kappa i}$ set the location of the potentials and the coupling $\lambda$ is initially taken to be small, $\lambda \ll 1$, so that a perturbative treatment can be applied. With $\lambda \rightarrow \infty$, infinite potential walls are imposed on the field, leading to Dirichlet boundary conditions. 

We consider the vacuum persistence amplitude $Z(t):= \langle 0 | \hat{U}_I(t) | 0\rangle$. With a perturbative treatment, we have:
\begin{align}
    Z(t) = 1+ Z^{(1)}(t) + Z^{(2)}(t) + \cdots,
\end{align}
where 
\begin{align}
    Z^{(n)}(t) :=  \frac{(-i)^{n}}{n!\, \hbar^n  c^n} 
    \langle 0 | \hat{T}  \int^{c t}_{0} dx^0_1 \, dx^0_2\, \ldots d x^0_n \hat{H}_{I} (x^0_1)\, \hat{H}_{I} (x^0_2) \cdots  \hat{H}_{I} (x^0_n) |0 \rangle,
\end{align}
with $\hat{H}_I(t)$ the interaction picture version of $\hat{H}_{int}$. From $Z(t)$, for example, the energy of the interacting vacuum can be extracted \cite{Peskin:1995ev}. We consider the case that the finite walls are in a superposition of locations with the matter objects, which are of size $2R$ and behind the walls, although just allowing for the idea that the boundaries obey quantum theory  would be sufficient. The vacuum persistence amplitude will then also be in a superposition, and thus generating entanglement in principle. 

To start with, we take $\bm{X}_{1i} = 0$ for simplicity and define $\bm{X}_{2j} = \bm{d}_{ij}$. Within the non-relativistic approximation used in the main text, $R \gg \hbar / (mc)$, up to fourth order, no entanglement is generated since  the vacuum persistence amplitudes scale as $e^{-2 m c d_{ij} /\hbar }$, and $d \geq 2R$. We can think of this as the virtual particles of the scalar field on average not having enough energy-momentum to  propagate between the walls. However, similar to the main text, processes occur at fourth order that  can  in principle generate entanglement. For example, consider the following process:
\begin{align}
    b^{(4)}_{ij} := \frac{4 m^6 \lambda^2}{\hbar^{10} c^2} \wick{\int_t d^2 z \,  d^2  w \,_{1i} \langle  0| \c1{\hat{\phi}}_{1i}\, \c2{\hat{\phi}}_{2j} \,   [\Phi(z)\,  \c1{\hat{\phi}}^{\dagger}(z) \c3{\hat{\phi}}(z)\,\Phi(w)\,  \c2{\hat{\phi}}^{\dagger}(w) \c4{\hat{\phi}}(w)]\,      \c4{\hat{\phi}}^{\dagger}_{1i}
\c3{\hat{\phi}}^{\dagger}_{2j} |0\rangle},
\end{align}
where $\int_t d^2 x := \int^t_0 dt \, \int d\bm{x}$. Here, we are considering how virtual particles from the classical gravitational interaction are propagating and interacting with the walls \cite{fn8}. The amplitude  is evaluated as follows:
\begin{align}\nonumber
    b^{(4)}_{ij} &=  \frac{4 m^6 \lambda^2}{\hbar^{10} c^2} \int_t d^2 x \,  d^2  y\, d^2 z \,  d^2  w\, [\delta(\bm{x}) \delta(\bm{y} - \bm{d}_{ij}) + \delta(\bm{y}) \delta(\bm{x} - \bm{d}_{ij})]\\
    &\times \Phi(z)\,\Phi(w)\, D_F(x-z)\, D_F(z-y)\, D_F(w-y)\, D_F(w-x),
\end{align}
where:
\begin{align}
D_F(x-y) := c \hbar \int \frac{d^2 q}{(2 \pi)^2} \frac{e^{iq.(x-y)}}{q^2 + m^2 c^2 / \hbar^2 + i\epsilon},
\end{align}
such that we have:
\begin{align}\nonumber
    b^{(4)}_{ij} =  \frac{4 m^6 c^2 \lambda^2}{\hbar^{6} } &\int_t d^2 x \,  d^2  y\, d^2 z \,  d^2  w\,  \frac{d s}{2\pi} \,  \frac{d u}{2 \pi} \, \Phi(s,\bm{z}) \, \Phi(u,\bm{w})\, [\delta(\bm{x}) \delta(\bm{y} - \bm{d}_{ij}) + \delta(\bm{y}) \delta(\bm{x} - \bm{d}_{ij})] e^{-i s z^0} e^{-i u w^0}\\\nonumber
    &\times \, 
\int \frac{d^2 p}{(2 \pi)^2} \frac{e^{ip.(x-z)}}{p^2 + m^2 c^2 / \hbar^2 + i\epsilon} \int \frac{d^2 q}{(2 \pi)^2} \frac{e^{iq.(z-y)}}{q^2 + m^2 c^2 / \hbar^2 + i\epsilon}  \\&\times\int\frac{d^2 k}{(2 \pi)^2} \frac{e^{ik.(w-y)}}{k^2 + m^2 c^2 / \hbar^2 + i\epsilon} \int \frac{d^2 l}{(2 \pi)^2} \frac{e^{il.(w-x)}}{l^2 + m^2 c^2 / \hbar^2 + i\epsilon},
\end{align}
where we have now taken $\Phi(x)$ to have a time dependence. This could be due to the matter objects moving through the experiment, oscillating in space in traps, or, theoretically, we could imagine  the mass of the objects changing with time. For convenience, we also choose $\Phi(x)$ to be defined only within the region $\bm{x}=0$ to $\bm{x}=d_{ij}$. Assuming a long time limit then allows the amplitude to be written as:
\begin{align}\nonumber
    I = \frac{8 m^4  \lambda^2}{\hbar^4} &\int^{d_{ij}}_0 d z\, d w\, \int \frac{ds}{2 \pi}\, \frac{d p^0}{2 \pi}\,  \Phi(z,s)\,\Phi(w,-s)\,  \frac{e^{- m c \, (w + z) \sqrt{f(p^0)}  / \hbar }}{ f(p^0)}  \frac{e^{- (2 d_{ij} - (w + z)) m c \sqrt{f(p^0+s)} / \hbar }}{ f(p^0+s)},
\end{align}
where:
\begin{align}
    f(p^0) &:= \sqrt{-(\hbar p^0 / (m c))^2 + 1  + i \epsilon}.
\end{align}
Assuming $d_{ij} \gg \hbar / (m c)$, for support for values of $s$ that are sufficiently large, the  integral does not necessarily exponentially decay with $d_{ij}$ to leading order. That is, in comparison to the case where $\Phi(\bm{x}) = 0$ and lower orders of perturbation theory, there can  now in principle be no exponential suppression of order $e^{ - 2 m c d_{ij} / \hbar}$ for entanglement, which is vanishingly small in the non-relativistic approximation $R \gg \hbar / ( m c)$ as discussed above. This can be viewed as virtual matter of the classical gravity interaction having enough energy-momentum to propagate between the walls.  Although a theoretical example, this process, which has connections to the dynamical Casimir effect, is another  illustration of how the classical gravity interaction can in principle lead to entanglement. It also illustrates that,  in theory at least, it is  not enough to just isolate the quantum fields of the matter objects themselves to prevent entanglement via the classical gravity interaction.

\subsection{Stochastic classical gravity} \label{sec:fundDec}

As argued in the main text, the existence of a classical gravitational interaction implies, when quantum matter is treated using QFT, the possibility of having a matter propagator, which  can generate entanglement regardless of the specific form of the classical gravity model. The virtual matter process considered in Section \ref{sec:CG}  is thus expected to exist in principle in  any   classical gravity theory. However, the size of the effect  will depend on how exactly gravity is sourced by quantum matter.  In  Section \ref{sec:CG}, we considered semi-classical Einstein gravity  where  gravity is sourced from the quantum expectation of the stress-energy operator of matter: $G_{\mu \nu} = \kappa \langle  \hat{T}_{\mu \nu}\rangle$, with $\kappa = 8 \pi G / c^4$, $G_{\mu \nu}$  the Einstein tensor, and the average depends on the particular theory chosen. For example, a straightforward Everettian interpretation where the expectation is over the global wavefunction is ruled out through experiment  \cite{PageExp}, but other theories, such as that with Copenhagen-like collapse, or with the average over the `local' matter states, have not been ruled out - see Appendix  \ref{sec:consistent} for more detail.

In this section, we consider theories of classical gravity that involve fundamental stochacisty. For example, classical gravity could be  sourced from stochastic fluctuations around  the quantum expectation of the stress-energy operator of matter \cite{DIOSI1987377,TilloyDiosi}: $G_{\mu \nu} = \kappa \langle  \hat{T}_{\mu \nu}\rangle + \delta T_{\mu \nu}$, where $\delta T_{\mu \nu}$ is a stochastic quantity. In the non-relativistic limit this becomes:
\begin{align} \label{eq:PhiClassical}
\Phi(\bm{x}) = - \frac{G}{c^2} \int d^3 \bm{x'}\, \frac{\langle \hat{T}_{00}(\bm{x'})\rangle + \delta T_{00} (\bm{x'})}{|\bm{x}-\bm{x'}|}. 
\end{align}
Theories based on this equation have been developed, with $T_{00}$ formally deriving from a continuous
measurement process \cite{TilloyDiosi}. A relativistic theory of stochastic gravity also reproduces this result in the  Newtonian limit \cite{oppenheim2018postquantum,layton2023weak}. 

We now consider the dynamics of these theories. Just before taking the full Newtonian limit of the relativistic theory of stochastic gravity \cite{oppenheim2018postquantum}, the dynamics  obey the following coupled equations when averaging over classical noise \cite{layton2023weak,Bassi_2017}:
\begin{align}
    \frac{d \Phi(t,\bm{x})}{dt} &= -\frac{1}{12} \partial_i n^i\\
    \frac{d \pi(t,\bm{x})}{dt} &= \frac{\nabla^2 \Phi(t,\bm{x}) }{4 \pi G} - \langle \hat{m} (\bm{x})\rangle\\
\hbar\frac{d \hat{\rho}(t)}{dt} &= - i [\hat{H}_0  + \hat{H}_{int}] + \frac{1}{2} \int d^3\bm{x} d^3 \bm{y} D(\Phi,\bm{x},\bm{y}) [ \hat{m}(\bm{x}), [\hat{\rho}(t), \hat{m}(\bm{y})]],    
\end{align}
where $\hat{\rho}(t)$ is the density operator for matter, $\hat{m}(\bm{x})$ is the mass density operator, $\pi$ is the conjugate momentum of the dynamical classical field $\Phi$, $n^i$ is the shift vector of the ADM decomposition \cite{layton2023weak}, $\hat{H}_0$ describes the free evolution, $D(\Phi,\bm{x},\bm{y})$ is a positive semi-definite kernel,  and $\hat{H}_{int}$ is the non-relativistic version of \eqref{eq:HCQ}:
\begin{align}
    \hat{H}_{int} = \int d^3 \bm{x} \,\hat{m}(\bm{x}) \Phi(\bm{x}).
\end{align}
The above coupled equations take the form of semi-classical Einstein gravity in the non-relativistic regime (see Section \ref{sec:CG}) but with a decoherence term.

For the full Newtonian limit, one must apply the constraint $\pi \approx 0$ \cite{oppenheim2018postquantum}. This results in the same dynamics for the potential and density operator as those derived using a formal continuos measurement process \cite{TilloyDiosi}. However, there is much freedom in these theories.  Taking the perspective of a formal continuous measurement process, this freedom comes from a free choice in the spatial resolution of a single detector and  the correlations of the outputs of the detectors \cite{TilloyDiosi}. However, by applying a principle of minimal decoherence \cite{LeastDecoh}, the latter freedom can be fixed, and the dynamical equation for $\hat{\rho}(t)$, after averaging over noise, becomes \cite{LeastDecoh}:
\begin{align}\label{eq:DP}
    \hbar\frac{d \hat{\rho}(t)}{dt } = - i \left[\hat{H}_0 + \frac{1}{2} \int d^3\bm{x}\, \hat{\Phi}(\bm{x}) \hat{m}(\bm{x}),\hat{\rho}(t)\right] + \frac{1}{2} \int d^3 \bm{x} \left[\hat{\Phi}(\bm{x}), [ \hat{m}(\bm{x}),\hat{\rho}(t)]\right],
\end{align}
with $\hat{\Phi}(\bm{x}) := - G \int d^3 \bm{y}\, \hat{m}(\bm{y}) / |\bm{x} - \bm{y}|$. 

This is the Di\'{o}si-Penrose model \cite{diosi1989models,DIOSI1987377,penrose1996gravity} but with a Newtonian quantum gravity unitary term - the first term on the right-hand side of \eqref{eq:DP}. Remarkably, although the gravitational potential is considered classical, see \eqref{eq:PhiClassical}, mathematically the evolution of the density operator is equivalent to there being (Newtonian) quantum gravity (where the potential is effectively operator-valued, $\hat{\Phi}(\bm{x})$) but with sufficient decoherence to prevent quantum communication through the gravitational potential.  There is still freedom in this theory due to the need to regularize the theory in order to keep it finite (for example, due to the infinite potential of point-like particles), which can be formally interpreted as describing a quantum system subjected to a continuous monitoring of its (smeared) mass density \cite{TilloyDiosi}.  It is possible that this can lead to non-locality sufficient to create entanglement \cite{jzht-fbwt,trillo2024di}. To avoid this, we instead  use smeared density functions with the step functions $\theta_{\kappa i}(\bm{x})$, as in the main text and Sections \ref{sec:QG} and \ref{sec:CG}, such that  there is no overlap in the density functions, avoiding any potential non-local effects.

We now apply the methodologies of Sections \ref{sec:QG} and \ref{sec:CG} to these theories in order to estimate the size of the entanglement effect through the virtual matter process considered in the main text. For this, we first upgrade \eqref{eq:DP} to relativistic complex scalar matter fields. This results in replacing $\hat{m}(\bm{x})$ with $\hat{\Pi}(\bm{x}) = \frac{4}{c^2} \left(\hat{\pi}(\bm{x}) \hat{\pi}^{\dagger}(\bm{x}) - \frac{m^2 c^2}{2 \hbar^2} \hat{\phi}^{\dagger}(\bm{x}) \hat{\phi}(\bm{x})\right) := \mathcal{\hat{T}}[\hat{\phi}^{\dagger}(\bm{x}) \hat{\phi}(\bm{x})] $ (see Section \ref{sec:CG}), such that we have: 
\begin{align}\label{eq:DPrel}
   \hbar \frac{d \hat{\rho}(t)}{dt } = - i \left[\hat{H}_0 + \frac{1}{2} \int d^3\bm{x}\, \hat{\Phi}(\bm{x}) \hat{\Pi}(\bm{x}),\hat{\rho}(t)\right] + \frac{1}{2} \int d^3 \bm{x} \left[\hat{\Phi}(\bm{x}), [ \hat{\Pi}(\bm{x}),\hat{\rho}(t)]\right].
\end{align}
Taking the non-relativistic limit $\hat{\phi}(\bm{x}) = \frac{\hbar}{\sqrt{2m}} \hat{\psi}(\bm{x}) e^{-i m c x^0 / \hbar}$, where $\hat{\psi}(\bm{x})$ is a non-relativistic quantum field,  results in \eqref{eq:DP} with $\hat{m}(\bm{x}) = m\, \hat{\psi}^{\dagger}(\bm{x}) \hat{\psi}(\bm{x})$. We next switch to the interaction picture: we insert $\hat{\rho}(t) = \hat{U}_0 \hat{\rho}_I(t) \hat{U}^{\dagger}_0$, where $\hat{U}_0 = e^{-i \hat{H}_0 t / \hbar}$, resulting in the removal of the free dynamics:
\begin{align}\label{eq:DP2}
    \frac{d \hat{\rho}_I(t)}{dt } &= -  \frac{1}{2 \hbar} i\int d^3\bm{x} \left[ \hat{\Phi}(x) \hat{\Pi}(x),\hat{\rho}_I(t)\right] + \frac{1}{2 \hbar} \int d^3 \bm{x} \left[\hat{\Phi}(x), [ \hat{\Pi}(x),\hat{\rho}_I(t)]\right]\\
    &:= \left(\hat{\mathcal{L}}_U +   \hat{\mathcal{L}}_D\right) \hat{\rho}_I(t)\\ \label{eq:Ldrhodt}
    &:= \hat{\mathcal{L}} \hat{\rho}_I(t),
\end{align}
where $\hat{\Pi}(x)$ and  $\hat{\Phi}(x)$ are the interaction picture versions of $\hat{\Pi}(\bm{x})$  and $\hat{\Phi}(\bm{x})$, and $ \hat{\mathcal{L}} \hat{\rho}:= \left(\hat{\mathcal{L}}_U + \hat{\mathcal{L}}_D\right) \hat{\rho}$, with $\hat{\mathcal{L}}_U \hat{\rho}$ and $\hat{\mathcal{L}}_D \hat{\rho}$ the superoperators:
\begin{align}
    \hat{\mathcal{L}}_U \hat{\rho}&:=-  \frac{1}{2 \hbar} i\int d^3\bm{x} \left[ \hat{\Phi}(x) \hat{\Pi}(x),\hat{\rho}\right],  \\
    \hat{\mathcal{L}}_D \hat{\rho}&:=\frac{1}{2\hbar} \int d^3 \bm{x} \left[\hat{\Phi}(x), [ \hat{\Pi}(x),\hat{\rho}]\right].
\end{align}
The solution to \eqref{eq:Ldrhodt} can be formally written as:
\begin{align}
    \hat{\rho}_I (t) &= \hat{T} e^{\int^t_0\,d \tau\,\hat{\mathcal{L}}(\tau)} \hat{\rho}(0) = \hat{T} e^{\int^t_0\,d \tau\,(\hat{\mathcal{L}}_U(\tau)+\hat{\mathcal{L}}_D(\tau))}  \hat{\rho}(0) \\ \label{eq:rhoexpansion}
    &=\hat{T}(1 + \int^t_0\,d \tau\,(\hat{\mathcal{L}}_U(\tau)+\hat{\mathcal{L}}_D(\tau)) \\\nonumber&\hspace{1cm}+ \frac{1}{2!} \int^t_0\,d \tau_1\,d \tau_2\,(\hat{\mathcal{L}}_U(\tau_1)+\hat{\mathcal{L}}_D(\tau_1)) (\hat{\mathcal{L}}_U(\tau_2)+\hat{\mathcal{L}}_D(\tau_2))\\\nonumber
    &\hspace{1cm}+ \dots)\hat{\rho}(0) .
    \end{align}
The Schr\"{o}dinger picture density operator is then obtained from $\hat{\rho}(t) = \hat{U}_0 \hat{\rho}_I(t) \hat{U}^{\dagger}_0$. Applying this to the experiment in Section \ref{sec:exp}, the initial and final (e.g.\ before a reverse Stern-Gerlach) density operators will be  the density operator versions of \eqref{eq:PsiInitial} and \eqref{eq:stateAfterG}. Ignoring the spin states, the density operator can then be written as a $4 \times 4$ matrix, with the rows and columns labelled by $\{|N\rangle_{1L},|N\rangle_{1R},|N\rangle_{2L},|N\rangle_{2R}\}$  and $\{\,_{1L}{\langle} N|, \,_{1R}{\langle} N|, \,_{2L}{\langle} N|, \,_{2R}{\langle} N|  \}$ respectively. Given the final density operator $\hat{\rho}(t)$, we can obtain the different entries $\rho_{ij, k l}$ of the matrix by $\rho_{ij ,k l}(t) =  \,_{1i}{\langle} N| \,_{2j}{\langle} N|\, \hat{\rho}(t) \, | N\rangle_{1k}|N\rangle_{2l}$. As discussed in Section \ref{sec:QG}, $\hat{U}_0$ only acts a global phase, and thus we only need to consider this for the interaction picture density operator:  $\rho_{ij, k l}(t) =  \,_{1i}{\langle} N| \,_{2j}{\langle} N|\, \hat{\rho}_I(t) \, | N\rangle_{1k}|N\rangle_{2l}$.

To begin with we consider the first order term in \eqref{eq:rhoexpansion}. To determine the density matrix, we then need to calculate the contractions:
\begin{align}
     & \,_{1i}{\langle} N| \,_{2j}{\langle} N|\, \int^t_0\,d \tau\,c \hbar\,\hat{\mathcal{L}}_U(\tau) \hat{\rho}(0) \, | N\rangle_{1k}|N\rangle_{2l} \\&= \int_\tau d^4 x \,_{1i}{\langle} N| \,_{2j}{\langle} N|\, \hat{\Phi}(x) \hat{\Pi}(x)  \hat{\rho}(0) \, | N\rangle_{1k}|N\rangle_{2l}- \int_\tau d^4 x  \,_{1i}{\langle} N| \,_{2j}{\langle} N|\, \hat{\rho}(0) \hat{\Phi}(x) \hat{\Pi}(x)   | N\rangle_{1k}|N\rangle_{2l}\\
     &= \int_\tau d^4 x  \,_{1i}{\langle} N| \,_{2j}{\langle} N|\, \hat{\Phi}(x) \hat{\Pi}(x)   \, | N\rangle_{1i}|N\rangle_{2j}- \int_\tau d^4 x  \,_{1k}{\langle} N| \,_{2l}{\langle} N|\,  \hat{\Phi}(x) \hat{\Pi}(x)   | N\rangle_{1k}|N\rangle_{2l}
\end{align}
and
\begin{align}
     &\,_{1i}{\langle} N| \,_{2j}{\langle} N|\, \int^t_0\,d \tau\,c\hbar\,\hat{\mathcal{L}}_D(\tau) \hat{\rho}(0) \, | N\rangle_{1k}|N\rangle_{2l} \\
     &= \int_\tau d^4 x  \,_{1i}{\langle} N| \,_{2j}{\langle} N|\, \hat{\Phi}(x) \hat{\Pi}(x)   \, | N\rangle_{1i}|N\rangle_{2j} + \int_\tau d^4 x  \,_{1k}{\langle} N| \,_{2l}{\langle} N|\,  \hat{\Pi}(x) \hat{\Phi}(x)    | N\rangle_{1k}|N\rangle_{2l}\\
     &-\int_\tau d^4 x  \,_{1k}{\langle} N| \,_{2l}{\langle} N|\,  \hat{\Pi}(x) | N\rangle_{1k}|N\rangle_{2l}\, \times \,_{1i}{\langle} N| \,_{2j}{\langle} N|\, \hat{\Phi}(x) | N\rangle_{1i}|N\rangle_{2j}\\
     &-\int_\tau d^4 x  \,_{1k}{\langle} N| \,_{2l}{\langle} N|\,  \hat{\Phi}(x) | N\rangle_{1k}|N\rangle_{2l}\, \times \,_{1i}{\langle} N| \,_{2j}{\langle} N|\,  \hat{\Pi}(x) | N\rangle_{1i}|N\rangle_{2j},
\end{align}
where we have used the orthonormality of the states in determining what components   of $\hat{\rho}(0)$ contribute a non-zero result. We are now able to apply the contractions developed in Sections \ref{sec:QG} and \ref{sec:CG}. For example, 
\begin{align}
   &\int_\tau d^4 x  \,_{1i}{\langle} N| \,_{2j}{\langle} N|\, \hat{\Phi}(x) \hat{\Pi}(x)   \, | N\rangle_{1i}|N\rangle_{2j} \\&= \int_t d^4 x   \, \left(\Phi_{1i} (\bm{x}) + \Phi_{2j} (\bm{x}) \right)     \,_{1i}{\langle} \wick{\c1 N| \, \,_{2j}{\langle}  N| \hat{\mathcal{T}}[\c1{\hat{\phi}}^{\dagger} (x) \c4{\hat{\phi}}(x) ]    |\c4 N \rangle_{1i} \,|  N \rangle_{2j}}\\
   &+\int_t d^4 x   \, \left(\Phi_{1i} (\bm{x}) + \Phi_{2j} (\bm{x}) \right)   \wick{  \,_{1i}{\langle}  N| \, \,_{2j}{\langle} \c1 N| \hat{\mathcal{T}}[\c1{\hat{\phi}}^{\dagger} (x) \c4{\hat{\phi}}(x) ]    | N \rangle_{1i} \,| \c4 N \rangle_{2j}}.
\end{align}
Using the contractions defined in \ref{sec:QG}, we can then determine the evolution of the density matrix for the matter system. We find that, to first order in \eqref{eq:rhoexpansion}:
\begin{align}\nonumber
&\rho(t) =\footnotesize \rho(0) + \frac{1}{4}\times \\ \label{eq:rhot1storder}
&{\footnotesize \left( \begin{array}{l} 
0 , \hspace{25pt} i  \Delta U_{LRLL}  - E_{G2}  ,\hspace{25pt} i \Delta U_{RLLL} - E_{G1} ,\hspace{40pt} i  \Delta U_{RRLL}- E_{GT} +
\Delta U_{LLLR} + \Delta U_{RRRL} \\
 -i  \Delta U_{LRLL} - E_{G2}  ,\hspace{25pt}0,\hspace{25pt}i \Delta U_{RLLR} - E_{GT} + \Delta U_{LRLL} + \Delta U_{RLLL} ,\hspace{35pt} i  \Delta U_{RRLR} - E_{G1}\\-i  \Delta U_{RLLL} - E_{G1} 
 ,\hspace{25pt} -i \Delta U_{RLLR} - (E_{GT} + 
 \Delta U_{LRLL} + \Delta U_{RLRR}) ,\hspace{25pt} 0 ,\hspace{30pt} i \Delta U_{RRRL} - E_{G2} \\
-i  \Delta U_{RRLL} - E_{GT} \Delta U_{LLLR} + \Delta U_{RRRL}  ,\hspace{25pt} -i  \Delta U_{RRLR} - E_{G1}  ,\hspace{25pt}  -i  \Delta U_{RRLL} - E_{G2} ,\hspace{20pt} 0 
\end{array}\right) t},
\end{align}
where:
\begin{align}
\Delta U_{ijkl}&:= U_{ij} - U_{kl}\\
 U_{ij} &:= \frac{G M^2}{\hbar \,d_{ij}},\\
 E_{GT}&:=E_{G1} + E_{G2}\\
 E_{G\kappa} &:= - \frac{M}{\hbar V} \int d^3 \bm{x} \, \Phi_{\kappa L}(\bm{x}) \left(\theta_{\kappa L}(\bm{x}) - \theta_{\kappa R}(\bm{x})\right),
\end{align}
with  $U_{ij}$ the gravitational interaction energy between the masses $1$ and $2$ in states $i$ and $j$ respectively, and $E_{G \kappa}$ is the gravitational self-energy of the difference between the mass distributions of the two states of the solid object $\kappa$. The latter, which is the same for both objects $\kappa$ in our case since  the objects are identical (and so we denote $E_{G}$), is the usual rate of collapse in the Di\'{o}si-Penrose model for a single object in a superposition and is given by 
\cite{Howl_2019}:
\begin{equation} \label{eq:EGUniformSphere}
E_{G} = \begin{cases} \frac{6 G M^2}{5R} \Big(  \frac{5}{3} \lambda^2 - \frac{5}{4} \lambda^3 + \frac{1}{6} \lambda^5\Big)  & \text{if}~ 0 \leq \lambda \leq 1,  \\
	\frac{6G M^2}{5R} \Big( 1 -  \frac{5}{12 \lambda}\Big) &\text{if}~  \lambda \geq 1,\end{cases},
\end{equation}
where $\lambda = \Delta x/ (2 R)$.  The density matrix \eqref{eq:rhot1storder} is in fact the first order of the matrix one would obtain by non-perturbatively solving the Di\`{o}si-Penrose model (with the unitary Newtonian quantum gravity term)  for a non-relativistic version of Feynman's experiment \cite{LeastDecoh,Howl_2019,fn11}:
\begin{align}\nonumber
&\rho(t) =\frac{1}{4} \times \\ \label{eq:rhot}
&\left( \begin{array}{l} 
1 , \hspace{25pt}e^{i t \Delta U_{LRLL} } e^{- E_{G2} t} ,\hspace{25pt} e^{i t \Delta U_{RLLL}} e^{- E_{G1} t} ,\hspace{40pt} e^{i t \Delta U_{RRLL}} e^{- (E_{GT} +
\Delta U_{LLLR} + \Delta U_{RRRL}) t} \\
 e^{-i t \Delta U_{LRLL} } e^{- E_{G2} t/ \hbar} ,\hspace{25pt}1,\hspace{25pt}e^{i t \Delta U_{RLLR} } e^{- (E_{GT} + \Delta U_{LRLL} + \Delta U_{RLLL}) t} ,\hspace{35pt} e^{i t \Delta U_{RRLR}} e^{- E_{G1} t} \\e^{-i t \Delta U_{RLLL} } e^{- E_{G1} t}
 ,\hspace{25pt} e^{-i t\Delta U_{RLLR}} e^{- (E_{GT} + 
 \Delta U_{LRLL} + \Delta U_{RLRR}) t} ,\hspace{25pt} 1 ,\hspace{30pt} e^{i t \Delta U_{RRRL} } e^{- E_{G2} t/ \hbar} \\
e^{-i t \Delta U_{RRLL} } e^{- (E_{GT} \Delta U_{LLLR} + \Delta U_{RRRL} ) t} ,\hspace{25pt} e^{-i t \Delta U_{RRLR}} e^{- E_{G1} t/ \hbar} ,\hspace{25pt}  e^{-i t \Delta U_{RRLL} } e^{- E_{G2} t/ \hbar} ,\hspace{20pt} 1 
\end{array}\right).
\end{align}
This illustrates that the methodology developed in Sections \ref{sec:QG} and \ref{sec:CG}   can be successfully applied to this class of classical gravity models.

With $d_{RL} \ll \Delta x$ and $\Delta x \gg R$, we can approximate \eqref{eq:rhot} by:
\begin{align}\nonumber
&\rho(t) =\frac{1}{4} 
\left( \begin{array}{cccc} 
1 & e^{- E_{G} t} &  e^{- E_{G} t} e^{i U_{RL} t} &  e^{- (2 E_{G}  - U_{RL}) t} \\
 e^{- E_{G} t} &1&  e^{- (2 E_{G} + U_{RL}) t} e^{ i U_{RL} t}&  e^{- E_{G} t} \\ e^{- E_{G} t} e^{- i U_{RL} t}
 &  e^{- (2 E_{G}  + U_{RL}) t}e^{- i U_{RL} t} & 1 &  e^{- E_{G} t}e^{- i U_{RL} t} \\
 e^{- (2 E_{G} - U_{RL}) t} &  e^{- E_{G} t} &   e^{- E_{G} t} e^{i U_{RL} t}& 1 
\end{array}\right),
\end{align}
with $E_G = 6 G M^2 / (5 R)$. In each off-diagonal entry of the density matrix, we have a decoherence term coming through $\sigma_G := E_{G} t$. Since this is always greater than $\varphi_{RL} = U_{RL} t$, which contributes a quantum phase, no entanglement is generated here.  That is, although we appear to have the quantum gravity-induced phase $\varphi_{RL}$, any entanglement it could generate gets cancelled by the decoherence process $\sigma_G$. As long as we keep to first order in the entanglement measure, the same applies to the matrix we directly calculated \eqref{eq:rhot1storder} - no entanglement is generated in this case. For a discussion on how  Tilloy-Di\'{o}si models of gravity can  create entanglement (through non-locality), see \cite{trillo2024di,jzht-fbwt}.

However, although $\sigma_{G}$ is greater than $\varphi_{RL}$, it is possible for $\sigma_{G}$ to be much smaller than $|\beta^{(4)}_{RL}|$ and its quantum-gravity version $|\kappa^{(4)}_{RL}|$ - \eqref{eq:alpha4RL}. This is to be expected since, as described in the main text, $|\beta^{(4)}_{RL}|$ can be much greater than $\varphi_{RL}$, and  $\sigma_G$ is of the same form as $\varphi_{RL}$ (same dependence on mass, Planck's constant etc.), and can be made close to $\varphi_{RL}$. For example, taking ytterbium objects  with masses even as large as $M = 10\,\mathrm{mg}$  (and $d_{RL} \approx 10 R$), $|\kappa^{(4)}_{RL}|$ is greater than $\sigma_G$ for times larger than just $t \approx 10^{-19}\mathrm{s}$ (it is $\kappa^{(4)}_{RL} = 16 \times \beta^{(4)}_{RL}$ that is of relevance here rather than $\beta^{(4)}_{RL}$ since mathematically the density matrix evolves as if there is quantum gravity with decoherence, rather than semi-classical Einstein gravity with decoherence). Thus it is possible for the rate of entanglement generation from the virtual matter process to overcome the rate of decoherence $\sigma_{G}$. However, to determine if entanglement definitively occurs in this model, we need to see if there are any other decoherence processes that have a rate that is  always greater than or equal to $|\kappa^{(4)}_{RL}|$. For this to occur, such a process must have the same dependence on experimental parameters as  $|\kappa^{(4)}_{RL}|$, for example the mass $m$. Therefore, we need to  consider processes involving virtual matter propagators. 

The first candidate for such a decoherence process occurs at second order in the perturbation series \eqref{eq:rhoexpansion} such that we need to consider terms of the form:
\begin{align}\nonumber
      \hat{T} \frac{1}{2!} \,_{1i}{\langle} N| \,_{2j}{\langle} N|\, &\int^t_0\,d \tau_1\, \,d \tau_2\, \Big[\hat{\mathcal{L}}_U(\tau_1) \hat{\mathcal{L}}_U(\tau_2) + \hat{\mathcal{L}}_U(\tau_1) \hat{\mathcal{L}}_D(\tau_2) \\ \label{eq:2ndorderL}&+ \hat{\mathcal{L}}_D(\tau_1) \hat{\mathcal{L}}_U(\tau_2) + \hat{\mathcal{L}}_D(\tau_1) \hat{\mathcal{L}}_D(\tau_2)\Big]
     \hat{\rho}(0) \, | N\rangle_{1k}|N\rangle_{2l}.
\end{align}
However, since we are only interested in contractions involving virtual matter propagators, we only need to consider those terms where $\hat{\rho}(0)$ is on the far right, far left, or where there is one or two $\hat{\Phi}$ terms on the far right or left. For example, considering the term $\,_{1i}{\langle} N| \,_{2j}{\langle} N|\, \int^t_0\,d \tau_1\, \,d \tau_2\, [\hat{\mathcal{L}}_U(\tau_1) \hat{\mathcal{L}}_U(\tau_2)]
     \hat{\rho}(0) \, | N\rangle_{1k}|N\rangle_{2l}$, with $\hat{\rho}(0)$ on the far right, we have the contractions \eqref{eq:contVirtWithin} discussed in Section \ref{sec:CG} but with $\Phi(\bm{x})$ replaced with $\hat{\Phi}(x)$:
\begin{align}\nonumber
         &\int_t d^4 x \, d^4  y \, \,_{1i}\langle \wick{\c1 N| \, _{2j}{\langle}  N|   \hat{\Phi}(y)  \hat{\Phi}(x)  \hat{\mathcal{T}}[\c1{\hat{\phi}}^{\dagger}(y) \c2{\hat{\phi}} (y)]
      \hat{\mathcal{T}}[\c2{\hat{\phi}}^{\dagger} (x) \c3{\hat{\phi}} (x) ] |\c3 N \rangle_{1i} \,|  N} \rangle_{2j}\\ \label{eq:2ndorderBeta}
      &+\int_t d^4 x \, d^4  y \, \,_{1i}\langle \wick{ N| \, _{2j}{\langle} \c1 N|   \hat{\Phi}(y)  \hat{\Phi}(x)  \hat{\mathcal{T}}[\c1{\hat{\phi}}^{\dagger}(y) \c2{\hat{\phi}} (y)]
      \hat{\mathcal{T}}[\c2{\hat{\phi}}^{\dagger} (x) \c3{\hat{\phi}} (x) ] | N \rangle_{1i} \,| \c3 N} \rangle_{2j},
\end{align}
whereas, with $\hat{\rho}(0)$ on the left, we have the same as above but with $1i$ replaced with $1k$. As discussed in Section \ref{sec:CG},  the corresponding contractions to the above - \eqref{eq:contVirtWithin} - in the classical gravity models considered there,  only contribute a relative phase and thus no entanglement. Even so,  it is possible that  they could contribute towards a decoherence process that can destroy any entanglement generated through the $\kappa^{(4)}_{ij}$ process. However, we show below that this is not the case.

The  contractions in \eqref{eq:2ndorderBeta} involve those that look like the virtual matter equivalent to the $E_G$ process above - the dependence on mass is the same as $\sqrt{\kappa^{(4)}_{ij}}$, but there are terms where we are integrating the potentials over their own source. For example:
\begin{align}
         &\frac{1}{4 \hbar^2 c^2} \int_t d^4 x \, d^4  y \, \,_{1R}\langle \wick{\c1 N| \, _{2L}{\langle}  N|   \Phi_{1L}(\bm{y})  \hat{\Phi}_{1L}(\bm{x})  \hat{\mathcal{T}}[\c1{\hat{\phi}}^{\dagger}(y) \c2{\hat{\phi}} (y)]
      \hat{\mathcal{T}}[\c2{\hat{\phi}}^{\dagger} (x) \c3{\hat{\phi}} (x) ] |\c3 N \rangle_{1R} \,|  N} \rangle_{2L}\\ \label{eq:selfInt} &= \frac{m^3 N t}{2 \pi \hbar^3 V} \int d^3 \bm{x} d^3 \bm{y} \frac{\Phi_{1L}(\bm{x}) \Phi_{1L}(\bm{y}) \theta_{1L}(\bm{x}) \theta_{1L}(\bm{y})}{|\bm{x}-\bm{y}|}.
\end{align}
For all terms in \eqref{eq:2ndorderL}, for $\hat{\rho}(0)$ on the right, we get positive contributions from  the commutations, such that there is an overall factor of $(1-i)^4 = -2 i$ to the above process. In contrast, when $\hat{\rho}(0)$ is on the left, we get negative contributions with $\hat{\mathcal{L}}_U \hat{\mathcal{L}}_D$ and $\hat{\mathcal{L}}_D \hat{\mathcal{L}}_U$, giving an overall contribution of $2i$ for $\hat{\rho}(0)$ on the left. The terms from $\hat{\rho}(0)$ on the left and on the right then cancel.  This leaves the terms with one and two $\hat{\Phi}$ on the left and right. Following the same analysis as above, most terms cancel, leaving  contributions that combine to provide a factor of $-4 i$ to the right-hand side of \eqref{eq:selfInt}. This `self-interaction' contraction thus does not contribute towards a decaying (decoherence) term for the $\rho_{RL,LL}$ entry, and we find that it does not contribute a decoherence term for any entry of the density matrix. Therefore, the analogue of the $E_G$ process with virtual matter propagators does not generate  a decoherence process.

As well as the `self-interaction' type term, there will also be other terms from \eqref{eq:2ndorderBeta}. Of these, the greatest involve integrals of the form:
\begin{align}\label{eq:2ndorder}
\int d^3 \bm{x} d^3 \bm{y} \frac{\Phi_{1i}(\bm{x}) \Phi_{2j}(\bm{y}) \theta_{1i}(\bm{x}) \theta_{1i}(\bm{y})}{|\bm{x}-\bm{y}|} \propto \frac{1}{d_{ij}},
\end{align}
which have the same proportionality with respect to $d_{ij}$ as $\sqrt{\kappa^{(4)}_{ij}}$. In the assumption that $d_{RL} \ll \Delta x $, we only need consider contractions  involving $1R$ and $2L$. Although this process contributes towards the first order of an exponential decay for some of the density matrix entries, unlike $E_G$ above, it does not enter  all off-diagonal entries. This can be seen, for example, by considering any density matrix entry that does not contain both $1R$ and $2L$  - in this case   there is no process that contributes a $1/d_{RL}$ scaling, and so any processes for this entry can be neglected. Overall, this process then does not create decoherence - it does not suppress entanglement generation.  

Finally, we now consider the fourth order of the expansion \eqref{eq:rhoexpansion}. The processes at this order will clearly involve the entanglement process considered in Section \ref{sec:CG} (in fact, they will involve the quantum gravity version - see Appendix \ref{sec:QGVirtMatter}).   For a decoherence process to contribute the same dependence with experimental parameters as $\kappa^{(4)}_{ij}$, there needs to be  two virtual matter propagators at this order. We have already seen that `self-interaction' terms \eqref{eq:selfInt} at second order do not contribute a decoherence process, and thus these do not need to be considered at fourth order. This then leaves only those terms in \eqref{eq:rhoexpansion} at fourth order where $\hat{\rho}(0)$ is on the far left or right, and when there are $\Phi$, $\Phi^2$, $\Phi^3$ and $\Phi^4$ on the right and left. With the assumption that $d_{RL} \ll \Delta x$, we also only need consider terms that will result in the expectations $\,_{1R}{\langle}N| \,_{2L}{\langle}N|\cdots |N\rangle_{1R} |N\rangle_{2L}$. Clearly, as at second order, any such decoherence process will not affect all off-diagonal entries.   If, however, it contributed to the first order of a exponential decay in the entries of the density matrix where we have the $\beta^{(4)}_{RL}$ process generating entanglement in the classical gravity theories of Section \ref{sec:CG} (and Appendix \ref{sec:QGVirtMatter}), then we might expect entanglement to never be generated in the theory considered in this section. This, however, is not the case. For example, considering $_{1L}{\langle} N| \,_{2L}{\langle} N |\hat{\rho}_I(t) |N\rangle_{1R} |N\rangle_{2L}$ at  fourth order, and applying the arguments  above,  in the approximation $d_{RL} \ll \Delta x$, there is just the contribution $i \vartheta_{RL}$, which does not contribute to an exponential decay. 

In summary, there is no entanglement-destroying processes besides the $\sigma_G$ process up to fourth order where the entanglement generation occurs. This means that if  $|\kappa^{(4)}_{RL}| \gg \sigma_G$, then entanglement occurs and with a rate that can be $16$ times larger than for the classical gravity theories considered in Section \ref{sec:CG} and the main text, since $\kappa^{(4)}_{RL} = 16 \times \beta^{(4)}_{RL}$. That entanglement in general  occurs in these theories is not unexpected: the decoherence in \eqref{eq:DPrel} due to stochasticity is, mathematically, to prevent virtual gravitons from generating quantum communication (since the gravitational field is classical and thus there cannot be quantum communication from gravitons),  but there is no reason why there needs to be a decoherence process in these theories to, mathematically, prevent quantum communication from  virtual matter since this is present even when gravity is classical (see the main text). As long then as the virtual matter process dominates over the (mathematical)  graviton process, which is of course the regime of relevance, then the entanglement from virtual matter overcomes the decoherence  preventing the, mathematical, virtual graviton exchange. 

Similar to virtual graviton exchange in quantum gravity, the process we have considered here for generating entanglement   can be viewed as coming from  local, quantum communication. It is also possible, depending on the chosen stochastic classical gravity model, that there could be an additional non-local process generating entanglement, as argued in \cite{TilloyDiosi,trillo2024di,jzht-fbwt}. This is a very different process to the one we have considered here since it is not through virtual matter exchange, is non-local and can be viewed as  due to pre-existing quantum correlations between  fictitious detectors that are continuously monitoring  the matter distributions \cite{TilloyDiosi}, as well as potential issues associated with modelling subsystems due to the spatial resolution of the detectors \cite{TilloyDiosi}.  As argued in the main text, on physical grounds, such non-local processes would generally want to be avoided in a realistic model of nature \cite{oppenheim2018postquantum}.

\section{\large Non-local, classical theories of gravity} \label{sec:nonlocalCG}

 As mentioned in the main text, there have been  several works that consider  whether the theorems for how entanglement evidences quantum gravity are violated \cite{anastopoulos2018commentaspinentanglement,Andersen_2019,Anastopoulos_2021,doner2022gravitational,Fragkos2022,Eduardo2023,kent2024necessarilytreatmasseslocalized,franzmann2024bewhere,marchese2024newtonslawsmotiongenerate}. Often inspired by   discussions at the Chapel Hill conference where Feynman first introduced his experiment \cite{FeynmanQG}, these works have generally focused on whether  classical gravity could  act through non-local  operations, violating the LO part of LOCC, which was also discussed in the original works on entanglement evidencing quantum gravity \cite{kafri2013noise,bose2017spin,marletto2017gravitationallyinduced}. For example, there are arguments that, since the experiments operate in the low-energy regime of quantum gravity, gravity in the real-world must or could act as Newtonian gravity \cite{FeynmanQG,Anastopoulos_2021,huggett2022quantum,marchese2024newtonslawsmotiongenerate} (despite  experiments evidencing the relativistic nature of gravity \cite{christodoulou2022gravity}), with Newtonian gravity  non-local; that general relativity  is non-local since there are gauges in which it can appear this way  \cite{FeynmanQG,anastopoulos2018commentaspinentanglement,Anastopoulos_2021,kent2024necessarilytreatmasseslocalized}; that we cannot be sure that there is a gravitational field such that gravity could act as a non-local constraint or a `quantum-controlled' field  similar to the absorber theory of electromagnetism by Feynman and Wheeler (there are no independent gravitational degrees of freedom) \cite{Fragkos2022,Eduardo2023,PhysRevD.111.085005}; that classical gravity could act through  de Broglie–Bohm theory \cite{Andersen_2019,doner2022gravitational}, which is inherently non-local; that classical gravity can be associated with a continuous monitoring of matter that is non-local \cite{TilloyDiosi,trillo2024di,jzht-fbwt};   or that there is no true meaning to the concept of subsystems and thus locality \cite{kent2024necessarilytreatmasseslocalized,franzmann2024bewhere}. 

In the process considered in the main text, the gravitational field remains the same in each branch. This can be seen, for example, in \eqref{eq:beta4ijFull} where $\Phi(\bm{x})$ is the same in each branch. Instead it is  the function $\theta_{1 i}(\bm{x}) \theta_{2 j}(\bm{y})$ coming from the matter sector that is different in each superposition branch,  such that $\theta_{\kappa i}(\bm{x})$  could  be considered to act as an implicit quantum operator. However, the function $\theta_{\kappa i}(\bm{x})$ being different in each branch is not enough to create entanglement - see e.g.\  \eqref{eq:phiki} - we also need the $|\bm{x}-\bm{y}|$ term from the virtual matter propagator connecting the two $\theta_{\kappa i}(\bm{x})$ functions (matter sectors). That is, what is generating entanglement in the process we consider is the local virtual matter interactions of classical gravity.

\section{\large Consistency of fundamentally classical gravity} \label{sec:consistent}

The theoretical consistency of fundamentally classical gravity has been much debated. In particular, three different works \cite{eppley1977necessity}, \cite{PageExp} and \cite{GISIN19901} have had a major impact. In   \cite{PageExp}, an experiment was performed that was designed to demonstrate that semi-classical Einstein gravity, where gravity is sourced by the expectation value of quantum matter \cite{ROSENFELD1963353,moller1962theories}, should be ruled out. However, this is only possible for the most straightforward approach of  coupling the many-worlds interpretation of quantum theory to classical gravity \cite{Kent_2018} and, therefore, the experiment does not rule out semi-classical Einstein gravity in general. Ref.\ \cite{eppley1977necessity}, argues that, in a fundamental theory of classical gravity, and in the Copenhagen interpretation of quantum mechanics \cite{Huggett_Callender_2001}, if a gravitational wave interacts with quantum matter then either the Heisenberg uncertainty relations must be violated, momentum conservation is violated or we must have superluminal signalling. Whether a violation of the  Heisenberg uncertainty relations or momentum conservation should rule out a theory is up for debate, however, in any case, it has been demonstrated \cite{mattingly2006eppley} that the argument for violation of the Heisenberg uncertainty relations (which is based on Heisenberg's controversial  `observer effect' interpretation of his relations \cite{RevModPhys.42.358,PhysRevLett.109.100404}) and momentum conservation in  \cite{eppley1977necessity} is false, and therefore the arguments in   \cite{eppley1977necessity} do not prove that fundamentally classical gravity leads to a violation in either the Heisenberg uncertainty relations, momentum conservation or the no-superluminal signalling principle of quantum mechanics. Furthermore,  \cite{mattingly2006eppley} showed that physically demonstrating any violation of the Heisenberg uncertainty relations, momentum conservation or no superluminal signalling as suggested in   \cite{eppley1977necessity}, would  not be physically possible such that, even if there were a violation in principle, it would not show up in practice. 

In  \cite{GISIN19901}, it was argued that  non-linear modifications to quantum mechanics generally lead to superluminal signalling. Fundamental classical gravity theories where gravity is sourced by the expectation of matter (without stochastic fluctuations) are examples of  non-linear modifications to quantum mechanics (see Appendix \ref{sec:linear}) - for example, see \cite{Bahrami_2014} for a discussion on how superluminal signalling could arise in this way in the Schrödinger–Newton equation. However,  \cite{kent1998causality} demonstrated that superluminal signalling does not necessarily lead to a contradiction, and it is also possible to avoid any signalling by generalizing the sourcing of gravity such that the expectation  value is taken over `local' states of matter \cite{Kent_2018,kent2005nonlinearity,Giulini2023}, where wavefunction collapse essentially becomes a real and relativistic process in the gravity sector \cite{Kent_2018,kent2005nonlinearity,Giulini2023,Helou_2017}, or self-consistently generalizing the measurement and states of the theory \cite{mielnik2000comments}.   It is also unclear if the superluminal signalling could be physically demonstrated \cite{mattingly2006eppley}. 

In the main text, we assumed semi-classical Einstein gravity (Schrödinger-Newton theory) where the Newtonian potential is sourced by the expectation of the matter states (see also Section \ref{sec:CG}). Using a perturbative analysis, we showed that entanglement can occur in this theory for versions of Feynman's experiment. In Appendix \ref{sec:linear}, we demonstrated that in this perturbative regime, the theory is linear. Therefore, the entanglement result is not coming from a superluminal signalling process, which would question the locality and significance of the effect. Furthermore, as detailed in Appendix \ref{sec:semiPhi}, the analysis is unchanged if we adopt the Newtonian potential as being sourced by `local' matter states \cite{Kent_2018,kent2005nonlinearity} or where wavefunction collapse is a relativistic process \cite{Helou_2017}  such that superluminal signalling can clearly not  occur \cite{kent2005nonlinearity,Helou_2017}. This is because, in determining whether the quantum systems become entangled, we only need consider the time evolution of the quantum system (due to the classical gravity interaction between the objects) after it is sent, for example, through the forward Stern-Gerlach devices and before  reverse devices, and no measurement processes occur during this period of Feynman's experiment. It is only at the end of the experiment that a measurement is performed, by which point the matter states are no longer in a superposition and entanglement can be transferred, for example, to the internal spin sector \cite{bose2017spin}. The experiment is also far from the regime of   \cite{eppley1977necessity} where gravitational waves are scattered off masses and then precisely analysed (indeed, as argued in \cite{mattingly2006eppley}, no physical experiment may be in this regime). 

Without updating the states or measurement process \cite{mielnik2000comments,kent2005nonlinearity,Helou_2017}, we can see how a superluminal signalling could, in theory and in principle, be  created using (a slightly modified version of) Feynman's experiment due to non-linearity, following a similar discussion in \cite{Bahrami_2014}. Consider that  two matter objects are sent through the Stern-Gerlach devices  so that they are in the superposition state as \eqref{eq:PsiInitial} in the main text:
\begin{align}\label{eq:psiAB}
    |\Psi\rangle = \frac{1}{2}  \left(|N\rangle_{1L} | \uparrow\rangle_1 + |N\rangle_{1R}  |\downarrow\rangle_1\right) 
    \otimes  \left( |N\rangle_{2L}   |\uparrow\rangle_2 +  |N\rangle_{2R} |\downarrow\rangle_2\right).
\end{align}
Immediately after this state is created, we assume that the two matter objects become  entangled through, for example, an added electromagnetic process (or potentially the classical gravity process considered in Section \ref{sec:CG}), placing them in the maximally entangled state:
\begin{align}\nonumber
    |\Psi_-\rangle = \frac{1}{2}  \Big(&|N\rangle_{1L} | \uparrow\rangle_1  |N\rangle_{2L}   |\uparrow\rangle_2 + |N\rangle_{1L} | \uparrow\rangle_1  |N\rangle_{2R}   |\downarrow\rangle_2\\
    &- |N\rangle_{1R}  |\downarrow\rangle_1 |N\rangle_{2L}   |\uparrow\rangle_2 + 
    |N\rangle_{1R}  |\downarrow\rangle_1
    |N\rangle_{2R} |\downarrow\rangle_2\Big).
\end{align}
Alice now looks after the matter object on the left, and Bob the one on the right. Alice immediately sends her object through the reverse Stern-Gerlach experiment, whereas Bob leaves his object alone, resulting in:
\begin{align}\nonumber
    |\Psi_-\rangle &= |N\rangle_{1C} \Big[ | \uparrow\rangle_1\, \frac{1}{2} \Big( \left(|N\rangle_{2L}   |\uparrow\rangle_2+ |N\rangle_{2R} |\downarrow\rangle_2\right)\Big) + | \downarrow\rangle_1\, \frac{1}{2} \Big(  - |N\rangle_{2L}   |\uparrow\rangle_2 + |N\rangle_{2R} |\downarrow\rangle_2\Big)\Big]\\\nonumber
    &\equiv|N\rangle_{1C} \frac{1}{\sqrt{2}}\Big[ | \leftarrow\rangle_1 |N\rangle_{2L}   |\uparrow\rangle_2 + | \rightarrow\rangle_1 |N\rangle_{2R}   |\downarrow\rangle_2\Big],
\end{align}
where $|\uparrow\rangle,|\downarrow\rangle = \frac{1}{\sqrt{2}} (|\rightarrow\rangle \pm |\leftarrow\rangle)$. Before the experiment, Alice tells Bob that as soon  as her object goes through the reverse Stern-Gerlach, she will perform a spin measurement in either the $\{\uparrow,\downarrow\}$ basis or the $\{\rightarrow,\leftarrow\}$ basis.  She will then immediately communicate to Bob which basis she decided on. Bob then performs a position measurement on his object  at a time $t_B$ just before the signal from Alice arrives. 

If Alice performs her measurement in the $\{\uparrow,\downarrow\}$ basis, then Bob's object is in  a superposition of left and right and, in the case that gravity is simply sourced by the quantum expectation of matter (and within the full non-perturbative, non-linear regime),  these states of matter will be `attracted' towards each other through a classical  gravitational force, as if there were  physically two matter objects at the left and right  positions - see $\Phi_{C2} (\bm{x})$ in \eqref{eq:semiclassicalPhi}. This then results in a change to the positions of the two states. If, on the other hand, Alice performs her measurement in the $\{\rightarrow,\leftarrow\}$ basis, then Bob's matter object will no longer be in a superposition and the gravitational potential is just that of an object at either the left or right position (depending on Alice's measurement result), such that there is no movement of the object. Therefore, by measuring the position of his matter object and checking if the object has shifted in position, Bob is, in principle, able to detect  which measurement basis Alice chose, and  before he receives Alice's signal.  Furthermore, although we are working in the non-relativistic limit of gravity here, this would still apply if we had taken a relativistic stance since wavefunction collapse is `instantaneous' in `standard' quantum mechanics \cite{Bahrami_2014}. 

Since the gravitational force between the states is small, we can approximate the  distance  which Bob's object moves  by $\delta x = G M t_B^2/(\Delta x)^2$, where we have assumed that the time at which Alice performs her measurement is not long after the state \eqref{eq:psiAB} has been created. If $t_B$ is just before the signal from Alice reaches Bob, we can take $t_B \approx d/c$, where $d$ is the distance between the centre of the two masses $1$ and $2$, resulting in  $\delta x \approx G M d^2/( c \Delta x)^2$. Taking some example experimental values \cite{bose2017spin},  $d = 450\,\mathrm{\mu m}$, $\Delta x = 250\,\mathrm{\mu m}$ and $M = 10^{-14}\,\mathrm{kg}$, we find $\delta x \approx 10^{-41}\,\mathrm{m}$. This is smaller than the Planck length and clearly beyond current or foreseeable technological abilities. It may also be theoretically impossible to observe \cite{mead1964possible}, pointing to the idea that there may be a principle that forbids using classical gravity to perform superluminal signalling \cite{mattingly2006eppley}. Here, we assumed that the Newtonian potential is being sourced by the expectation of the quantum matter states. If instead, we chose them to be  sourced by the expectation of the `local matter states' \cite{kent2005nonlinearity} then we are effectively making the measurement process for gravity a relativistic process \cite{Helou_2017}. That is, when Alice performs her measurement, this no longer immediately updates the gravitational potentials $\Phi_{\kappa i}(\bm{x})$ because they are now defined as coming from the expectation of the matter states that are only updated by measurement processes in the past light cone of the matter objects: 
\begin{align}\label{eq:localPhi}
  \Phi_{\kappa}(\bm{x}) = - \frac{G}{c^2} \int d^3 \bm{y} \frac{\mathrm{Tr} \left(\rho^{(\kappa)}_{loc}\,\hat{T}_{00} (\bm{y}) \right)}{|\bm{x}-\bm{y}|}, 
\end{align}
where $\rho^{(\kappa)}_{loc}$  is defined by taking the joint state of the system but allowing for only measurements in the past light cone of  object $\kappa$ and then tracing out the other object \cite{kent2005nonlinearity,Kent_2018}. In this case, irrespective of the measurement basis  Alice uses, the effective gravitational potential of Bob's object remains the same up until time $d/c$, and so it  feels a force and is displaced from its original position independent of Alice's measurement basis (up until time $d/c$).  Therefore, even in principle, Alice and Bob are not able to perform superluminal signalling.  As argued above, using a relativistic wavefunction collapse prescription in the gravity sector does not change the results of Section \ref{sec:CG} since  no measurement is performed before the matter states are brought back together, e.g.\ with  reverse Stern-Gerlachs as in   \cite{bose2017spin}. Measurements are only carried out at the end of the experiment to determine if the objects are entangled. In considering local matter states, we have assumed that it is only in the gravity sector that the collapse process is relativistic (it only changes how we update the gravitational potentials), and so  this does not affect the final measurements on the spins and the deduction of whether the objects are entangled. It could be possible to extend this so that the collapse process is always relativistic such that we have causal quantum theory \cite{kent2005nonlinearity,kent2018testing}. In this case, the final measurements would have to be performed outside of each others' light cones to test for entanglement, since collapse is now always a relativistic process and not just in the gravity sector. There are questions over whether casual quantum theory, where normal quantum mechanics is updated with a relativistic collapse process, is consistent with  current experiments \cite{kent2018testing}. In contrast, no experiment has tested relativistic collapse in just the gravity sector \cite{kent2005nonlinearity,Kent_2018,Helou_2017}. Yet another option to avoid superluminal signalling, at least in the above example, would be  to  define the potentials of each object in terms of just the standard reduced states of the respective objects rather than the full state vector: 
\begin{align} \label{eq:reducedPhi}
  \Phi_{\kappa}(\bm{x}) = - \frac{G}{c^2} \int d^3 \bm{y} \frac{\mathrm{Tr} \left(\rho_{\kappa}\,\hat{T}_{00} (\bm{y}) \right)}{|\bm{x}-\bm{y}|},  
\end{align}
where $\rho_\kappa := \mathrm{Tr}_{\lambda \neq \kappa} (\rho)$ with $\kappa,\lambda \in {1,2}$. Then there is no need to introduce a `relativistic' version of collapse.

The stochastic theories of classical gravity considered in Appendix \ref{sec:fundDec} are linear theories and so the arguments for superluminal signalling processes occurring in non-linear theories do not apply. These theories, however,  tend to come with violations of energy conservation, which is not considered a theoretical inconsistency unless there is experimental evidence to rule this out \cite{donadi2021underground,oppenheim2023gravitationally}. Interestingly, the consistency of these theories in explaining 
simple classical observations, such as spacecrafts undergoing slingshot manoeuvrers, has also been questioned recently by considering scattering processes in these theories \cite{carney2024classical}. Furthermore, to avoid infinite divergences, the decoherence processes involved in the theories also  need to effectively `smear' matter \cite{TilloyDiosi,layton2023weak} (there is continuous monitoring of a smeared mass density), which, together with quantum correlations of fictitious detectors,  can lead to non-local processes \cite{TilloyDiosi,layton2023weak,trillo2024di,jzht-fbwt}. However, as discussed in Appendix \ref{sec:fundDec}, when discounting these non-local effects, the virtual matter process considered here still results in entanglement. There is also a debate on the full-consistency of a relativistic stochastic theory due to potential open problems associated with reconciling Markovian decoherence and diffusion with relativity \cite{diosi2024classicalquantumhybridcanonicaldynamics,PhysRevD.106.L051901,grudka2024renormalisation}.

We should also contrast open problems with classical theories of gravity with those of  quantum gravity. The historical approach to quantizing the other interactions, electromagnetism and  the weak and strong interactions (perturbative QFT), fails with gravity, leading to a non-renormalizable theory. This full theory is thus self-inconsistent. However, at low energy scales we can use the prescription of effective QFT, and all realistic theories of quantum gravity  are thought to approximate perturbative quantum gravity in such a regime. This  was used in Section \ref{sec:QG} since the experiment is very much working within the low-energy regime. There is a  connection here with classical gravity: although it has been  argued that certain fundamental classical gravity theories involve potential self-inconsistencies, such as superluminal signalling, we have seen how these effects are beyond the regime of the experiment and do not affect physical predictions in the regime of  the experiment. Currently, there are many ideas and theories for a fully self-consistent quantum gravity theory, with no general consensus on what the true quantum gravity theory should be, if indeed gravity is quantized. The two most prominent theories are String Theory and Loop Quantum Gravity. As with classical gravity theories, the correctness of these theories as theories of  nature, has also been questioned. For example, String Theory has received questions on the difficulty of formulating a fully background-independent version, and the prediction of extra dimensions \cite{PenroseFashion,smolin2007trouble,woit2006not,schwarz1997status}. Similarly, question marks over Loop Quantum Gravity include a  lack of clear emergence of classical general relativity, and the uniqueness of the theory \cite{Nicolai_2005}. 

\subsection{`Local' semi-classical Einstein gravity} \label{sec:semiPhi}

As discussed above, there is a debate on whether traditional semi-classical Einstein gravity is inconsistent  due to the possibility of introducing  superluminal signalling processes. This is known to occur in non-linear modifications of quantum mechanics, which  traditional semi-classical Einstein gravity is an example of. However, since  our results are within the linear regime of the theory, as discussed in the previous section, such processes do not contribute to the entanglement effect we are considering in Section \ref{sec:CG}, which would question the locality of the process. It was also shown   that such superluminal signalling processes are far beyond the regime of future experiments (and in fact may not be possible in any physical experiment \cite{mattingly2006eppley}). However,  as we saw above, it is also possible to avoid superluminal signalling processes outright by small modifications to traditional semi-classical Einstein gravity. Here, we discuss in more detail how the results of Section \ref{sec:CG} are unchanged if we modified traditional semi-classical Einstein gravity to avoid superluminal signalling by (i) making the collapse postulate relativistic \cite{Helou_2017,kent2005nonlinearity}, (ii) making the collapse postulate relativistic in just the gravity sector \cite{Helou_2017}, (iii) sourcing gravity from `local' states of matter \cite{kent2005nonlinearity,Kent_2018},  or (iv) souring gravity from the reduced density matrices of matter.

It is clear that the results are unmodified by options (i) and (ii) since no measurement process occurs for the period of time where we analyse whether entanglement is created in  Section \ref{sec:CG}. Measurements only occur at the end of (many runs of) the experiment to determine whether there is entanglement between the objects. In the case of (i), this would mean that the experimentalist must make sure to perform the measurements on the objects inside each other's light cones \cite{kent2018testing} such that there is sufficient time for the measurement result of one object to `propagate' to the other.  This is  not an issue for future experiments since this level of control will be  outside current technology \cite{LocallyHowl}. For option (ii), no such requirement on the final measurements is needed since at the end of the experiment the entanglement is transferred solely to the spin sector and the masses are not in position superposition states.  

We now  discuss option (iv). In traditional semi-classical Einstein gravity, the Newtonian potential is defined by \eqref{eq:semiclassicalPhi}, where $\hat{T}_{00} = \hat{\pi}  \hat{\pi}^{\dagger} - \partial_i \hat{\phi} \partial^i \hat{\phi} + m^2 c^2 \hat{\phi}^{\dagger} \hat{\phi} / \hbar^2$. As described in Appendix \ref{sec:linear}, we only need consider the initial state $|\psi\rangle$ in calculating $\Phi(\bm{x})$, which is given by \eqref{eq:PsiInitial}  in the main text (\eqref{eq:psiAB} above). That is, using the orthonormality of the states, we have:
\begin{align}\label{eq:avT00}
    \langle \hat{T}_{00} \rangle = \frac{1}{4} \sum_{i,j} \,_{1i}{\langle} N| \,_{2j}{\langle} N| \hat{T}_{00} |N\rangle_{1i} |N\rangle_{2j}. 
\end{align}
Then, given  the non-relativistic approximation $R \gg \hbar / (mc)$, we can use \eqref{eq:WickN} (without the time dependence)  and discount the spatial derivatives in $\hat{T}_{00}$. The first and third terms in $\hat{T}_{00}$ then contribute the same and we end up with: 
\begin{align}\label{eq:avT00}
    \langle \hat{T}_{00} \rangle =  \frac{1}{2} \frac{2 m^2 c^2}{\hbar^2} \sum_{\kappa,i} \,_{\kappa i}{\langle}  \wick{ \c1 N| \c1{\hat{\phi}}^{\dagger} \c2{\hat{\phi}} |\c2  N\rangle_{\kappa i} }. 
\end{align}
Plugging in  the contractions \eqref{eq:WickN}, we have \eqref{eq:semiclassicalPhi}:
\begin{align} 
    \Phi(\bm{x}) = \Phi_{C1}(\bm{x}) + \Phi_{C2}(\bm{x}),
\end{align}
where
\begin{align} 
    \Phi_{C\kappa}(\bm{x}) &:= \frac{1}{2} \left(\Phi_{\kappa L}(\bm{x}) + \Phi_{\kappa R}(\bm{x})\right),
\end{align}
which is the result expected for traditional semi-classical Einstein gravity - the potential for each matter object is the average of its potentials for the two superposition states. 

In option (iv), we replace the definition of $\Phi(\bm{x})$ with $\Phi(\bm{x}) = \Phi_{1}(\bm{x}) + \Phi_{2}(\bm{x})$ where $\Phi_{\kappa}(\bm{x})$ is given by \eqref{eq:reducedPhi} above. Using  \eqref{eq:WickN} again  and  $R \gg \hbar / (mc)$, we end up with the same expression for $\Phi(\bm{x})$  as in traditional semi-classical Einstein gravity. This is easy to see  from the fact that the initial state is a product state: $|\psi\rangle = |\psi\rangle_1 \otimes |\psi\rangle_2$ and so $\mathrm{Tr}(\rho_1 \hat{T}_{00}) + \mathrm{Tr}(\rho_2 \hat{T}_{00}) = \, _{1}{\langle} \psi |  \hat{T}_{00} |\psi\rangle_1 + \, _{2}{\langle} \psi |  \hat{T}_{00} |\psi\rangle_2$. Here, $|\psi\rangle_{\kappa}$ is the state vector for object $\kappa$, i.e. $|\psi\rangle_{\kappa} = \frac{1}{\sqrt{2}} (|N\rangle_{\kappa L} + |N\rangle_{\kappa R})$. Then, from the orthonormality of the states, we clearly end up with \eqref{eq:avT00} above and thus \eqref{eq:semiclassicalPhi}. In general, in fact, since $_{\kappa i}{\langle} N  |\hat{T}_{00} | N \rangle_{\kappa j} = 0$ for $i\neq j$, we should expect the only difference between the definitions of $\Phi(\bm{x})$ for this `local' case and traditional semi-classical Einstein gravity to occur when there is a measurement process, and as described above, this only occurs right at the end of the experiment and so does not affect  the entanglement calculation in Section \ref{sec:CG}. Option (iii) is to change the definition of $\Phi_{\kappa}(\bm{x})$ to be \eqref{eq:localPhi}.  The  only difference is to trace out all measurement outcomes outside of the object's light cone. Therefore, since there is no measurement until the end of the experiment, again, nothing changes for the calculation in Section \ref{sec:CG}. 

It is not surprising that the entanglement effect is the same in traditional semi-classical Einstein gravity as with  the above modifications since the entanglement process is not associated with a superluminal signalling effect, and is in fact a local process.  Furthermore, the entanglement is not coming from the form of $\Phi(\bm{x})$, which cannot be in a superposition in classical gravity, and instead from the fact that there is a superposition of virtual matter propagators from the classical gravity interaction.

\section{Alternative signatures of quantum gravity to entanglement} \label{sec:alt}

That there are ways for entanglement to be generated from non-quantum and local theories of gravity raises the question of whether there are any signatures that can only ever be generated by quantum gravity theories. Alternative signatures to entanglement for evidencing quantum gravity include a measurement inequality \cite{Lami:2023gmz}, but since this inequality derives from considering classical gravity as LOCC, the virtual matter process considered here would also be expected to violate the inequality in general. Another alternative that has been considered is non-Gaussianity or Wigner negativity \cite{howl2021nongaussianity}. This is thought not to be based on LOCC and instead relies on the idea that, when the gravitational field has no associated quantum operator, the theory preserves Gaussianity since the Hamiltonian is quadratic in quantum field matter operators. Since the entangling virtual matter process considered in the main text  derives from a Hamiltonian that is quadratic, increases in non-Gaussianity could be used as a signal of quantum gravity in this context. However, the work here demonstrates that we do not need general proofs that certain signatures can only ever be associated with quantum gravity, we just need \emph{strong evidence}, as discussed in the main text.

\end{document}